\documentclass[acmlarge]{acmart}
\AtBeginDocument{%
  }

\usepackage{booktabs} 
\usepackage{subfig}
\usepackage{color}
\usepackage[ruled]{algorithm2e} 
\usepackage{graphicx}
\usepackage{caption}
\usepackage{wrapfig}
\usepackage{url}
\usepackage{enumerate}
\usepackage{gensymb}
\usepackage{color}
\usepackage{amsmath}
\usepackage{dcolumn}
\usepackage{multirow}
\usepackage{cancel}

\usepackage{enumitem}

\usepackage{balance}
\usepackage{stfloats}

\setcopyright{none}
\renewcommand\footnotetextcopyrightpermission[1]{}
\def\systemname{PnPSelect}
\begin{document}

\title{\systemname: Plug-and-Play IoT Device Selection Using Ultra-Wideband Signals}

\author{Zhaoxin Chang}
\affiliation{%
  \institution{Institut Polytechnique de Paris}
  \department{SAMOVAR, Telecom SudParis}
  \city{Palaiseau}
  \country{France}
}
\email{zhaoxin.chang@telecom-sudparis.eu}

\author{Fusang Zhang}
\affiliation{%
  \institution{Beihang University and Institute of Software, Chinese Academy of Sciences}
  \city{Beijing}
  \country{China}}
\email{fusang@iscas.ac.cn}

\author{Jie Xiong}
\affiliation{%
  \institution{Nanyang Technological University}
  \city{Singapore}
  \country{Singapore}
}
\email{jie.xiong@ntu.edu.sg}

\author{Ziyu Li}
\affiliation{%
 \institution{Institut Polytechnique de Paris}
 \city{Palaiseau}
 \country{France}}
\email{liziyu0104@gmail.com}

\author{Badii Jouaber}
\affiliation{%
  \institution{Institut Polytechnique de Paris}
  \department{SAMOVAR, Telecom SudParis}
  \city{Palaiseau}
  \country{France}}
\email{badii.jouaber@telecom-sudparis.eu}

\author{Daqing Zhang}
\affiliation{%
  \institution{Institut Polytechnique de Paris}
  \department{SAMOVAR, Telecom SudParis}
  \city{Palaiseau}
  \country{France}}
\affiliation{%
  \institution{Peking University}
  \city{Beijing}
  \country{China}}
\email{daqing.zhang@telecom-sudparis.eu}

\renewcommand{\shortauthors}{Chang et al.}

\begin{abstract}
In recent years, the number of Internet of Things~(IoT) devices in smart homes has rapidly increased. A key challenge affecting user experience is how to enable users to efficiently and intuitively select the devices they wish to control. This paper proposes \systemname, a plug-and-play IoT device selection solution utilizing Ultra-Wideband (UWB) technology on commercial devices. Unlike previous works, \systemname{} does not require the installation of dedicated hardware on each IoT device, thereby reducing deployment cost and complexity, and achieving plug-and-play functionality. To enable intuitive device selection, we introduce a pointing direction estimation method that utilizes UWB readings from a single anchor to infer the user’s pointing direction. Additionally, we propose a lightweight device localization method that allows users to register new IoT devices by simply pointing at them from two distinct positions, eliminating the need for manual measurements. We implement \systemname{} on commercial smartphones and smartwatches and conduct extensive evaluations in both controlled laboratory settings and real-world environments. Our results demonstrate high accuracy, robustness, and adaptability, making \systemname{} a practical and scalable solution for next-generation smart home interactions.
\end{abstract}



\thanks{This work is supported by the European Union through the Horizon EIC Pathfinder Challenge project SUSTAIN (No. 101071179) and the Innovative Medicines Initiative 2 Joint Undertaking project IDEA-FAST (No. 853981).}


\maketitle

\section{Introduction}
With the rapid advancement of IoT technology, an increasing number of devices can now be connected to the Internet. A prominent example is smart home appliances, such as lights, air conditioners, and TVs, which enable remote control and automation. This trend has revolutionized how we interact with our surroundings, enhancing convenience, comfort, and efficiency in daily life. Unlike traditional devices that are typically controlled individually (e.g., through remote controls and switches), IoT devices within the same environment can be managed through a unified user interface. Currently, in many smart home environments, smartphones or voice assistants serve as primary interfaces for controlling IoT devices. For instance, users often select devices from a list displayed on a smartphone. However, this method can be time-consuming and inconvenient, particularly as the number of smart devices increases and similar devices (e.g., multiple lamps in the same room) become more common. On the other hand, voice assistant-based solutions offer convenience but face challenges such as limited working range, disruption in quiet environments, and susceptibility to noise interference. For example, voice-based interaction requires users to speak commands aloud, which can be disruptive or socially inappropriate in quiet environments such as bedrooms, offices, or meeting rooms.

Recent research has explored alternative interaction methods that utilize common household devices, such as smartphones, smartwatches, and earphones. These methods allow users to select an IoT device by looking at or pointing toward it, offering a more intuitive and user-friendly experience. Several solutions~\cite{de2016snap,chen2018snaplink,mayer2020enhancing,qin2023selecting} leverage smartphone cameras for IoT device selection, using computer vision (CV) algorithms to recognize objects in the device’s field of view~(FoV). However, camera-based methods may struggle in low-light conditions and are limited to recognizing devices present in predefined training datasets. Moreover, distinguishing between devices of the same type (e.g., two identical lamps in different locations) remains challenging for deep-learning-based CV algorithms~\cite{strecker2023mr}.

\begin{figure}[!b]
    \centering
     \includegraphics[width=2.7in]{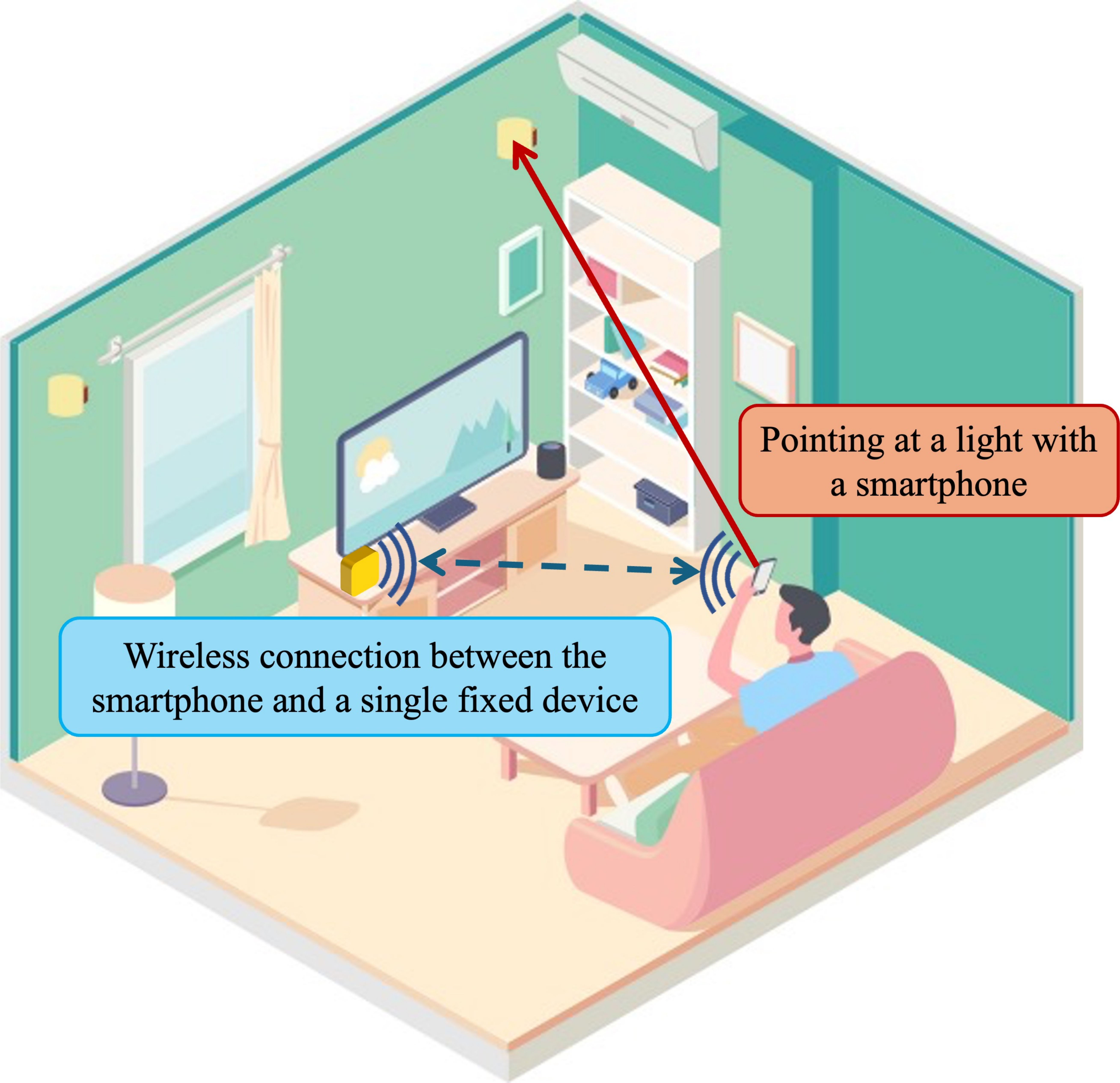}
    \caption{The envisioned IoT device selection application scenario.}
    \label{fig:fig1_1}
\end{figure}

Recent research has taken a different approach by leveraging information from wireless signals for IoT device selection. The key insight behind wireless signal-based solutions is that, by installing dedicated wireless modules on each IoT device, a wireless channel can be established between the user device and IoT devices. The device being pointed at or looked at can then be determined by extracting signal features such as Time-of-Flight (ToF) and Angle-of-Arrival (AoA). For example, ToF estimation allows the system to measure distance variations between the user device and IoT devices during a pointing gesture. The device that exhibits the most significant distance change is likely the one being pointed at (Figure~\ref{fig2_1a}). Similarly, AoA measurements help determine the direction of IoT devices relative to the orientation of the user device, allowing the system to identify the selected device based on which AoA value is closest to zero (Figure~\ref{fig2_1b}). Unlike camera-based approaches, wireless signal-based solutions maintain high accuracy regardless of lighting conditions. Additionally, similar devices can be distinguished more easily because the user device establishes a distinct wireless link with each IoT device. Various wireless technologies, including UWB~\cite{alanwar2017selecon}, acoustic signals~\cite{wang2022faceori}, and Bluetooth~\cite{zhang2023bleselect}, have been explored for this purpose. While promising, a significant limitation of existing wireless signal-based solutions is the requirement that each IoT device be equipped with a dedicated wireless module, such as a UWB transceiver or a Bluetooth module. However, most smart home appliances do not currently include UWB transceivers, and Bluetooth-based solutions require compatibility with specific Bluetooth protocols (e.g., Bluetooth 5.1), which limits applicability. Therefore, the requirement for each IoT device to be equipped with dedicated hardware presents a major barrier to achieving a true plug-and-play solution for real-world deployments.

In this paper, we propose PnPSelect, an IoT device selection solution that addresses this limitation, bringing wireless signal-based systems one step closer to real-world deployment. Specifically, we ask a question: \textit{\textbf{Can multiple IoT devices be selected using only a single dedicated wireless device?}} Figure~\ref{fig:fig1_1} illustrates the envisioned application scenario. The user can simply point at the IoT device they wish to control using their device (e.g., a smartphone). By analyzing the connection between the user device and a single wireless device, the pointing direction of the user can be estimated to select a specific IoT device. Compared to existing solutions, this approach significantly reduces deployment costs and complexity. To realize this vision, the first step is to determine the most suitable wireless signal for device selection. While various wireless technologies (Wi-Fi, Bluetooth, 4G/5G, and UWB) are available on commodity devices, most are primarily designed for communication purposes. In contrast, UWB technology is inherently designed for precise ranging, and using it for device selection does not interfere with its original function. Notably, UWB modules are now integrated into a wide range of consumer devices, including those from Apple, Google, Huawei, Samsung, and Xiaomi. Therefore, we choose to utilize UWB signals available on commercial devices for IoT device selection.
 
Achieving IoT device selection with a single UWB device in the environment presents several challenges. The first one is estimating the pointing direction of the user device. Since IoT devices are not equipped with UWB transceivers, the user device lacks a direct connection to them and cannot rely on conventional ToF or AoA measurements. Instead, we leverage a key observation: when a user performs a pointing gesture, UWB signals can be used to track the movement of their device. By analyzing the trajectory of this movement, we can infer the pointing direction. Building on this insight, we develop an algorithm that estimates the pointing direction using UWB readings from a single anchor. The second challenge is mapping the estimated pointing direction to specific IoT devices, which requires knowledge of their locations relative to the anchor. A straightforward approach would be to manually measure the position of each IoT device, but this is impractical, especially in a 3-D space where precise measurements can be cumbersome. To address this, we propose a user-friendly localization method: when a new IoT device is introduced, our design enables the user to ``tell'' the system where this device is located by simply pointing at it twice at two distinct locations. By jointly analyzing the two pointing lines, we estimate the device location as the midpoint of their shortest connecting segment. This approach significantly reduces setup effort while ensuring adaptability to real-world deployments.

To validate our approach, we implement PnPSelect on several commercial devices and conduct comprehensive experiments. We benchmark the accuracy of pointing direction estimation and IoT device selection in controlled lab settings. We then deploy our system in real-world environments and conduct user studies to evaluate its practical effectiveness. The main contributions of this work are summarized below:
\begin{itemize}
\item We present PnPSelect, an IoT device selection solution that eliminates the need for dedicated hardware on each IoT device, enabling true plug-and-play functionality.
\item We propose a pointing direction estimation algorithm that allows users to select IoT devices using a single UWB anchor. We develop a user-friendly IoT device localization approach, requiring only two pointing gestures to estimate device locations accurately.
\item We prototype our design and showcase the effectiveness of the proposed system with both benchmark and real-world experiments. Comprehensive experiments demonstrate high device-selection accuracy and robustness across diverse deployment conditions.
\end{itemize}
\section{Background and Related Work}

In this section, we first provide background on UWB technology and its use in commodity devices. We then review existing IoT device selection solutions and highlight the differences between PnPSelect and related work.

\subsection{Background of UWB}
\label{sec21}

UWB is a wireless technology characterized by its large frequency bandwidth, typically exceeding 500~MHz or 20\% of its center carrier frequency~\cite{sabath2005definition}. Its large bandwidth provides fine temporal resolution, allowing UWB devices to resolve closely spaced multipath components and enabling accurate Time-of-Arrival (ToA) measurements. Thus, compared to narrowband signals~(e.g., Wi-Fi and Bluetooth), UWB is more suitable for precise device localization even in a multipath-rich environment~\cite{ma2024push}. Currently, three methodologies can be used for UWB-based localization~\cite{yang2022vuloc}, including ToF~\cite{6012487,neirynck2016alternative,ma2024push}, Time-Difference-of-Arrival (TDoA)~\cite{xu2006position,ledergerber2015robot,kempke2016harmonium}, and AoA~\cite{heydariaan2020anguloc,zhao2021uloc,arun2023xrloc}. ToF-based localization estimates the distance between the target device and multiple anchors based on ToA measurements. By trilaterating these distances, the device’s position can be determined. TDoA-based approaches utilize the ToA differences between multiple anchors to compute the device’s position. It should be noted that ToF- and TDoA-based positioning generally requires at least three anchors for 2-D localization and additional anchors for unconstrained 3-D localization. AoA-based methods, in contrast, use an antenna array to measure the Phase-Difference-of-Arrival (PDoA), enabling angle estimation. By combining distance and angle measurements, AoA-based localization can accurately determine a device’s position with just a single anchor.

In recent years, advancements in UWB technology, coupled with reductions in hardware cost and size, have led to its widespread integration into consumer electronic devices. Major manufacturers, including Apple~\cite{nearbyapi}, Samsung~\cite{Samsung2021}, Xiaomi~\cite{xiaomi2020}, and Google~\cite{pixel6}, have incorporated UWB modules into their smartphones to enable precise device positioning. Additionally, UWB has been integrated into other consumer devices, including selected Apple Watch and HomePod models and the Charging Case for AirPods Pro. Thanks to its high-precision distance and angle estimation capability~\cite{heinrich2023smartphones}, UWB has been widely adopted for applications including tracking~\cite{yang2022vuloc,zhou2012ultra,ma2022involving}, item positioning~\cite{findmy,wang2018research}, and digital keys~\cite{zheng2023nn}. A well-known smart home application involves using a UWB-equipped iPhone to locate an AirTag, where the phone precisely measures both the distance and angle of the tag for object tracking~\cite{findmy}. Beyond traditional positioning tasks, recent research has explored UWB’s potential in contactless sensing~\cite{chen2021movi,chen2021octopus,li2021fine,zheng2020v2ifi,zhang2022mobi2sense,zhang2023embracing}. Furthermore, UWB hardware manufacturers have released development kits capable of establishing UWB connections with both Apple and Android smartphones~\cite{qm33120w,type2bp}, offering new possibilities for customized UWB-based applications. In this paper, we leverage UWB-equipped smartphones and smartwatches as user devices to implement our IoT device selection system, enabling intuitive and seamless interaction without requiring dedicated UWB or device-selection hardware on individual IoT devices.

\subsection{IoT Device Selection Techniques}
\begin{figure}[!t]
        \centering
        \subfloat[Distance-change/Doppler-base solutions.]{
    \includegraphics[height=1.5in]{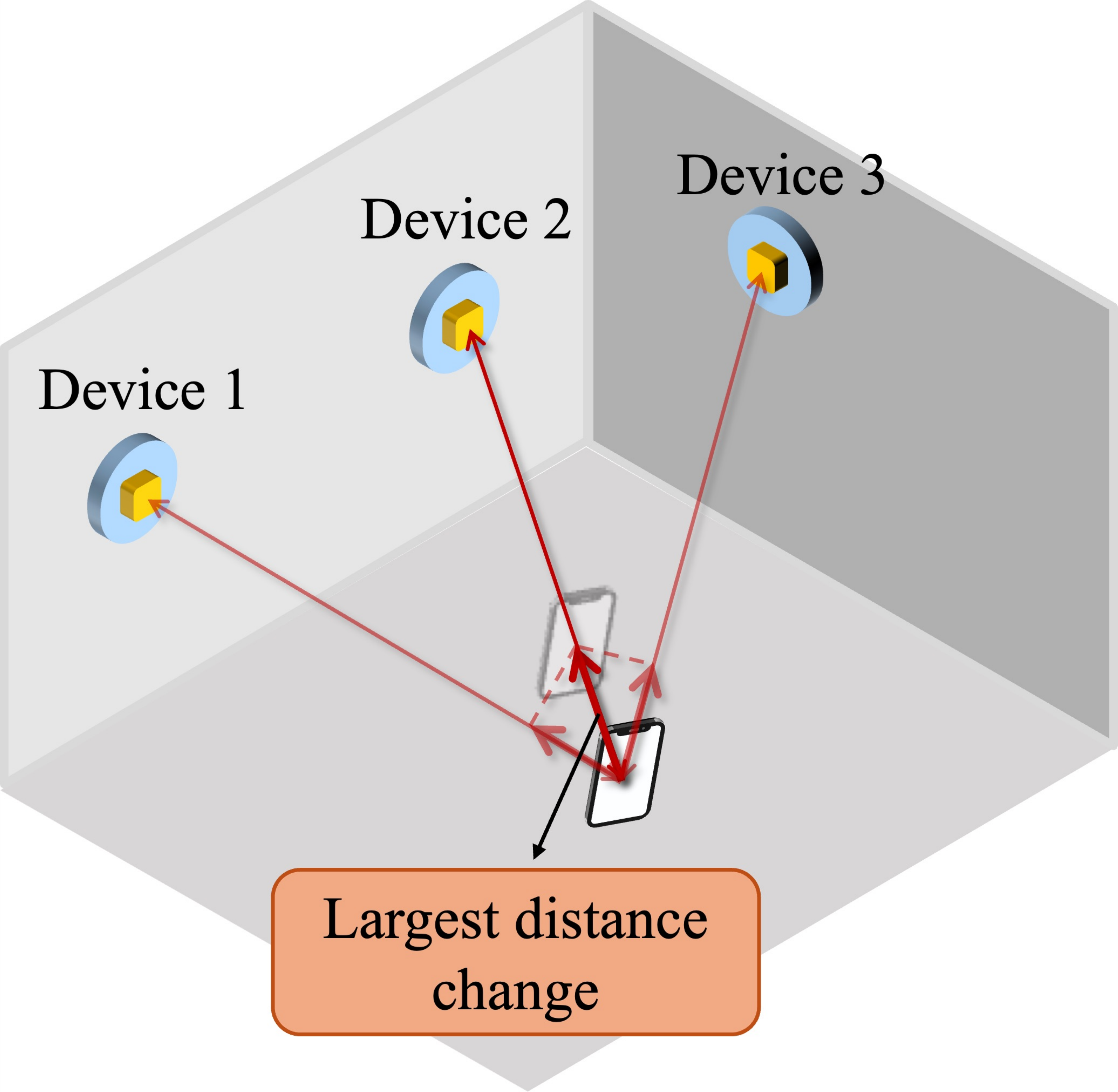}
            \label{fig2_1a}
        }
        \hspace{1in}
        \subfloat[Direction/AoA-based solutions.]{
    \includegraphics[height=1.5in]{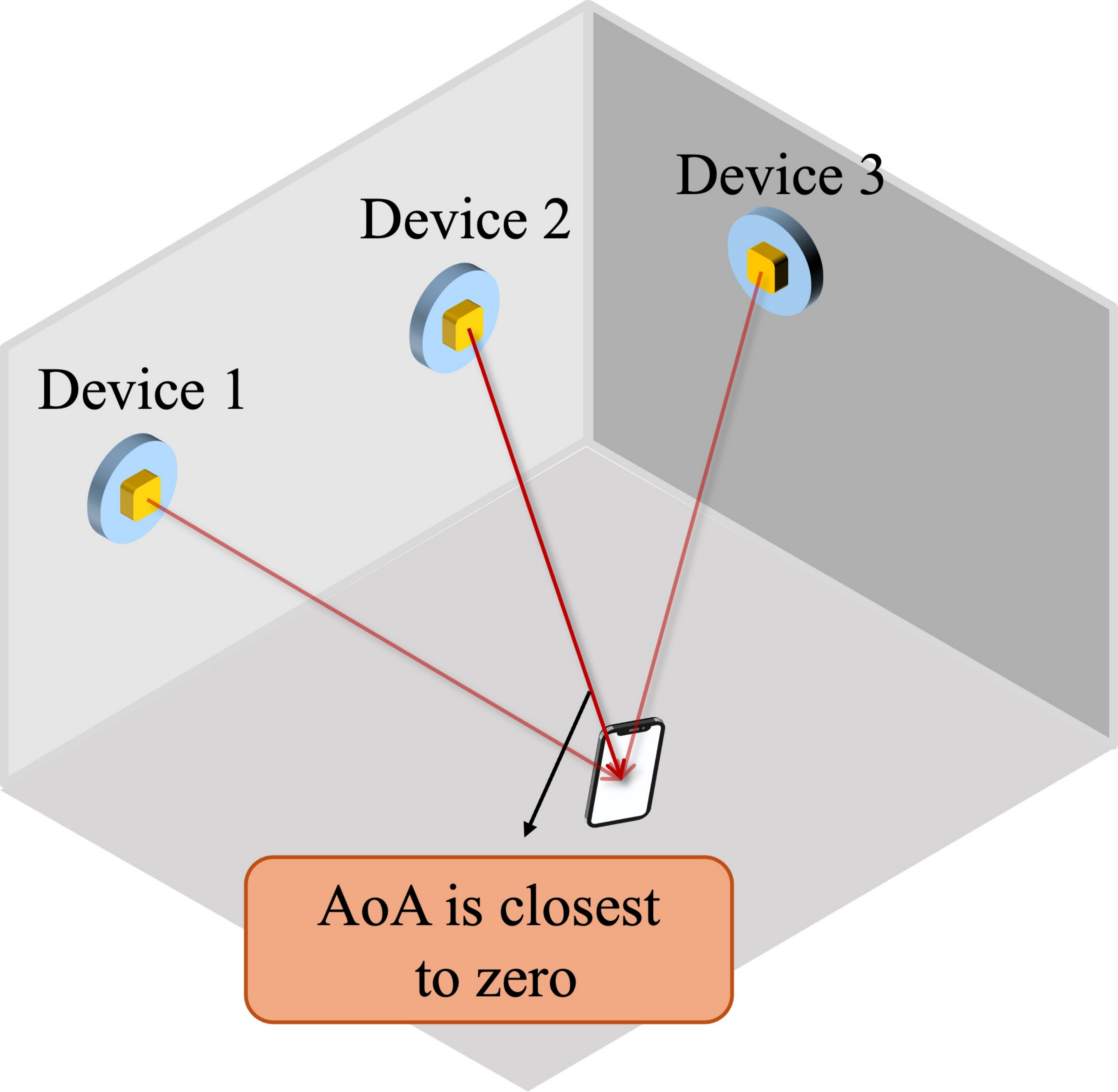}
            \label{fig2_1b}
        }
        \caption{Illustration of existing wireless signal-based IoT device selection techniques. Both solutions require each target device to provide a compatible signal source or transceiver.}
        \label{fig2_1}
\end{figure}

With the rapid expansion of IoT ecosystems, researchers have explored various solutions to simplify IoT device selection and control. Some approaches~\cite{xiao2017deus,zhang2019facilitating,zhang2018tap,verweij2017smart} rely on gesture-based interactions, requiring users to perform specific movements to select and control devices. However, as the number of IoT devices in an environment increases, memorizing a large set of gestures becomes impractical. To enhance usability, alternative methods focus on more intuitive interactions, allowing users to select a device simply by pointing at or looking at it. Various devices have been leveraged for this purpose, including smartwatches~\cite{alanwar2017selecon}, earphones~\cite{wang2022faceori}, smart glasses~\cite{zhang2023bleselect}, and smartphones~\cite{qin2023selecting,mayer2020enhancing}. Among these, camera-based approaches enable selection by capturing an image of the target device~\cite{chen2018snaplink,de2016snap} or tapping on a previewed object on a screen~\cite{boring2010touch,vincent2013precise}. Other methods utilize eye gaze tracking~\cite{mayer2020enhancing,mayer2018effect}, while recent work~\cite{qin2023selecting} combines front-camera-based iris localization with rear-camera scene understanding to select the object occluded by the phone from the user’s perspective. However, camera-based solutions may experience degraded performance in low-light conditions and may have difficulty distinguishing visually similar devices, limiting their practicality~\cite{strecker2023mr}. To overcome these challenges, wireless signal-based approaches have been proposed, leveraging signal properties such as ToF and AoA measurements to enable device selection. These methods utilize signals such as Bluetooth~\cite{zhang2023bleselect}, UWB~\cite{alanwar2017selecon}, and acoustic waves~\cite{wang2022faceori,aumi2013doplink,sun2013spartacus}. In general, wireless signal-based selection techniques fall into two main categories: distance-change/Doppler-based solutions~\cite{alanwar2017selecon,aumi2013doplink,sun2013spartacus} and direction/AoA-based solutions~\cite{zhang2023bleselect,wang2022faceori}. As shown in Figure~\ref{fig2_1a} and~\ref{fig2_1b}, both categories require each target device to provide a compatible signal source or transceiver so that its distance or direction relative to the user device can be estimated.

\textbf{Distance-Change/Doppler-Based Solutions.} These methods measure how the distance between the user device and each IoT device changes during the pointing gesture. As illustrated in Figure~\ref{fig2_1a}, the key insight is that the distance to the target device exhibits the most significant change, allowing it to be identified. For example, SeleCon~\cite{alanwar2017selecon} measures distance variations between a smartwatch and IoT devices using UWB signals. Alternatively, Doppler-based approaches infer motion by measuring frequency shifts in signals caused by movement. Because both distance variation and Doppler shift capture radial motion between the user device and a target device, the two approaches exploit a similar principle. Doplink~\cite{aumi2013doplink} and Spartacus~\cite{sun2013spartacus} leverage acoustic signals to measure Doppler shifts, enabling smartphone-based IoT device selection.

\textbf{Direction/AoA-Based Solutions}. AoA-based techniques extract the angle of incoming signals transmitted by IoT devices to determine which device the user is pointing at. As shown in Figure~\ref{fig2_1b}, the AoA of the target device is expected to be closest to zero relative to the forward direction of the user device, making it possible to identify the intended selection. BLEselect~\cite{zhang2023bleselect} designs an antenna array on smart glasses to estimate the AoA of Bluetooth signals emitted by IoT devices in the environment, determining which IoT device the user is pointing at or looking at. FaceOri~\cite{wang2022faceori} uses multiple microphones on the user’s earphones to measure ultrasonic ranges to speakers on nearby devices and infer which device the user is facing.

\textbf{UWB-Based Products and Solutions}. Recent UWB-based remote-control products support pointing interactions for TV control. They allow users to aim a handheld controller at a TV and perform on-screen operations such as pointing, selecting, dragging, and issuing control commands~\cite{huayuen_hybi305,calterah_airmice_2025,gizchina_huawei_smart_screen_v5}. However, if extended from TV-centric control to more general multi-device selection and control scenarios, such designs would typically require additional receiving infrastructure or dedicated support on the target-device side. For example, one recent UWB remote-control solution explicitly states that smart-home devices such as air conditioners, lights, and speakers can be controlled when they are ``attached with a receiver''~\cite{huayuen_hybi305}. Recent work, UWBOri~\cite{chang2024uwbori}, demonstrates that it is feasible to estimate a device's static orientation through mutual localization between a smartphone and a UWB anchor, and further shows the potential of using such orientation estimation for IoT device selection. However, this solution relies on a restrictive practical assumption: the anchor needs to remain within the effective view of the smartphone in order for the orientation estimation to work. Since the effective view of commercial smartphones is typically limited in practice, this requirement is difficult to satisfy in many real-world scenarios.

\textbf{Limitations of Existing Wireless Signal-Based Methods.} Although existing designs differ in sensing mechanisms, they generally share a stronger deployment assumption: to control multiple appliances, the target devices are typically required to provide a compatible signal source or transceiver, such as an acoustic speaker, a BLE radio, or a UWB transceiver. Although BLE connectivity is relatively common in smart devices, this does not directly enable BLE AoA-based device selection~\cite{zhang2023bleselect}. Such methods require compatible antenna-array hardware and access to angle measurements, which are not uniformly supported across commodity smartphones and heterogeneous household appliances. Although UWB is increasingly available in mobile consumer devices, it has not been widely adopted in ordinary household appliances such as air conditioners, fans, and lights. Therefore, the above approaches are still difficult to deploy in a plug-and-play manner in current smart homes. Our method addresses this gap by requiring only a single UWB anchor in the environment, enabling room-level IoT device selection without modifying existing appliances. By removing the need to install additional hardware on every target device, our system provides a more scalable and practical solution for current smart-home deployments.

\section{Principle of Single-Anchor-Based Device Selection}\label{sec3}
\begin{figure}[htbp]
    \centering
     \includegraphics[height=1.8in]{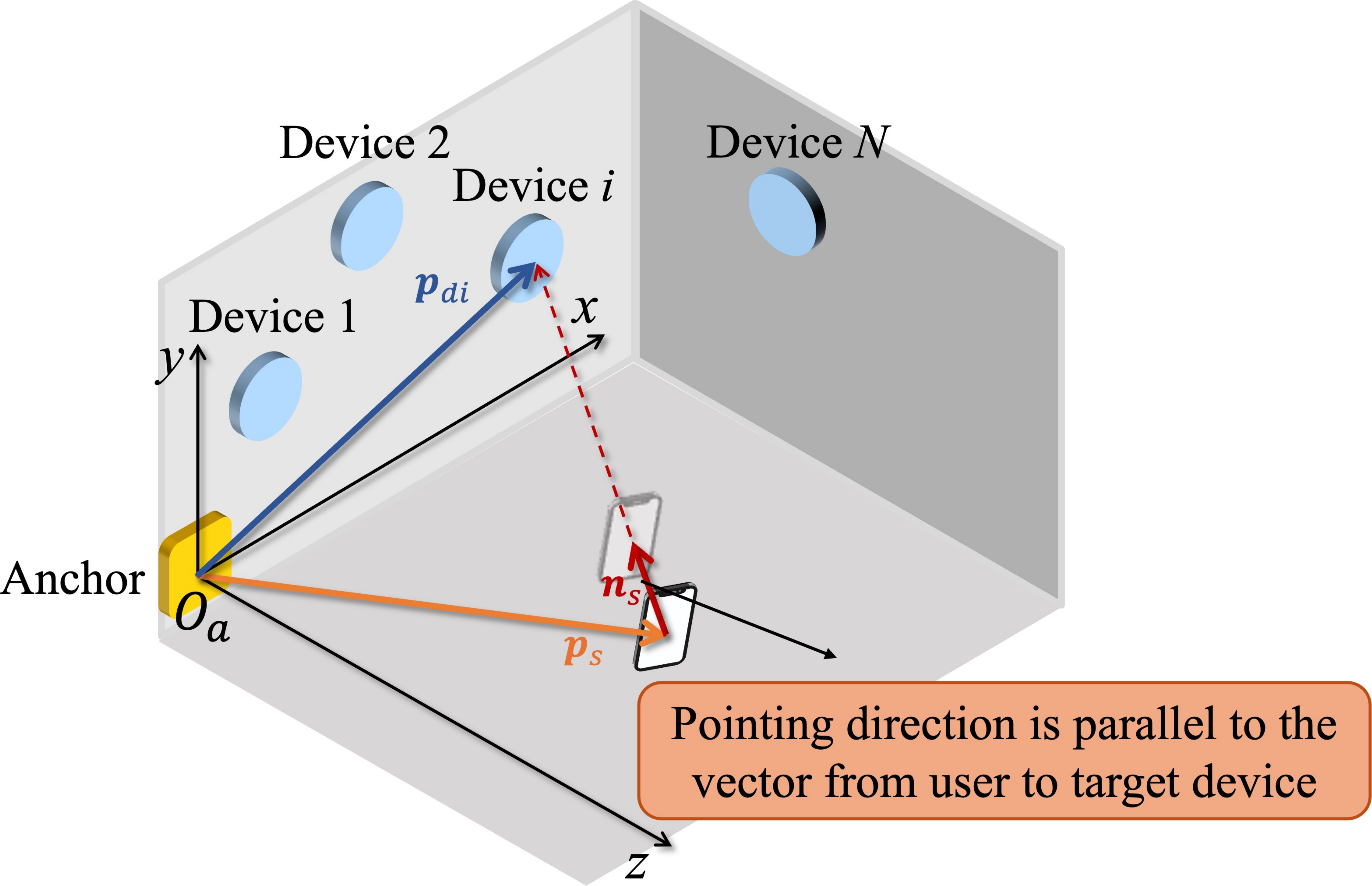}
    \caption{Illustration of \systemname. Only one anchor is required in the environment.}
    \label{fig2_1c}
\end{figure}

In this section, we describe the principle of IoT device selection using a single UWB anchor. Unlike the existing solutions shown in Figure~\ref{fig2_1}, \systemname{} does not require dedicated UWB hardware to be installed on each IoT device. Instead, as shown in Figure~\ref{fig2_1c}, only one UWB-equipped device is deployed in the environment as an anchor.

\subsection{Fundamental Principle of \systemname}
\label{sec31}

Figure~\ref{fig2_1c} illustrates the core concept behind our single anchor-based IoT device selection design. A user selects the desired device by moving their smartphone or smartwatch toward it in a natural pointing gesture. The key insight is that the direction of the pointing gesture should closely align with the direction from the user device to the target IoT device. In other words, the pointing direction should align with the vector from the user device to the target device. Based on this principle, \systemname{} follows two essential steps for device selection. 

The first step estimates the pointing direction of the user device. Since the system does not rely on dedicated transceivers on IoT devices, we develop a UWB-based pointing direction estimation algorithm using the readings reported by a single anchor. As shown in Figure~\ref{fig2_1c}, we define an anchor coordinate system ($C_a$) bound to the anchor fixed in the environment. The pointing-direction vector is denoted by $\mathbf{n_{s}}$. We will introduce pointing direction estimation in Section~\ref{sec32}.

The second step identifies the target IoT device by comparing the estimated pointing direction with the direction vectors from the user device to all registered IoT devices. The device with the best directional match is selected as the target. As shown in Figure~\ref{fig2_1c}, the direction vector from the user device (e.g., a smartphone) at $\mathbf{p_{s}}$ to the $i$-th IoT device is $\mathbf{p_{di}}-\mathbf{p_{s}}$, where $\mathbf{p_{di}}$ denotes the location of the $i$-th IoT device in $C_a$. Thus, a crucial requirement for the second step is knowledge of each IoT device’s location ($\mathbf{p_{di}}$), which presents a non-trivial challenge in real-world deployments. We address this issue by introducing a lightweight and user-friendly device localization method, detailed in Section~\ref{sec33}.

\begin{figure}[htbp]
    \centering
     \includegraphics[height=2in]{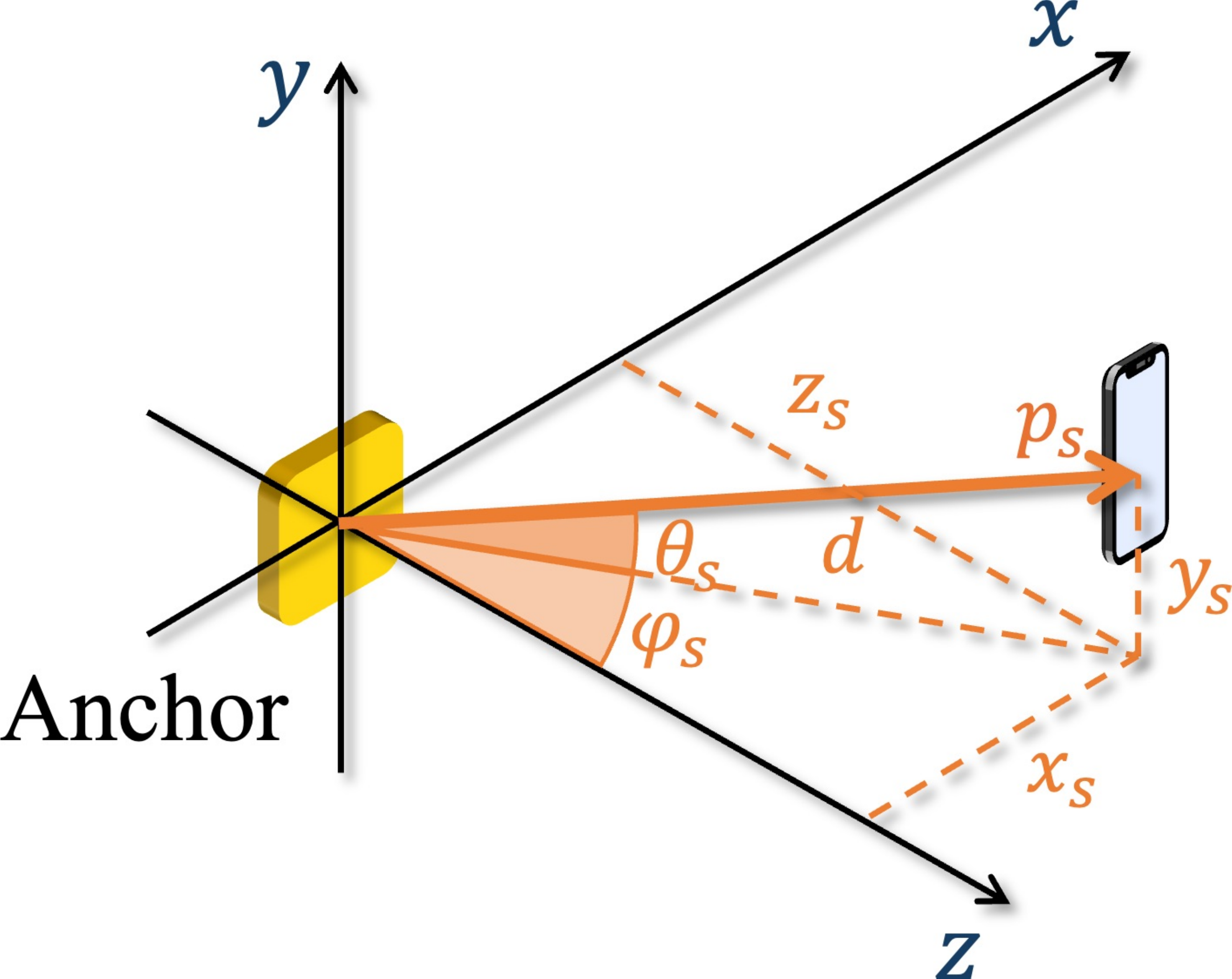}
    \caption{Location estimation with a single anchor.}
    \label{fig3_1}
\end{figure}

\subsection{Pointing Direction Estimation}
\label{sec32}

In our single-anchor pointing-direction estimation problem, two types of information are needed: the position of the user device in the global anchor/environment coordinate system, and the pointing direction of the user device in the same global coordinate system. Traditional methods for estimating a user’s pointing direction often rely on Inertial Measurement Unit (IMU) sensors, including accelerometers, gyroscopes, and magnetometers. Among them, accelerometers and gyroscopes can be used to estimate the device’s motion and trajectory, and thus may help infer a pointing direction. However, they do not sense the absolute orientation of the smartphone in the environment. Therefore, the pointing direction estimated from them is still expressed in the device’s own coordinate system, rather than in the global environment coordinate system. To transform the direction estimated from accelerometer and gyroscope measurements into the environment coordinate system, the system still needs to estimate the device’s absolute orientation. This typically requires a magnetometer, because it provides a reference to the Earth’s magnetic field and thus serves as an external reference rather than a device-local reference. However, magnetometer measurements are highly susceptible to interference from ferromagnetic objects in the indoor environment as well as ferromagnetic components inside smartphones, resulting in inaccurate orientation estimation~\cite{gao2022mom}. Even when deep learning-based approaches are employed to compensate for such distortions, the orientation estimation error can still remain as high as 8.02\degree{} in one of its evaluated buildings~\cite{sun2021idol}. In dense IoT environments, where precise pointing direction estimation is required, such an error margin can be problematic. For example, two devices separated by 30 cm at approximately the same depth and viewed from a distance of 3 m subtend an angle of about 5.7\degree. An orientation error of this magnitude can therefore make it difficult to reliably distinguish the intended target, leading to ambiguous or incorrect selections.

The key insight behind our solution is that when a user performs a pointing gesture, the trajectory of the device’s movement can be captured through UWB-based localization. By analyzing this trajectory, we can estimate the pointing direction of the user device. First, we describe how the trajectory is acquired using UWB localization. As introduced in Section~\ref{sec21}, three schemes, i.e., ToF, TDoA, and AoA, can be used for localization. Among these, AoA-based methods are particularly suitable for our design since they require only one anchor. Accordingly, we employ AoA-based location estimation in our system. The distance ($d$) between the user device and the anchor is estimated using two-way-ranging (TWR) techniques~\cite{6012487,neirynck2016alternative}. Simultaneously, the azimuth angle ($\varphi_s$) and elevation angle ($\theta_s$) of the user device relative to the anchor are measured using multiple antennas on the anchor~\cite{heydariaan2020anguloc,zhao2021uloc}. As illustrated in Figure~\ref{fig3_1}, these measurements allow us to compute the vector from the origin to the location of the user device in the anchor coordinate system $C_a$ as follows:
\begin{equation}\label{eq3_1}
\mathbf{p_s} =(x_s, y_s, z_s)^T  =(d \cos \theta_s \sin \varphi_s, d \sin \theta_s, d \cos \theta_s \cos \varphi_s)^T.
\end{equation}

\begin{figure}[t]
    \centering
     \includegraphics[height=1.8in]{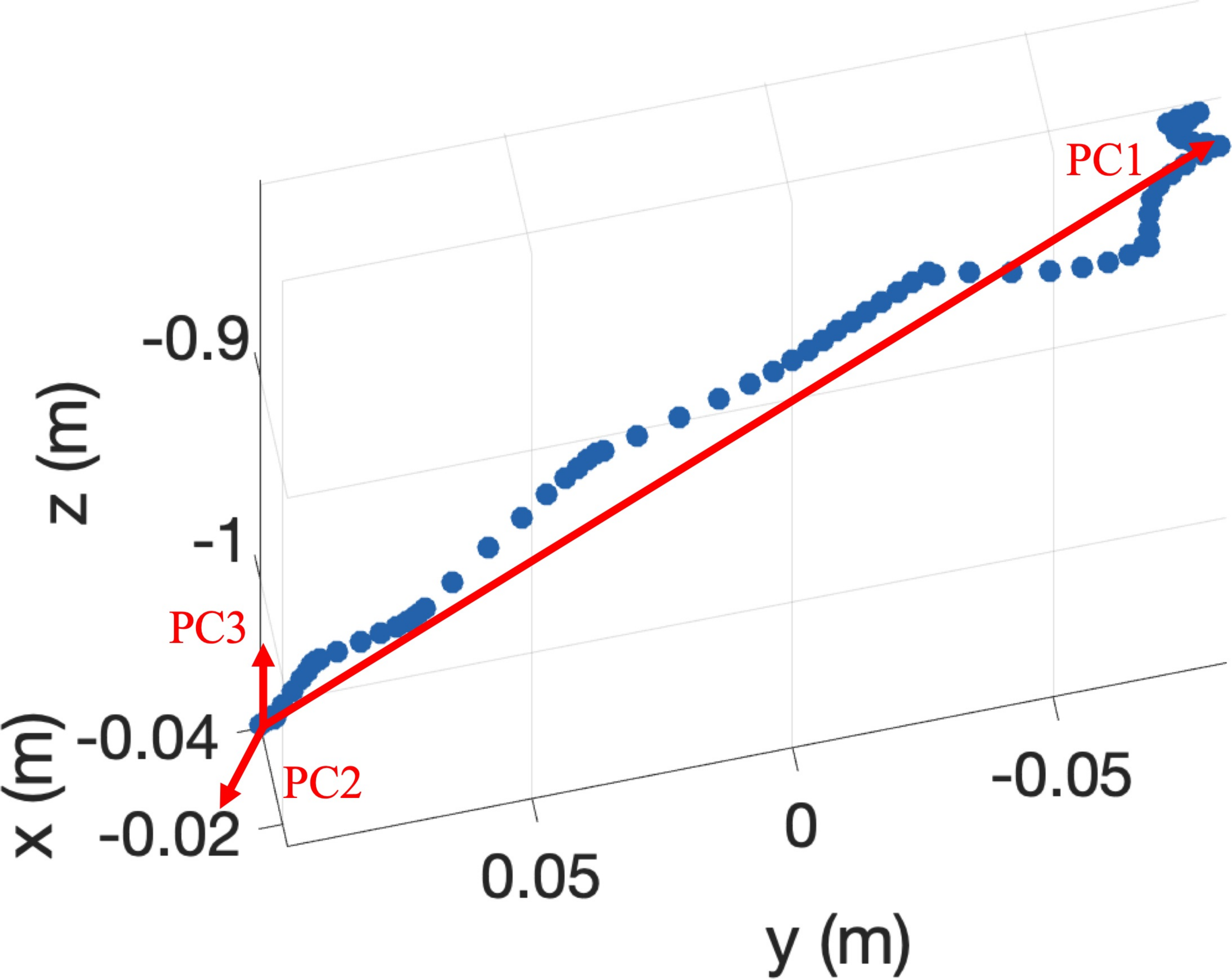}
    \caption{A user-device trajectory during a pointing gesture. PCA is used to estimate the pointing direction.}
    \label{fig3_3}
\end{figure}

By continuously estimating the user device’s location, we obtain a series of position samples over time, forming a trajectory in 3-D space. Figure~\ref{fig3_3} illustrates an example trajectory recorded when the user points at an IoT device. The dominant movement direction along this trajectory corresponds to the user’s pointing direction. A straightforward way to estimate the motion direction is by computing displacement vectors between consecutive location samples and averaging them. However, this method is susceptible to random errors arising from two main sources. The first arises from involuntary hand movements, which cause slight deviations from the intended pointing direction. The second is the UWB localization error, which introduces small inaccuracies in location estimation. To address these issues, we first apply a Kalman filter~\cite{kalman1960new} to smooth the trajectory, mitigating fluctuations and improving robustness. We then formulate the pointing direction estimation as an optimization problem:
\begin{equation}\label{eq3_2}
\mathop{\arg\max}\limits_{\mathbf{n_s}} \sum^{M}_{m=1}[\mathbf{n_s}\cdot \mathbf{p_s}(m)]^2,
\end{equation}
where $M$ is the number of location samples in the trajectory, and $\mathbf{p_s}(m)$ represents the location vector of the $m$-th sample. This optimization finds the direction vector $\mathbf{n_s}$ that maximizes the variance when the trajectory is projected onto it, ensuring that the estimated direction aligns with the dominant movement trend. To solve this problem, we employ Principal Component Analysis (PCA), which uses an orthogonal transformation to identify the direction along which the data exhibits the largest variance. The first principal component (PC1) is extracted as the pointing direction, while PC2 and PC3, which correspond to noise and other deviations, are discarded. Because PCA determines an axis with an arbitrary sign, we orient PC1 according to the temporal direction of the pointing trajectory, ensuring that it points from the beginning toward the end of the gesture. As illustrated in Figure~\ref{fig3_3}, this approach ensures that only the primary motion trend is retained, leading to accurate and stable pointing direction estimation.

\subsection{IoT Device Selection}\label{sec33}

After estimating the pointing direction, the next step is to identify the selected IoT device. However, before identifying the specific device, we must first obtain the locations of all IoT devices in the environment. Assume there are $N$ IoT devices, and the location vector of the $i$-th IoT device in the anchor coordinate system $C_a$ is denoted by $\mathbf{p_{di}}=(x_{di}, y_{di}, z_{di})^T$. For device selection to be feasible, these locations must be known in advance.

\subsubsection{IoT Device Location Estimation}
\label{sec331}

A straightforward approach is to manually measure and record each IoT device’s position. However, this is impractical and labor-intensive, particularly since it requires measuring the relative 3-D positions of all IoT devices with respect to the anchor coordinate system $C_a$. An alternative approach is to simply place the user device on the new IoT device and record its position using UWB localization. In this case, the measured location of the user device would correspond to the IoT device’s position. However, this method has several limitations due to the varying spatial relationships between IoT devices and the anchor. In many scenarios, an IoT device may be located outside the FoV of the anchor, making direct localization inaccurate or impossible. For instance, if the anchor is mounted on a wall, IoT devices installed on the same wall may fall outside the anchor's typical 120$\degree$ FoV. Similarly, if the anchor is embedded in a smart TV, IoT devices positioned behind the TV would be out of the FoV. Moreover, in some cases, direct line-of-sight (LoS) between the anchor and the IoT device may be obstructed by objects in the environment, leading to significant localization errors.

\begin{figure}[!t]
        \centering
        \subfloat[ Pointing at the device at two locations.]{
         \includegraphics[width=1.6in]{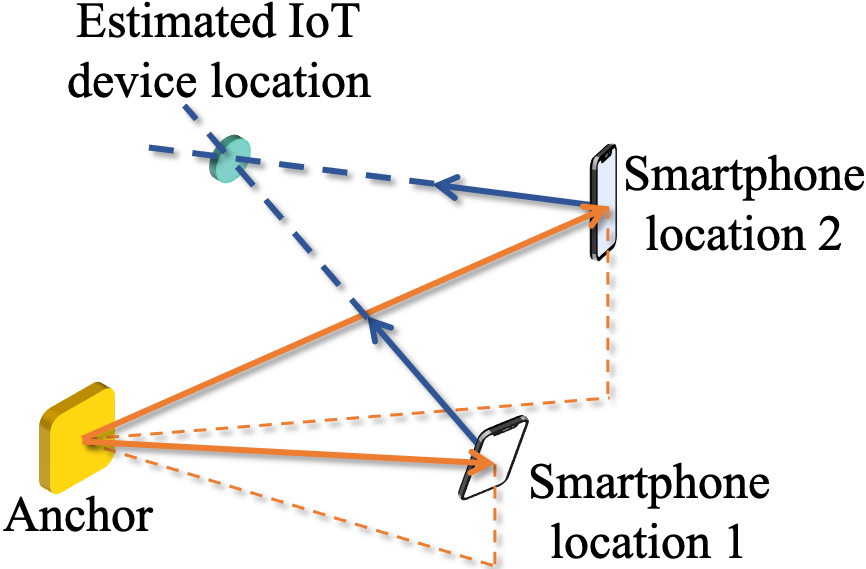}
            \label{fig:fig5_1a}
        }
        \hspace{0.2in}
        \subfloat[Two directions do not intersect.]{
         \includegraphics[width=1.6in]{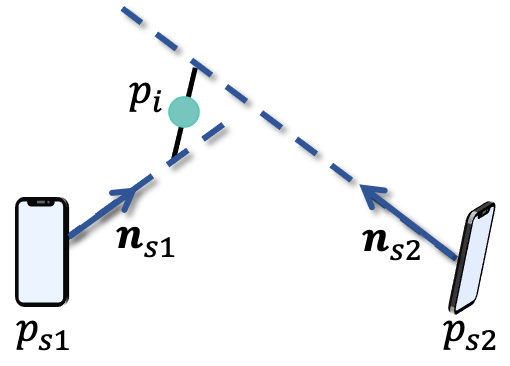}
            \label{fig:fig5_1b}
        }
        \vspace{-0in}
        \caption{Illustration of IoT device location estimation. The user can point at the device from two locations to ``tell'' the system where the device is. If the two pointing rays do not intersect, the midpoint of the shortest segment between them is used as the estimated device location.}
\end{figure}

To address these limitations, we introduce a registration procedure that estimates an IoT device’s position without requiring the user device to be placed on it. The user device must nevertheless remain within reliable UWB coverage at both pointing locations. Our method is designed to be robust to different placement conditions, ensuring reliable localization in real-world environments. Intuitively, the key idea of our approach is to let the user point at the device to ``tell'' the system where it is. As shown in Figure~\ref{fig:fig5_1a}, the smartphone points at an IoT device whose location is unknown. Then, it can be determined that this device is on the ray of the pointing direction. However, this is not sufficient, because any location on this ray is a possible location for this device. To this end, we let the user point at the device again at another location to obtain another ray containing the possible location. The intersection of the two rays is thus the location of the IoT device. Because of pointing errors, UWB localization errors, and the finite dimensions of the target device, the two rays may not intersect exactly in 3-D space, as shown in Figure~\ref{fig:fig5_1b}. We compute the shortest line segment between the two rays and use the midpoint of this segment as the estimated device location. Let $\mathbf{p_{s1}} = (x_{s1}, y_{s1}, z_{s1})^T$ and $\mathbf{p_{s2}} = (x_{s2}, y_{s2}, z_{s2})^T$ denote the smartphone locations when the user points at the \(i\)-th device from two different positions, and let $\mathbf{n_{s1}} = (e_{x1}, e_{y1}, e_{z1})^T$ and $\mathbf{n_{s2}} = (e_{x2}, e_{y2}, e_{z2})^T$ denote the corresponding pointing-direction vectors. Then, the equations of the ray in which the two directions the smartphone is pointing at are:
\begin{equation}\label{eq:Eq5_1}
\begin{split}
\mathbf{p_{1}} = \mathbf{p_{s1}} + t_1 \cdot \mathbf{n_{s1}}, \\
\mathbf{p_{2}} = \mathbf{p_{s2}} + t_2 \cdot \mathbf{n_{s2}},
\end{split}
\end{equation}
where $t_1$ and $t_2$ are variable parameters, and different values can be substituted to obtain different locations~(e.g., $\mathbf{p_1}$ and $\mathbf{p_2}$) on the rays. To calculate the location of the intersection of two rays or the point closest to the two rays, the following equation can be solved using the least squares method~\cite{han2010nearest}:
\begin{equation}\label{eq:Eq4_2}
\left[\begin{array}{l}
t_1 \\
t_2
\end{array}\right]=\left(M^T M\right)^{-1} M^T B,
\end{equation}
where $M=\left[\begin{array}{ll}e_{x1} & -e_{x2} \\ e_{y1} & -e_{y2} \\ e_{z1} & -e_{z2}\end{array}\right]$ and $B=\left[\begin{array}{ll}x_{s2}  -x_{s1} \\ y_{s2}  -y_{s1} \\ z_{s2}  -z_{s1}\end{array}\right]$. 

Consequently, the estimated location of an IoT device can be obtained by substituting $t_1$ and $t_2$ into Equation~\ref{eq:Eq5_1} as:
\begin{equation}\label{eq:Eq4_3}
\mathbf{p_{di}}=\frac{1}{2}(\mathbf{p_1}+\mathbf{p_2})=\frac{1}{2}(\mathbf{p_{s1}} + t_1 \cdot \mathbf{n_{s1}}+\mathbf{p_{s2}} + t_2 \cdot \mathbf{n_{s2}}).
\end{equation}
Therefore, whenever a new IoT device joins the system, we can estimate the location of that device using the above approach. Meanwhile, if a registered device is moved, the user can repeat the same procedure to update its stored location.

\subsubsection{Identifying Target IoT Device}
\label{sec332}
\begin{figure}[!t]
    \centering
     \includegraphics[width=2.6in]{
        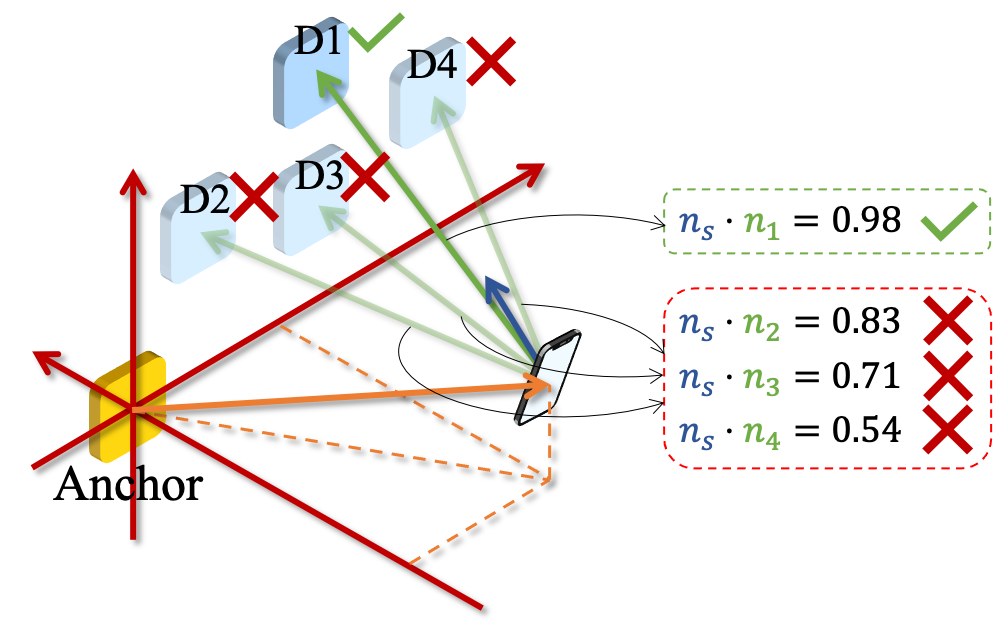}
    \caption{Illustration of device selection based on pointing direction estimation.}
    \label{fig:fig4_2}
\end{figure}

At this stage, we have obtained the 3-D location of each IoT device relative to the anchor. It is important to note that IoT device location estimation is only required when a new device is introduced into the environment or when its placement changes. Once an IoT device’s position has been estimated, it is recorded and reused for subsequent pointing-based selection, eliminating the need for repeated localization. The fundamental principle behind IoT device selection is that the direction vector of the user device’s pointing gesture should align with the direction vector from the user device to the target IoT device. Based on this, we first compute the unit direction vectors from the user device to all IoT devices. The direction vector to the $i$-th IoT device is expressed as:
\begin{equation}\label{eq3_3}
\begin{split}
    &\mathbf{n_i}(m)=(e_{ix},e_{iy},e_{iz})^T =\mathbf{p_{di}}-\mathbf{p_s}(m) \\ &=\frac{(x_{di}-x_s(m),y_{di}-y_s(m),z_{di}-z_s(m))^T}{\sqrt{(x_{di}-x_s(m))^2+(y_{di}-y_s(m))^2+(z_{di}-z_s(m))^2}},
\end{split}
\end{equation}
where $m$ represents the index of a location sample in the trajectory. As illustrated in Figure~\ref{fig:fig4_2}, when the user device is pointing at the $i$-th IoT device, the pointing direction vector ($\mathbf{n_s}$) is expected to be identical to the unit vector $\mathbf{n_i}(m)$. This condition can be expressed as $\mathbf{n_s}\cdot \mathbf{n_i}(m)=1$. However, in practice, pointing-direction errors, UWB localization errors, device-registration errors, and the finite dimensions of the target device may cause the directional similarity to be lower than 1, as shown in Figure~\ref{fig:fig4_2}. To account for this, we systematically search through all IoT devices to identify the one most likely being pointed at by the user. Since multiple location samples are collected while the user performs the pointing gesture, we incorporate all trajectory samples to enhance robustness against random localization errors. The selection process is formulated as the following minimization problem:
\begin{equation}\label{eq3_4}
\arg\min_{i} \sum_{m=1}^M |1-\mathbf{n_s}\cdot \mathbf{n_i}(m)|/M.
\end{equation}
By minimizing this expression, we identify the IoT device whose direction vector best matches the user’s pointing direction, effectively achieving robust and accurate selection based on pointing gestures.

\section{Implementation}

To validate our design, we implement our system on both Apple and Android platforms. For evaluations involving Apple devices, we employ an iPhone 11 as a fixed anchor within our test environments. Users interact with the system using either an iPhone 12 Pro Max or an Apple Watch Series 6 to select IoT devices. To facilitate UWB-based measurements, we develop an iOS application utilizing Apple’s Nearby Interaction framework~\cite{nearbyapi}, which enables peer-to-peer UWB communication between two iPhones or between an iPhone and an Apple Watch. On the iPhone, the Nearby Interaction framework reports the distance and relative direction of the peer iPhone or Apple Watch. These measurements are used for pointing-direction estimation. In our measurements, the Nearby Interaction framework reports UWB measurements at an average rate of approximately 55 Hz. The extracted distance and angle information from the Nearby Interaction framework is transmitted via a TCP connection to a server laptop (a MacBook Pro equipped with an Apple M3 Max CPU and 32GB RAM) for further processing and evaluation. In addition to Apple’s ecosystem, Android also supports UWB-based measurements through its dedicated UWB API~\cite{androidapi}. To evaluate our system on the Android platform, we utilize two Xiaomi Mix 4 smartphones, where one acts as the anchor and the other serves as the user device. During the pointing gesture, the anchor continuously measures the distance and angle of the user device. These measurements are stored in a local file on the anchor and later transmitted to the laptop for further data processing and analysis.

\section{Evaluation}

In this section, we first conduct controlled experiments to characterize the accuracy of pointing-direction estimation and examine the effects of pointing direction, user-device location, gesture displacement, gesture velocity, and NLoS obstruction. We then evaluate spatial resolution and device-selection accuracy, followed by experiments in real-world environments.

\subsection{Accuracy of Pointing Direction Estimation}
\subsubsection{Experiment Setting}
A core component of PnPSelect is estimating the user device’s pointing direction. We therefore conduct controlled experiments to evaluate estimation accuracy under different parameter settings. We fix the anchor in the environment and then vary the location and pointing direction of the user device (i.e., a smartphone). The smartphone is mounted on a sliding track, the displacement and velocity of which can be precisely controlled. This setup allows us to use the sliding track to manage the smartphone’s movements, thereby simulating the pointing gesture. All experiments are conducted in an open indoor hall. The environment is relatively empty, and the devices are placed on the ground during data collection.

\subsubsection{Impact of User Device Pointing Direction} 
\begin{figure*}[!t]
    \centering
        \subfloat[Experiment setup.]{
        \includegraphics[height=1.4in]{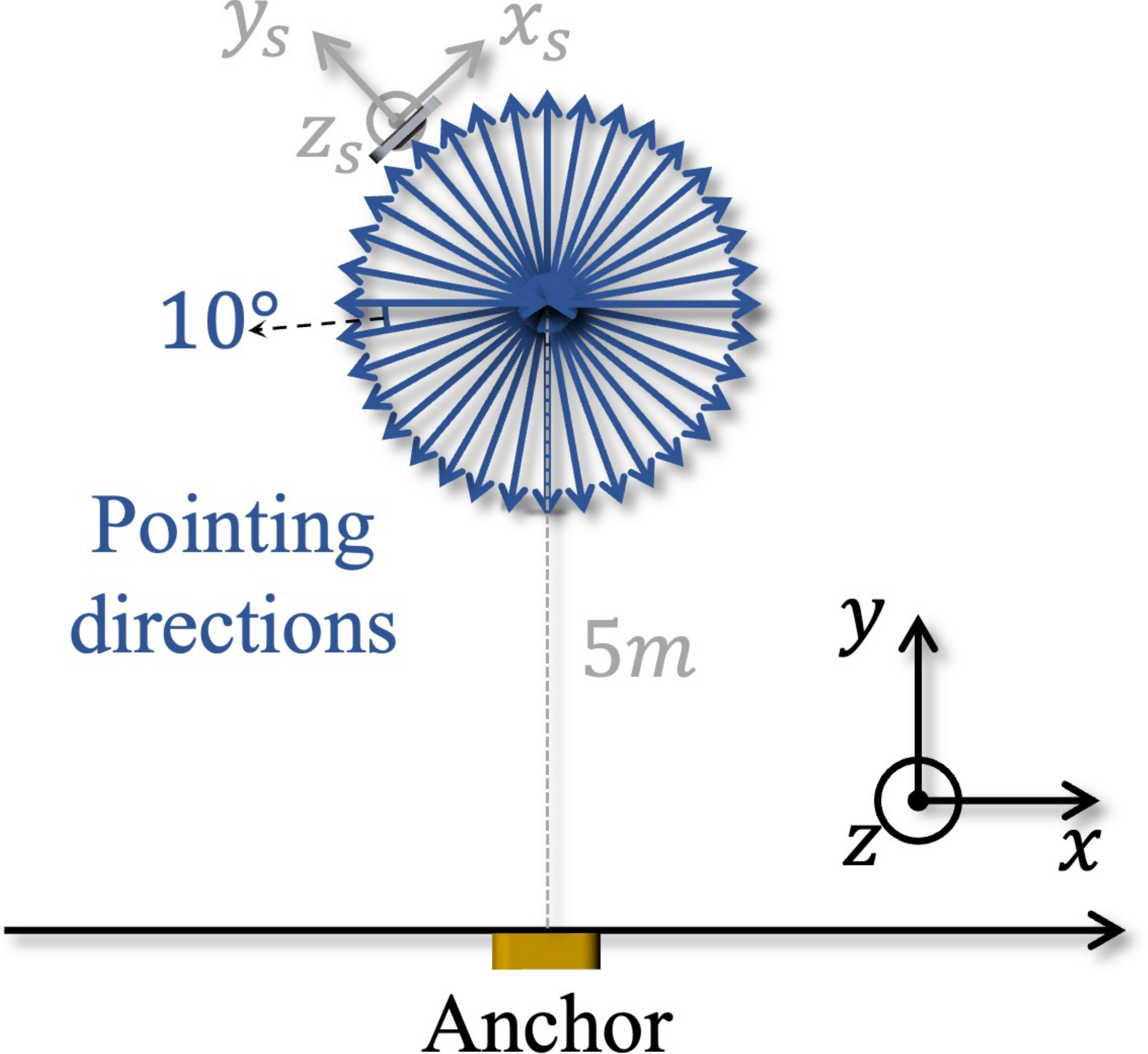}
        \label{fig5_1a}
        \hspace{0in}
    }
        \subfloat[Overall performance.]{
        \includegraphics[height=1.4in]{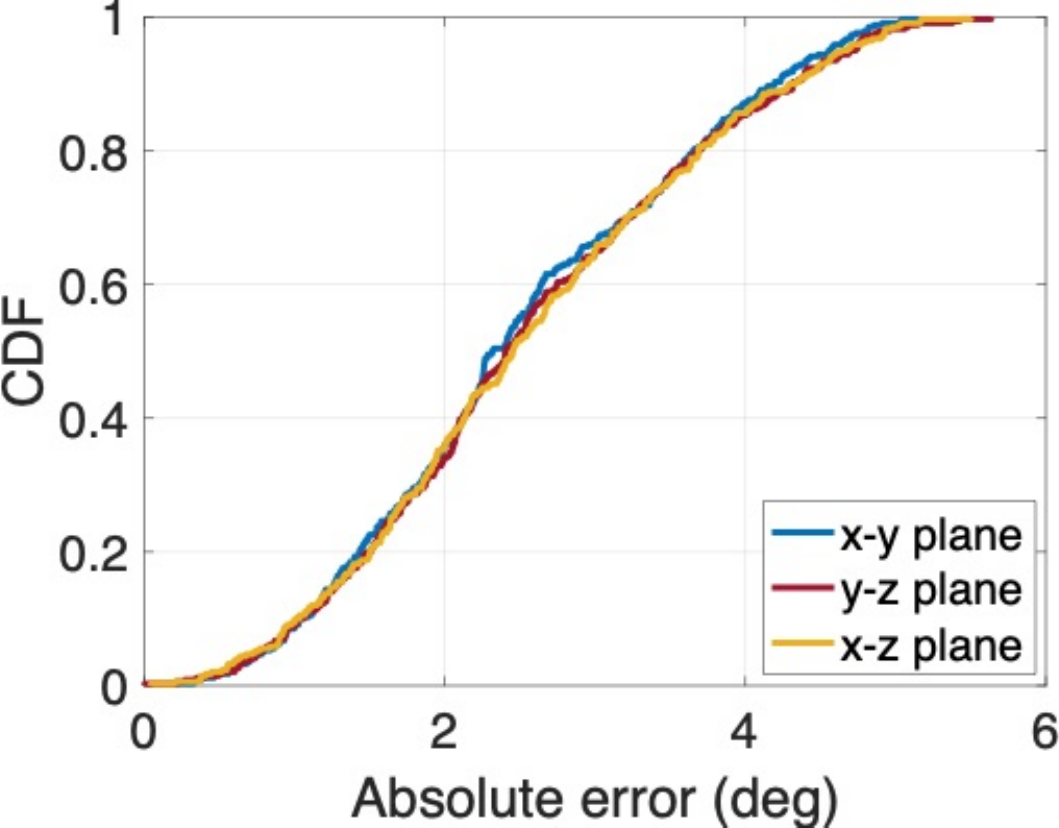}
        \label{fig5_1b}
        \hspace{0in}
    }
        \subfloat[Different pointing directions.]{
        \includegraphics[height=1.4in]{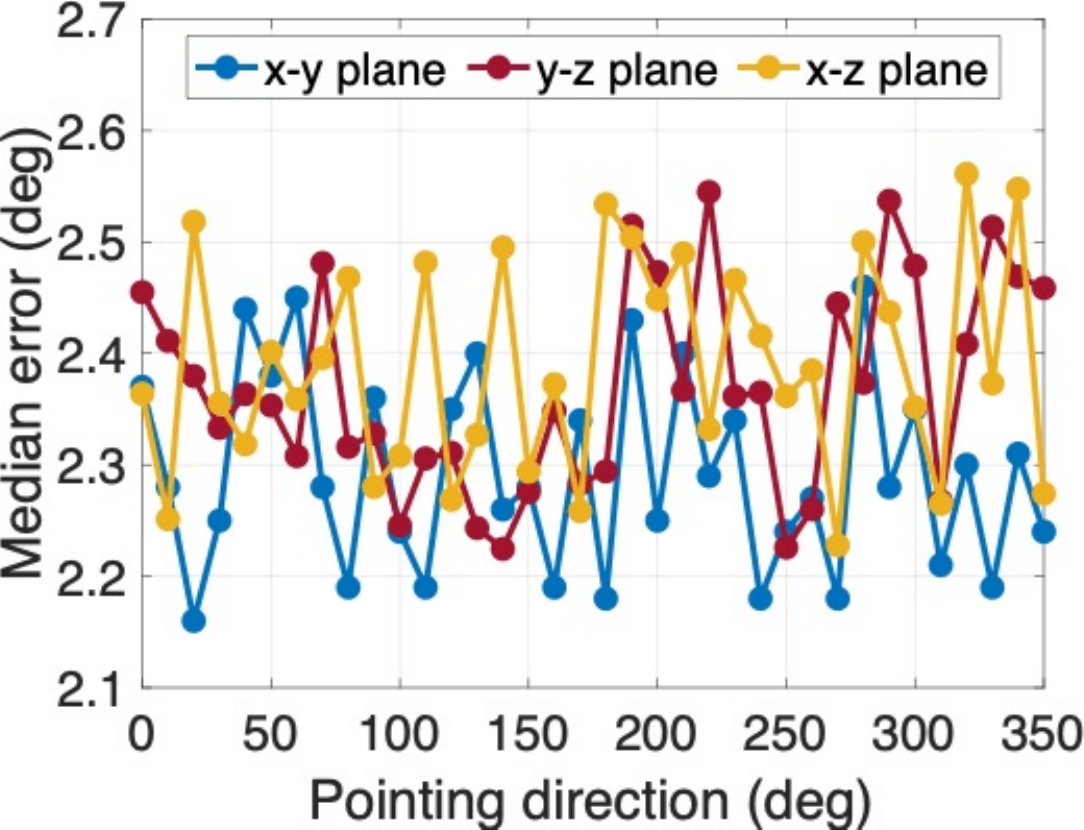}
        \captionsetup{justification=centering}
        \label{fig5_1c}
        \hspace{0in}
    }
    \caption{Overall performance and impact of pointing direction.}
    \label{fig5_1}
\end{figure*}

We evaluate the estimation accuracy under different pointing directions. As illustrated in Figure~\ref{fig5_1a}, the initial distance between the user device and the anchor is set at 5~m. We then use the sliding track to move the user device at a speed of 0.1~m/s over a distance of 20~cm to simulate the ``pointing'' gesture. We traverse a full 360$\degree$ of pointing directions at a step of 10$\degree$ and use our proposed algorithm to estimate the direction. The 10\degree{} angular steps are controlled by a rotation platform with a control accuracy of 0.05\degree{}. It should be noted that the pointing direction vector is 3-D in space. Thus, besides the setting shown in Figure~\ref{fig5_1a}, in which the pointing direction is in the x-y plane, we also repeat the experiment when it is in the y-z and x-z planes. Figure~\ref{fig5_1b} shows the overall direction estimation errors using the data collected under all parameters. For the movement in the x-y plane, the median error and the 90th-percentile error of our design are $2.33\degree$ and $4.20\degree$, respectively. For the movement in the y-z plane, the median error and the 90th-percentile error are $2.39\degree$ and $4.32\degree$, respectively. For the movement in the x-z plane, the median error and the 90th-percentile error are $2.44\degree$ and $4.34\degree$, respectively. Figure~\ref{fig5_1c} shows the median estimation error for each direction in these settings. It is evident that there is no clear variation among different directions for each setting.  Meanwhile, the performance under the three settings is similar, indicating that PnPSelect achieves comparable pointing-direction estimation accuracy across tested movement planes. 

\begin{figure*}[!t]
    \centering
        \subfloat[Experiment setup.]{
        \includegraphics[height=1.3in]{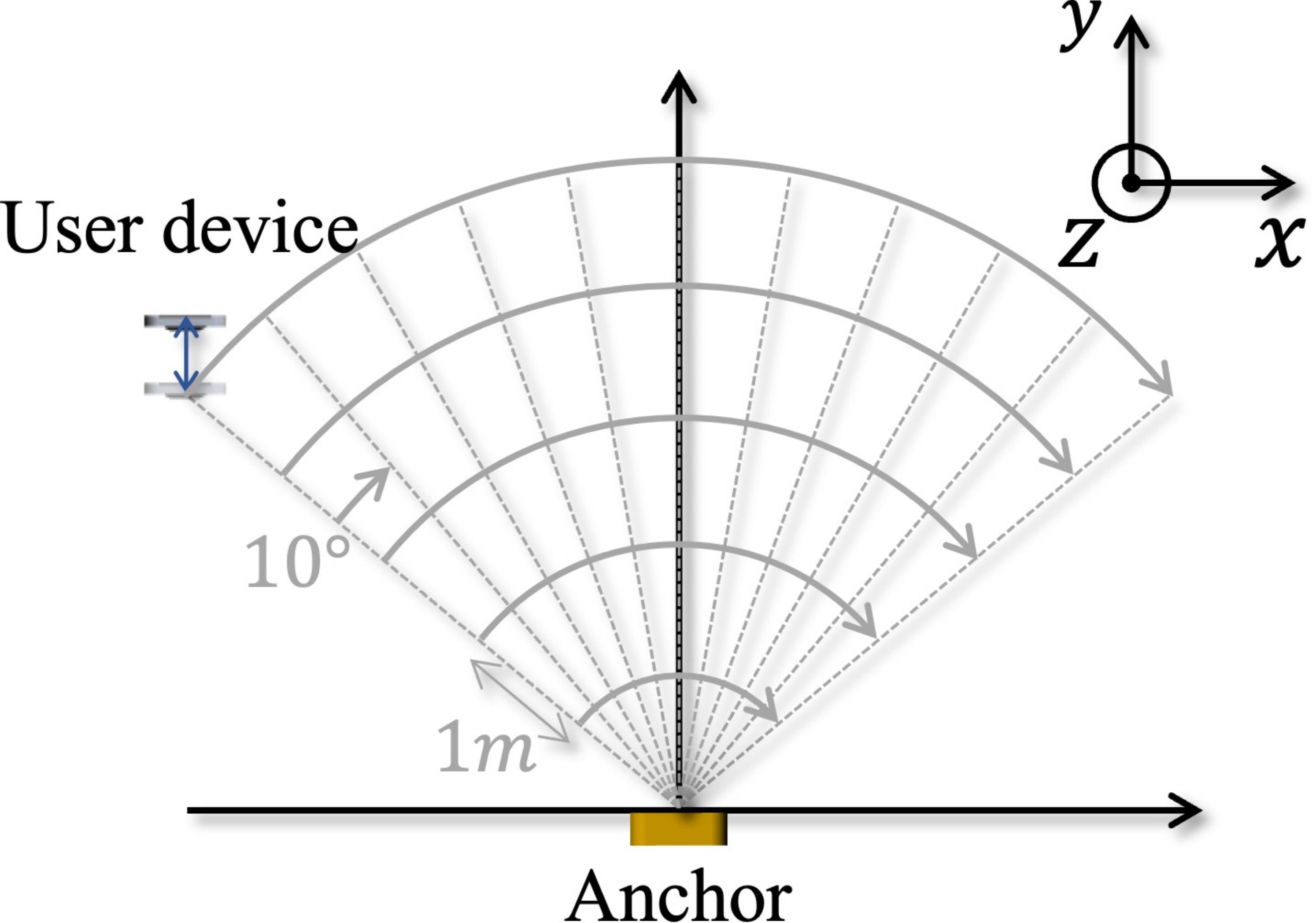}
        \label{fig5_2a}
        \hspace{0.1in}
    }
        \subfloat[Different distances between user device and anchor.]{
        \includegraphics[height=1.3in]{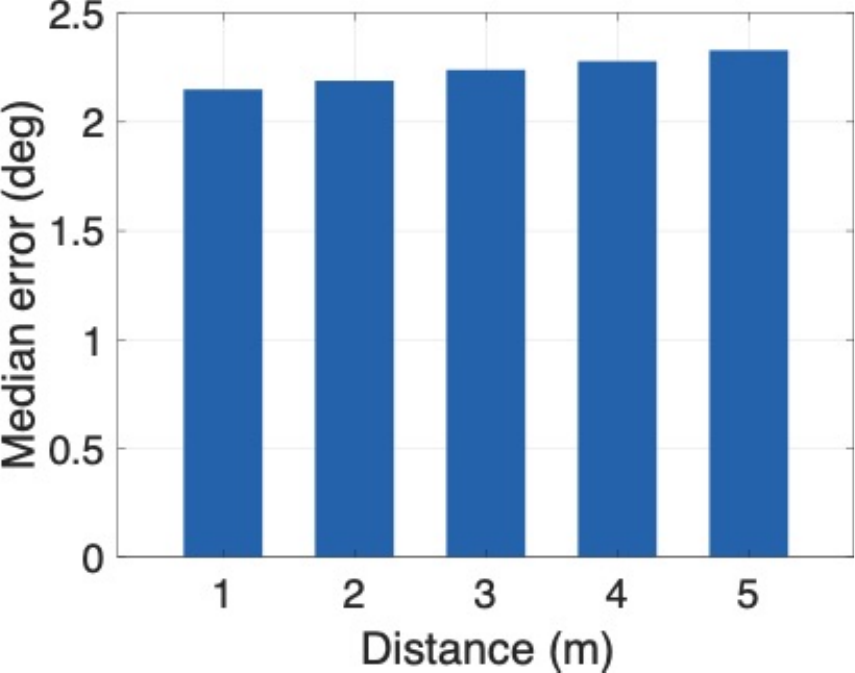}
        \label{fig5_2b}    
    }\hspace{0.3in}
        \subfloat[Different angles between user device and anchor.]{
        \includegraphics[height=1.3in]{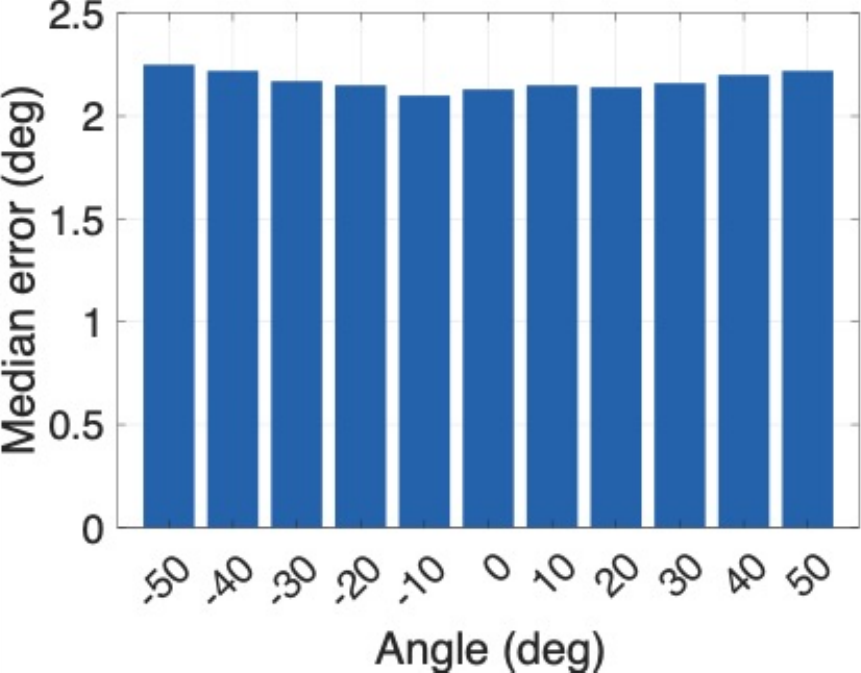}
        \captionsetup{justification=centering}
        \label{fig5_2c}
        \hspace{0in}
    }
    \caption{Impact of user device location.}
    \label{fig5_2}
\end{figure*}

\subsubsection{Impact of User Device Location}\label{sec513}

In this experiment, we evaluate the impact of user device location on pointing direction estimation. Note that since Section 5.1.2 shows similar performance trends across different movement planes, we use the x-y plane as a representative movement plane in this experiment and focus on evaluating the effect of position. Specifically, we characterize the user-device location using two parameters: its distance from the anchor and its angular position relative to the anchor. As shown in Figure~\ref{fig5_2a}, we vary the distance between the user device and the anchor from 1~m to 5~m at a step of 1~m. At each distance, we move the user device from $-50\degree$ to $50\degree$ in increments of $10\degree$. The 10\degree{} angular settings are implemented by marking predefined positions on the floor. Specifically, we first determine the desired angular positions according to the experimental geometry and convert them into corresponding floor locations based on the measured distances. We then mark these locations before the experiment and place the user device at the corresponding marked positions during data collection. Note that we observe that Apple's Nearby Interaction API stops reporting angle estimates when the angle exceeds 55$^\circ$. Therefore, we limit the tested angle range to within $\pm50^\circ$ to ensure that the evaluation is conducted within the reliable angle-estimation range of the device. We calculate the absolute error between the estimated value and the ground truth. Figure~\ref{fig5_2b} and~\ref{fig5_2c} show the impact of distance and angle between the user device and anchor. Although we find that the pointing direction error slightly increases with the angle and distance of the user device with respect to the anchor, for all locations of the user smartphone, the median direction estimation error and the 90th-percentile error are at most $2.33\degree$ and $4.92\degree$, respectively. These results show that our design can achieve a high direction estimation accuracy in typical home-sized environments.

\begin{figure}[!b]
    \centering
    \subfloat[Setup for evaluating displacements.]{
        \includegraphics[height=1.25in]{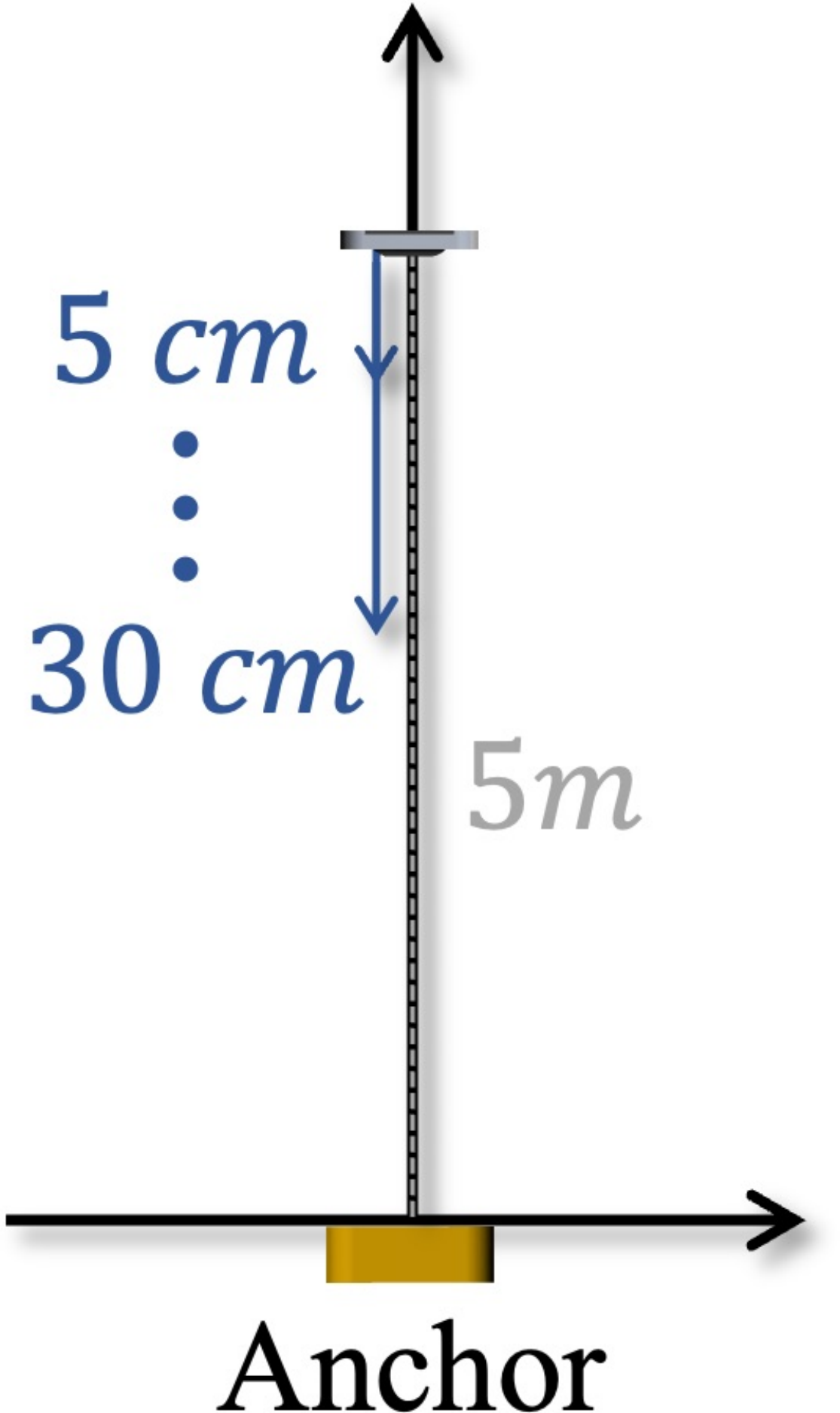}
        \label{fig5_3c}
    }
    \hspace{0.1in}
        \subfloat[Impact of different displacements.]{
        \includegraphics[height=1.25in]{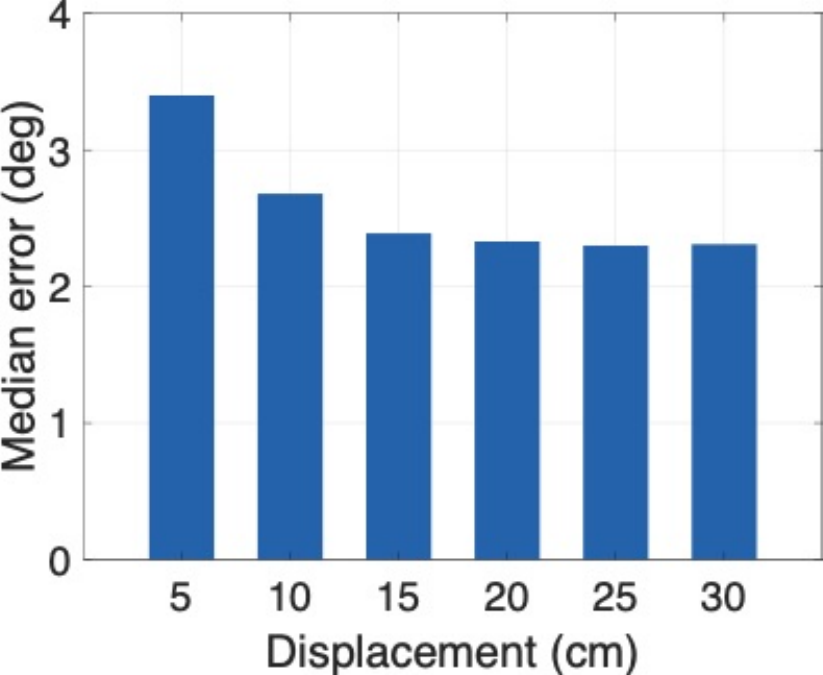}
        \label{fig5_3a}
    }
    \hspace{0.1in}
    \subfloat[Setup for evaluating velocities.]{
        \includegraphics[height=1.25in]{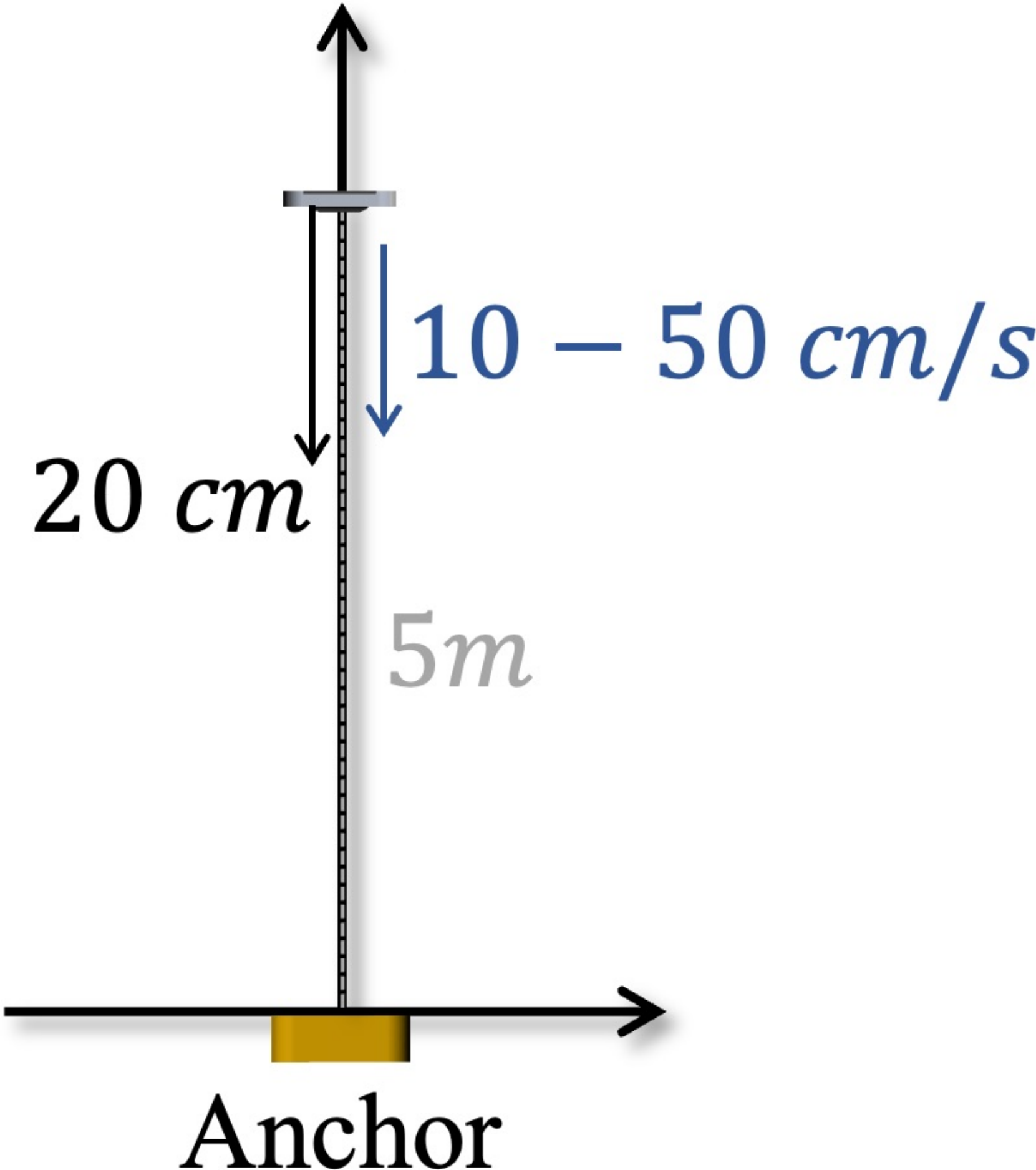}
        \label{fig5_3d}
    }
    \hspace{0.1in}
        \subfloat[Impact of different velocities.]{
        \includegraphics[height=1.25in]{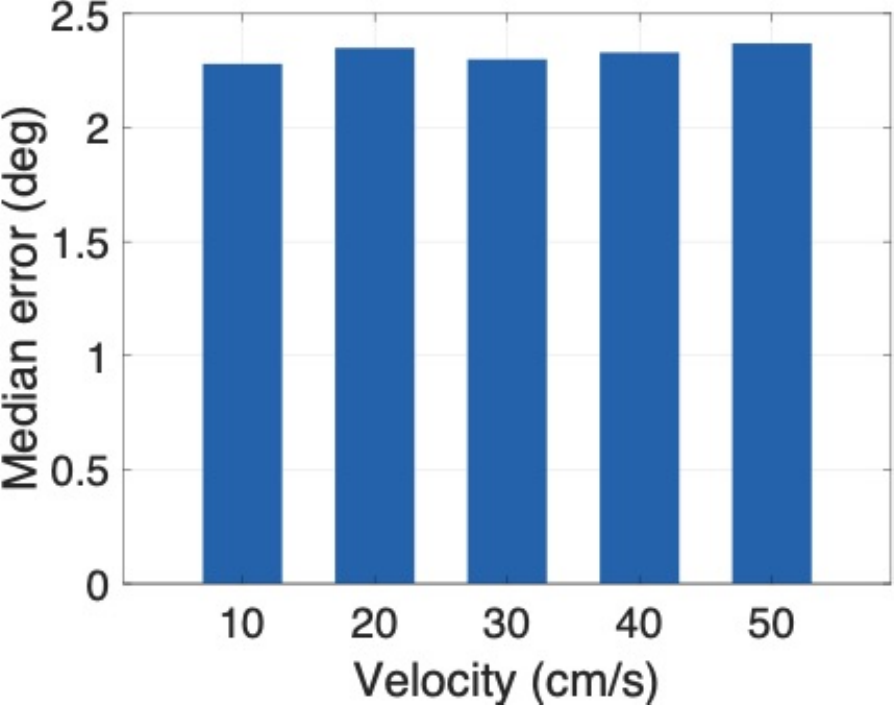}
        \label{fig5_3b}
        \hspace{0in}
    }
    \vspace{-0.0em}
    \caption{Impact of pointing displacement and velocity.}
    \label{fig5_3}
\end{figure}

\subsubsection{Impact of Pointing Displacement}\label{sec514}

In reality, different users may exhibit various pointing gesture patterns, leading to distinct displacements while performing the gesture. Moreover, even the same individual is likely to experience variations in displacement during usage. Therefore, in this experiment, we evaluate the impact of different pointing gesture displacements on direction estimation. As shown in Figure~\ref{fig5_3c}, we set six different displacements ranging from 5 cm to 30 cm at a step of 5 cm. As shown in Figure~\ref{fig5_3a}, when the displacement is 5 cm, the median direction estimation error and the 90th-percentile error are 3.40$\degree$ and 6.91$\degree$, respectively. When the displacement is 10 cm or more, the median error and the 90th-percentile error decrease and stabilize at below 2.68$\degree$ and 5.60$\degree$, respectively. This indicates that a longer displacement improves the signal-to-noise ratio of the trajectory and reduces the relative influence of hand jitter and UWB localization errors. Notably, as long as the displacement exceeds 10 cm, the system can ensure an error sufficiently small, and a movement of 10 cm is easily achievable. Section~\ref{sec533} further examines whether users naturally produce pointing displacements above this threshold. 

\subsubsection{Impact of Pointing Velocity}

Similar to the displacements evaluated in the previous section, pointing velocity may vary across users and gestures. Therefore, in this experiment, we evaluate the impact of different pointing gesture velocities on direction estimation. As shown in Figure~\ref{fig5_3d}, we evaluate five different velocities ranging from 10 cm/s to 50 cm/s, at a step of 10 cm/s. As illustrated in Figure~\ref{fig5_3b}, for all velocities, the median estimation error and the 90th-percentile error remain below 2.37$\degree$ and 4.93$\degree$, respectively, and there is no clear difference in errors across different velocities. This indicates that varying gesture speeds have a minimal impact on direction estimation. The distribution of speeds of multiple users in real-world conditions is also presented in Section~\ref{sec533}.

\subsubsection{Impact of NLoS Conditions}

PnPSelect relies on UWB localization of the user device to estimate its motion trajectory. Therefore, NLoS conditions between the anchor and the user device may affect pointing-direction estimation. To evaluate this impact, we test several common indoor obstruction conditions between the anchor and the user device: an office chair with a mesh back, a 1~cm-thick wooden cabinet door, a 10~cm-thick interior partition wall, a transparent glass panel, a laptop, complete body blockage when the user holds the phone with the body blocking the LoS path to the anchor, and direct hand blockage of the UWB antenna area near the camera module of the iPhone~\cite{heinrich2023smartphones}. In each case, the obstruction was placed along the LoS path, 30~cm away from the anchor, and the user device was placed 3~m away on the other side of the obstruction.

\begin{figure}[t]
    \centering
     \includegraphics[height=1.6in]{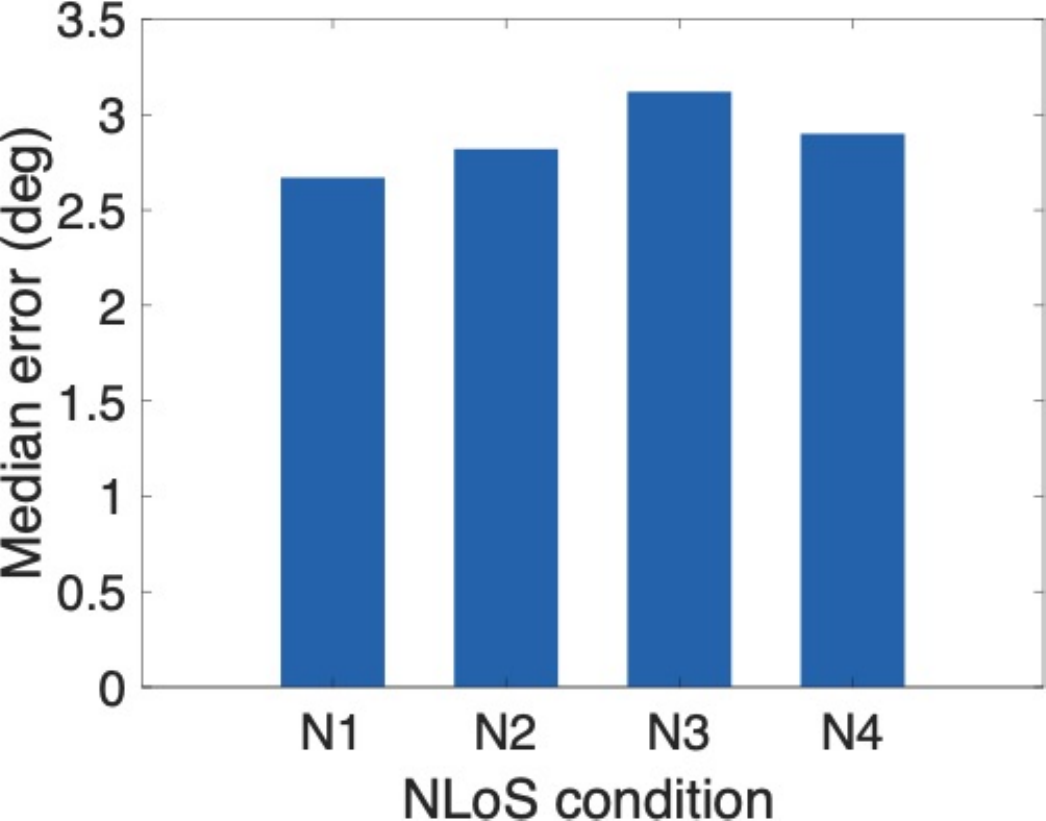}
    \caption{The impact of mild NLoS conditions.}
    \label{fig5_n}
\end{figure}

Figure~\ref{fig5_n} shows the pointing-direction estimation error under different obstruction conditions. The results show that PnPSelect can still operate under mild NLoS conditions, including the chair (N1), wooden cabinet door (N2), interior wall (N3), and glass panel (N4), where the anchor can still obtain sufficiently stable UWB measurements. Under these mild NLoS conditions, the maximum median error and the maximum 90th-percentile error are 3.12\degree{} and 5.28\degree{}, respectively. However, stronger obstruction, including the laptop, complete body blockage, and direct hand blockage of the UWB antenna area, causes significant localization drift even when the phone is kept static. Under these conditions, PnPSelect cannot reliably estimate the pointing direction.

These results indicate that PnPSelect does not require perfect optical LoS, but it does require reliable UWB localization coverage between the anchor and the user device. In practical deployments, the anchor should be placed to cover the main interaction area, and users should avoid covering the UWB antenna region during interaction. A practical implementation could monitor UWB localization quality and prompt the user to adjust their posture or position when unstable measurements are detected.

\subsection{Evaluation of Spatial Resolution and Device Selection Accuracy}

In real-world smart home environments, IoT devices may be densely placed. Therefore, a crucial factor in determining the effectiveness of the system is its spatial resolution. Suppose there are two closely positioned IoT devices. The key question is the minimum distance at which the system can accurately discern and select one device when the user points at it. We define this minimum distance as the spatial resolution. Intuitively, this spatial resolution is closely related to the accuracy of the pointing direction estimation. In this section, we validate the spatial resolution through several experiments to confirm its alignment with theoretical predictions.

\begin{figure}[b]
    \centering
        \subfloat[Experiment setup.]{
        \includegraphics[height=1.1in]{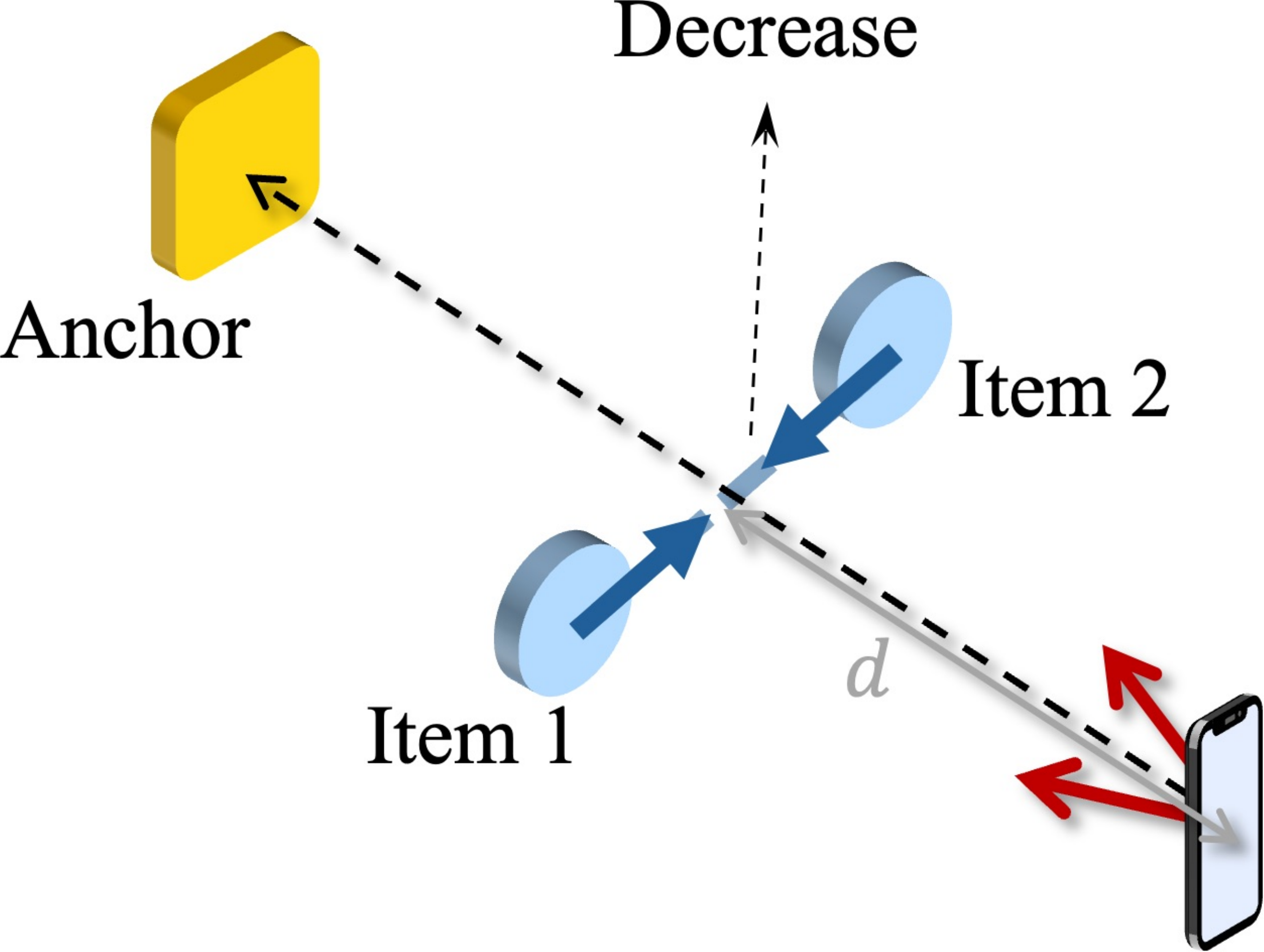}
        \label{fig5_5a}
    }
    \hspace{0.3in}
        \subfloat[Spatial resolutions at different distances.]{
        \includegraphics[height=1.1in]{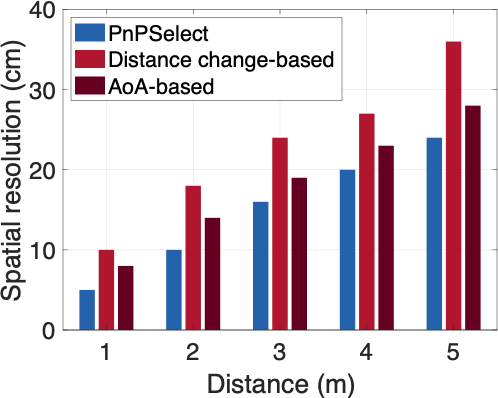}
        \label{fig5_5b}
        \hspace{0in}
    }
    \vspace{-0.0em}
    \caption{Impact of distance between user device and IoT device on spatial resolution. }
    \label{fig5_5}
\end{figure}

\begin{figure*}[!b]
    \centering
        \subfloat[Experiment setup. L indicates the locations of the items.]{
        \includegraphics[height=1.4in]{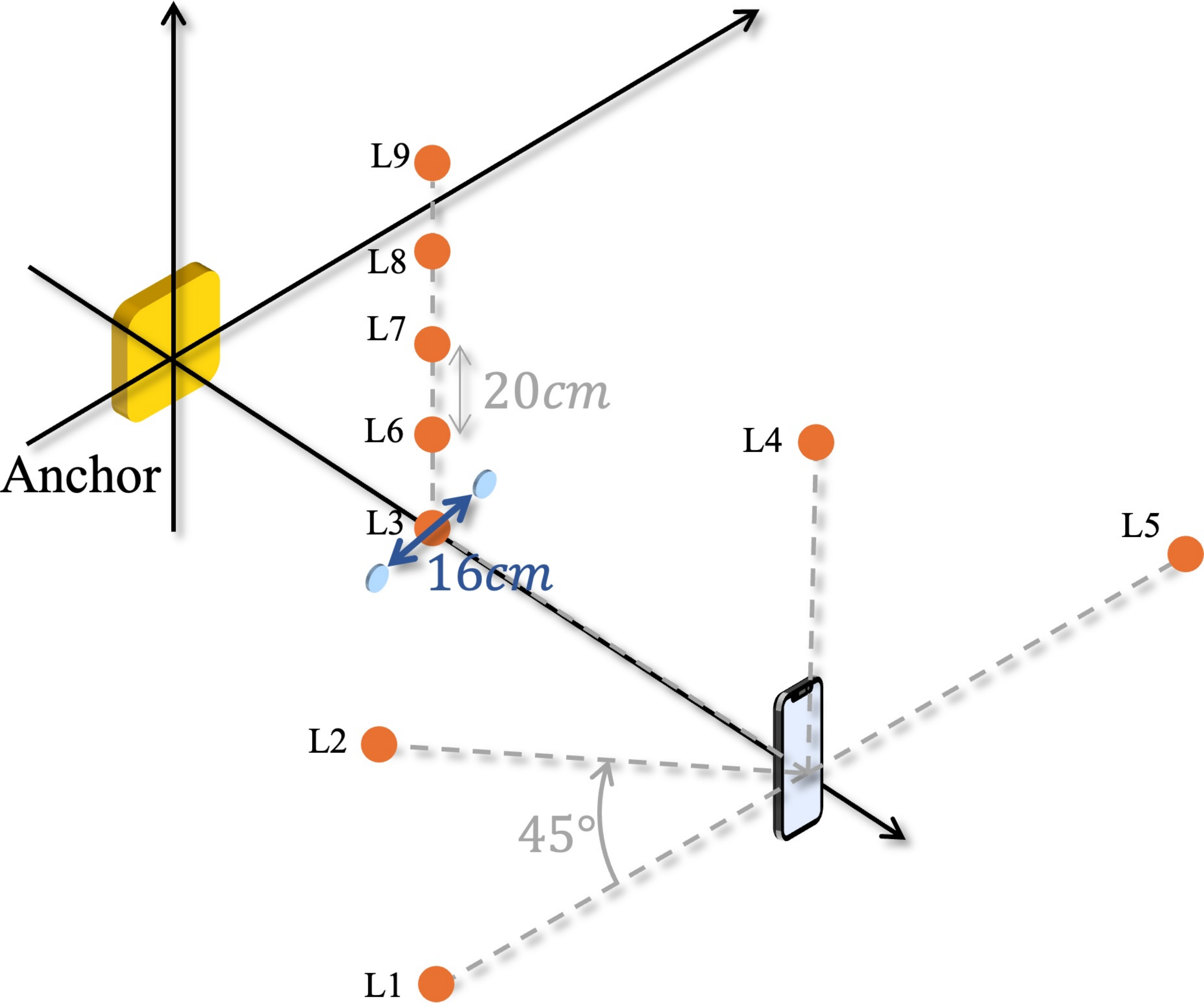}
        \label{fig5_6a}
        \hspace{0in}
    }
        \subfloat[Impact of angle.]{
        \includegraphics[height=1.4in]{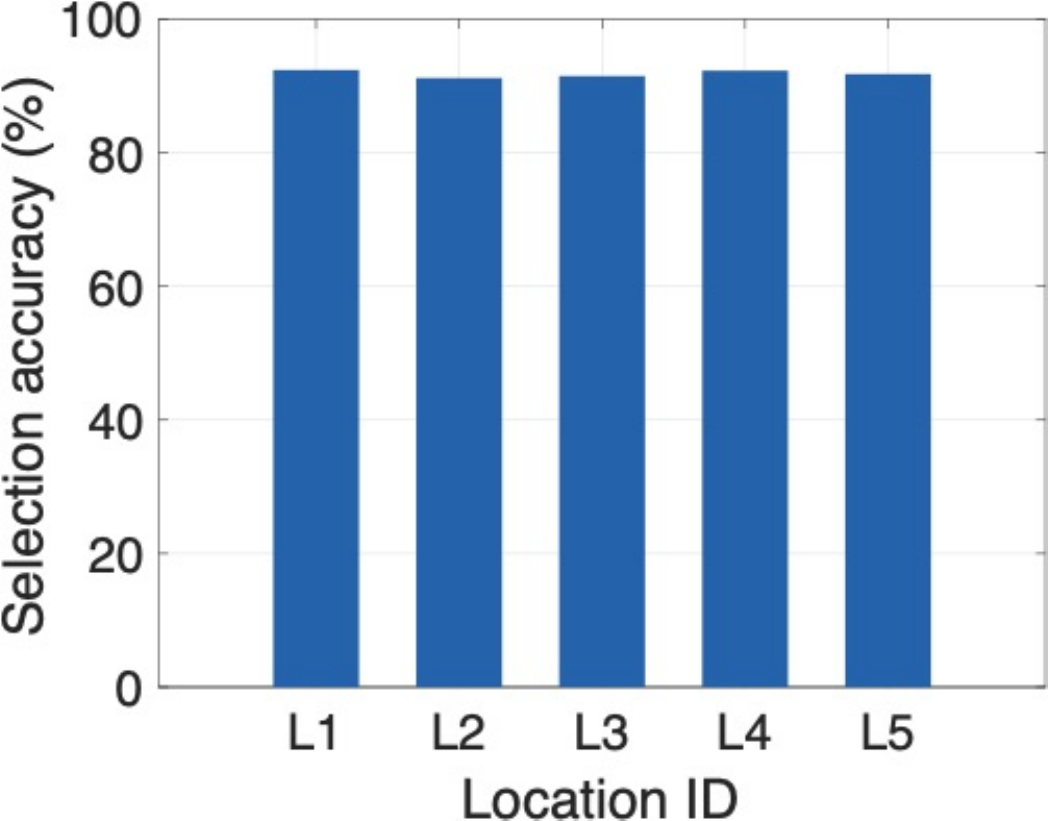}
        \label{fig5_6b}
        \hspace{0in}
    }
        \subfloat[Impact of height.]{
        \includegraphics[height=1.4in]{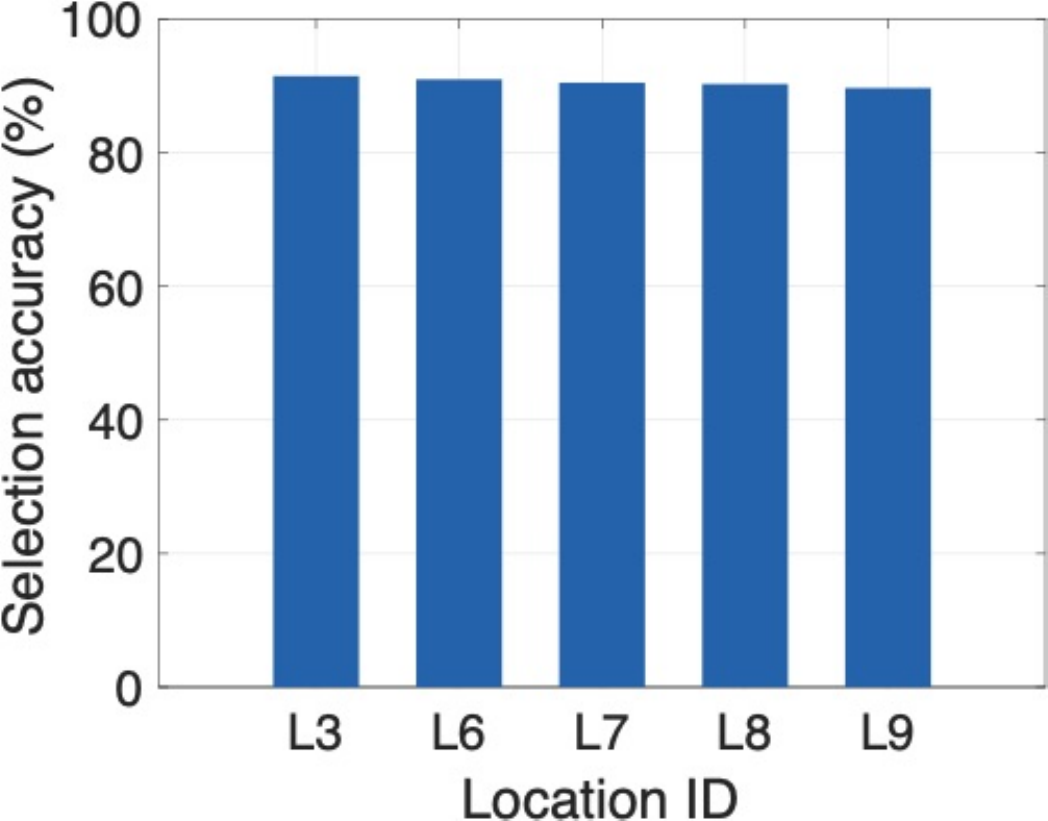}
        \captionsetup{justification=centering}
        \label{fig5_6c}
        \hspace{0in}
    }
    \vspace{-0.0em}
    \caption{Impact of IoT device location on device selection accuracy.}
    \label{fig5_6}
\end{figure*}

\subsubsection{Impact of Distance Between User Device and IoT Device}

In this section, we evaluate the spatial resolution of our system under various settings. As shown in Figure~\ref{fig5_5a}, the distance between the user device and the anchor is fixed at 5 m. We vary the distance between the user device and the IoT device (i.e., $d$) from 1 m to 5 m at a step of 1 m. At each distance, two items, each with a size of $3 cm \times 3 cm \times 6 cm$, are used to represent two IoT devices. We gradually reduce the separation between the two items. The spatial resolution is defined as the minimum separation at which the average selection accuracy remains at or above 90\%. Simultaneously, we compare our system with two baseline methods. The first one is Selecon~\cite{alanwar2017selecon}, a distance change-based solution. Note that acoustic pointing direction estimation systems such as DopLink~\cite{aumi2013doplink} and Spartacus~\cite{sun2013spartacus} also exploit Doppler shifts or distance-change profiles caused by the user's pointing motion to infer pointing intent, and are thus conceptually related to the distance change-based baseline. It should be noted that due to the attenuation and environmental sensitivity of acoustic signals, DopLink and Spartacus mainly operate within a short range, typically within about 3 m. The second baseline is BLEselect~\cite{zhang2023bleselect}, an AoA-based solution. Note that BLEselect~\cite{zhang2023bleselect} is implemented using Bluetooth signals. However, Bluetooth-based systems typically exhibit higher inaccuracies in AoA estimation compared to UWB-based systems, since Bluetooth works with a much smaller bandwidth than UWB. Thus, to ensure a fair comparison, we implement the AoA-based method using UWB instead of Bluetooth. In this way, the hardware and the UWB data used for our design, the distance change-based solution, and the AoA-based solution are identical. The only difference lies in the processing approach of the UWB data for device selection. Meanwhile, UWB anchors must be attached to both items when implementing the baseline approaches, while \systemname{} only requires one anchor in the environment. 

Figure~\ref{fig5_5b} illustrates the spatial resolutions of the three approaches at varying distances. At the five different distances, the spatial resolutions of our method are 5 cm, 10 cm, 16 cm, 20 cm, and 24 cm, respectively. The spatial resolutions for the distance change-based approach are 10 cm, 18 cm, 24 cm, 27 cm, and 36 cm, while those for the AoA-based method are 8 cm, 14 cm, 19 cm, 23 cm, and 28 cm. These results show that PnPSelect achieves finer spatial resolution than the two evaluated baselines while requiring only one environmental anchor.

\subsubsection{Impact of the IoT Device Location}

In practical deployments, IoT devices may be positioned throughout various locations in a room. To this end, our experiment evaluates the impact of IoT device location on spatial resolution and device selection accuracy. As shown in Figure~\ref{fig5_6a}, we fix the distance between the anchor and the user device at 5 m. We evaluate nine different IoT device locations in total. Initially, we fix the distance between two items (simulating IoT devices) and the user device at 3 m. The distance between the two items is fixed at 16 cm, which is the spatial resolution at 3 m. We begin by changing the angle of the two items relative to the user device in increments of 45$\degree$ (L1 - L5). Subsequently, using L3 as a reference, we change the heights of the two items by moving the two items upwards in increments of 20 cm (L6 - L9). We calculate the device selection accuracy of two items at each location. As illustrated in Figure~\ref{fig5_6b}, at five different angles, the device selection accuracy exceeds 91.2\%. At five different heights, the device selection accuracy remains above 89.7\% as shown in Figure~\ref{fig5_6c}. These results demonstrate that PnPSelect maintains stable device-selection accuracy across the evaluated IoT device placements.

In conclusion, spatial resolution determines how far apart two IoT devices need to be at different user device-IoT device distances. For example, when the two devices are 3~m away from the user device, our results show that a 16~cm separation is sufficient for reliable selection.

\subsection{Evaluation in Real-World Environments}\label{sec53}

In this section, we evaluate our system in four real-world environments, including two bedrooms, a classroom, and a meeting room. In each environment, we select 10 common pieces of furniture or appliances as potential IoT devices. We first evaluate the accuracy of the proposed IoT device location estimation method, which is essential for real-world system development. Then, we evaluate the device selection accuracy within each setting. Additionally, we investigate the impact of several realistic parameters on accuracy, including different users, locations, pointing displacements and velocities, as well as various types of user devices.

\begin{figure*}[!t]
    \centering
        \subfloat[S1: a bedroom.]{
        \includegraphics[height=1.3in]{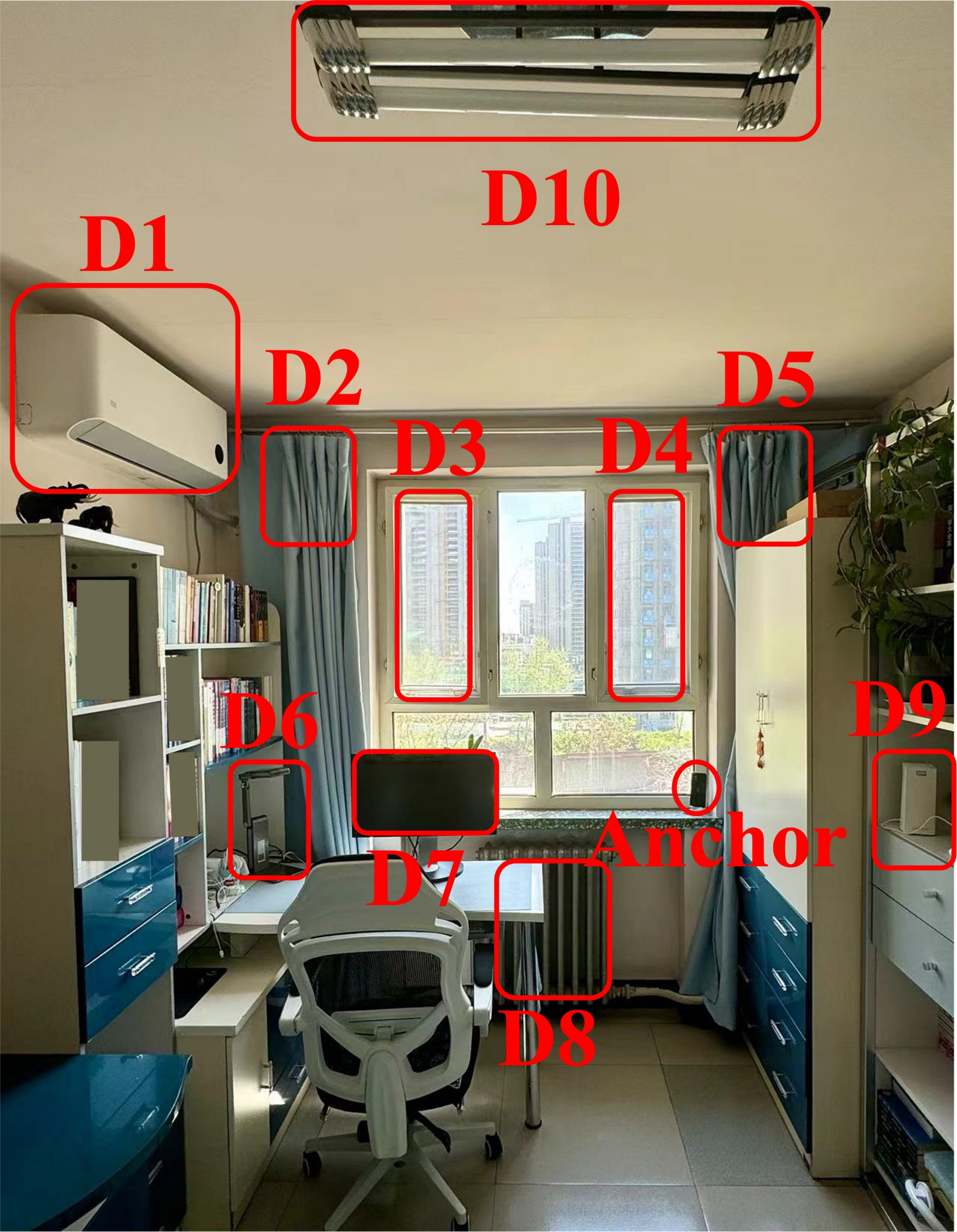}
        \label{fig5_7a}
        \hspace{0in}
    }
        \subfloat[S2: a bedroom.]{
        \includegraphics[height=1.3in]{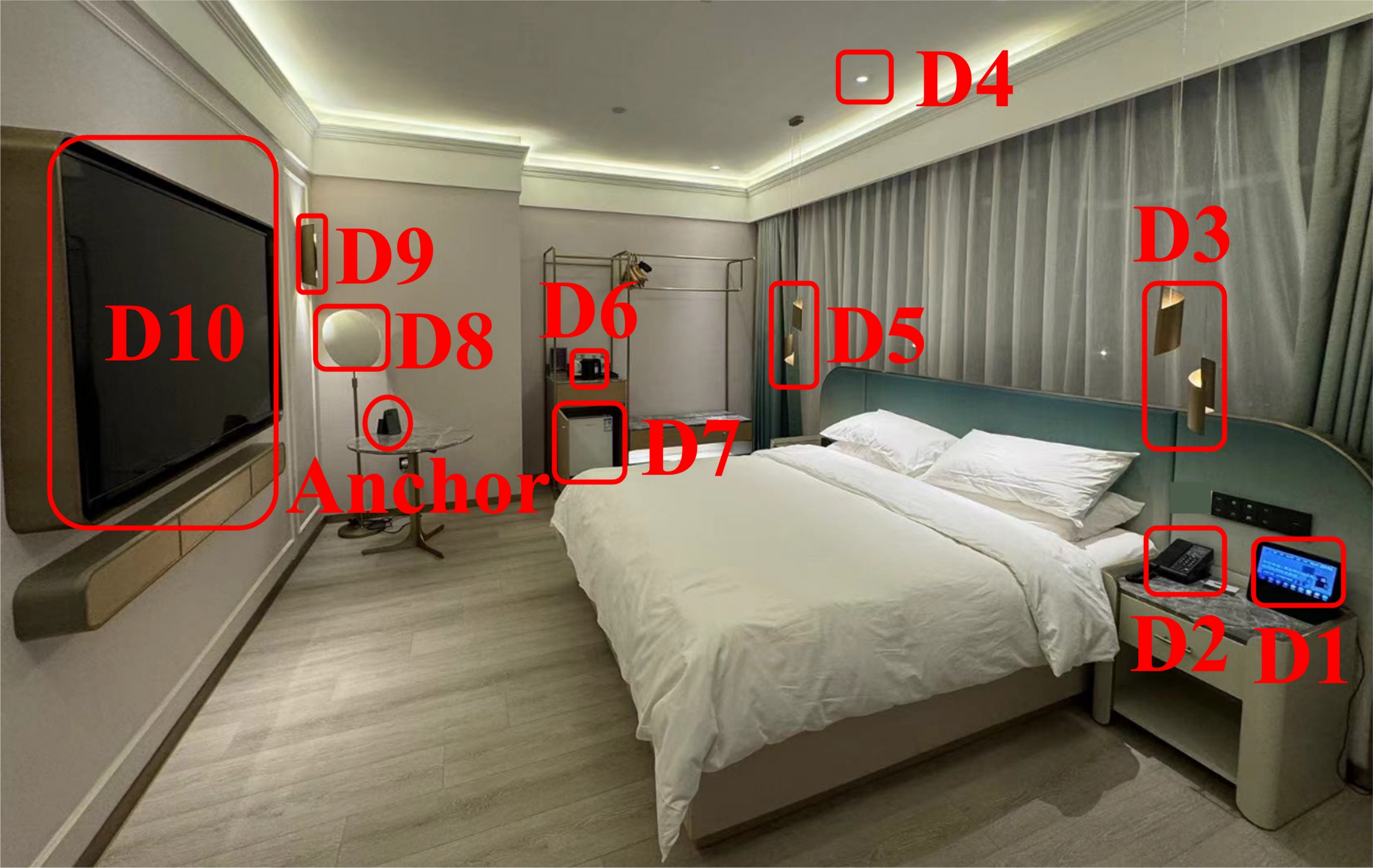}
        \label{fig5_7b}
        \hspace{0in}
    }\\
        \subfloat[S3: a classroom.]{
        \includegraphics[height=1.3in]{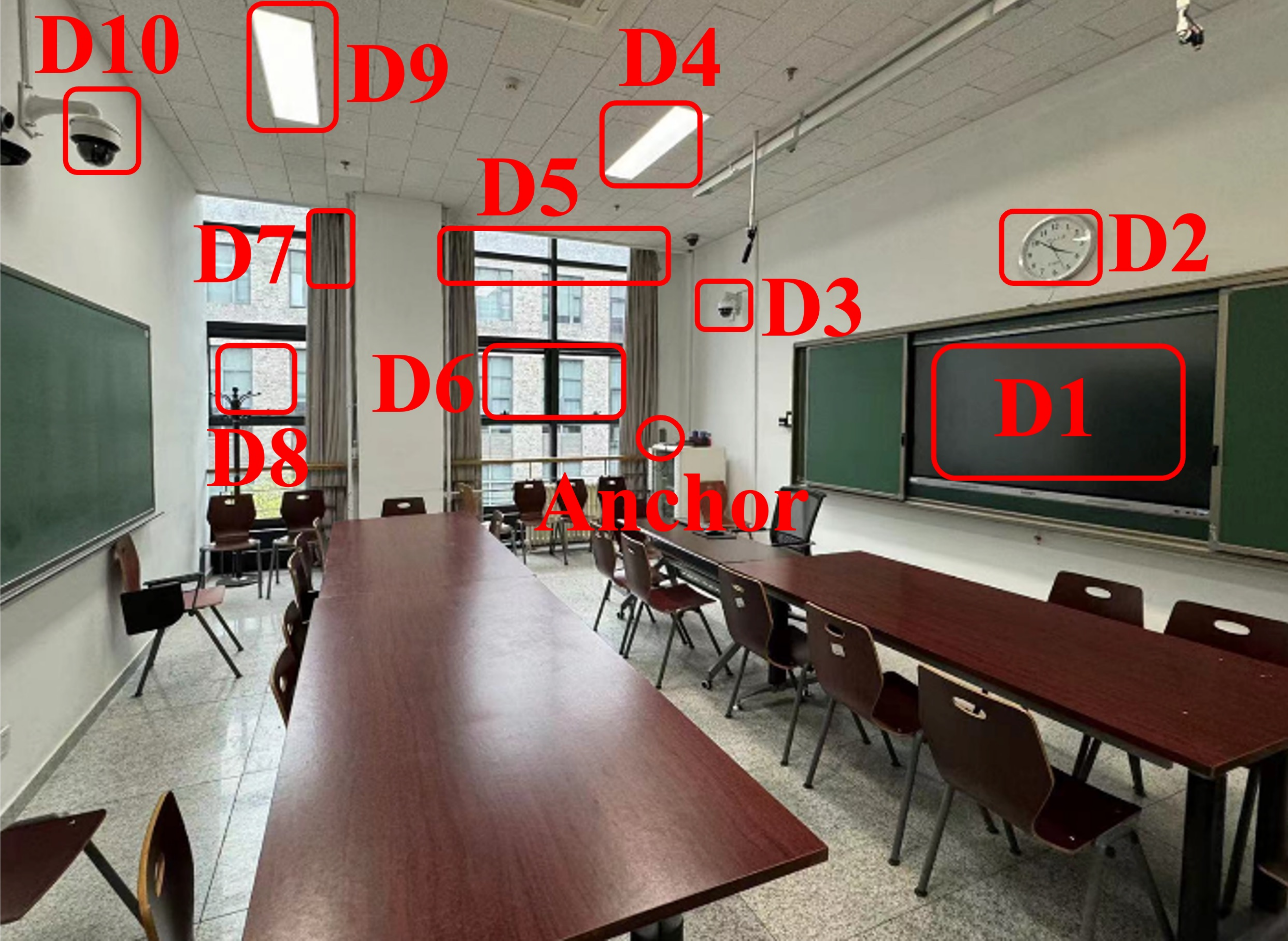}
        \label{fig5_7c}
        \hspace{0in}
    }
    \subfloat[S4: a meeting room.]{
        \includegraphics[height=1.3in]{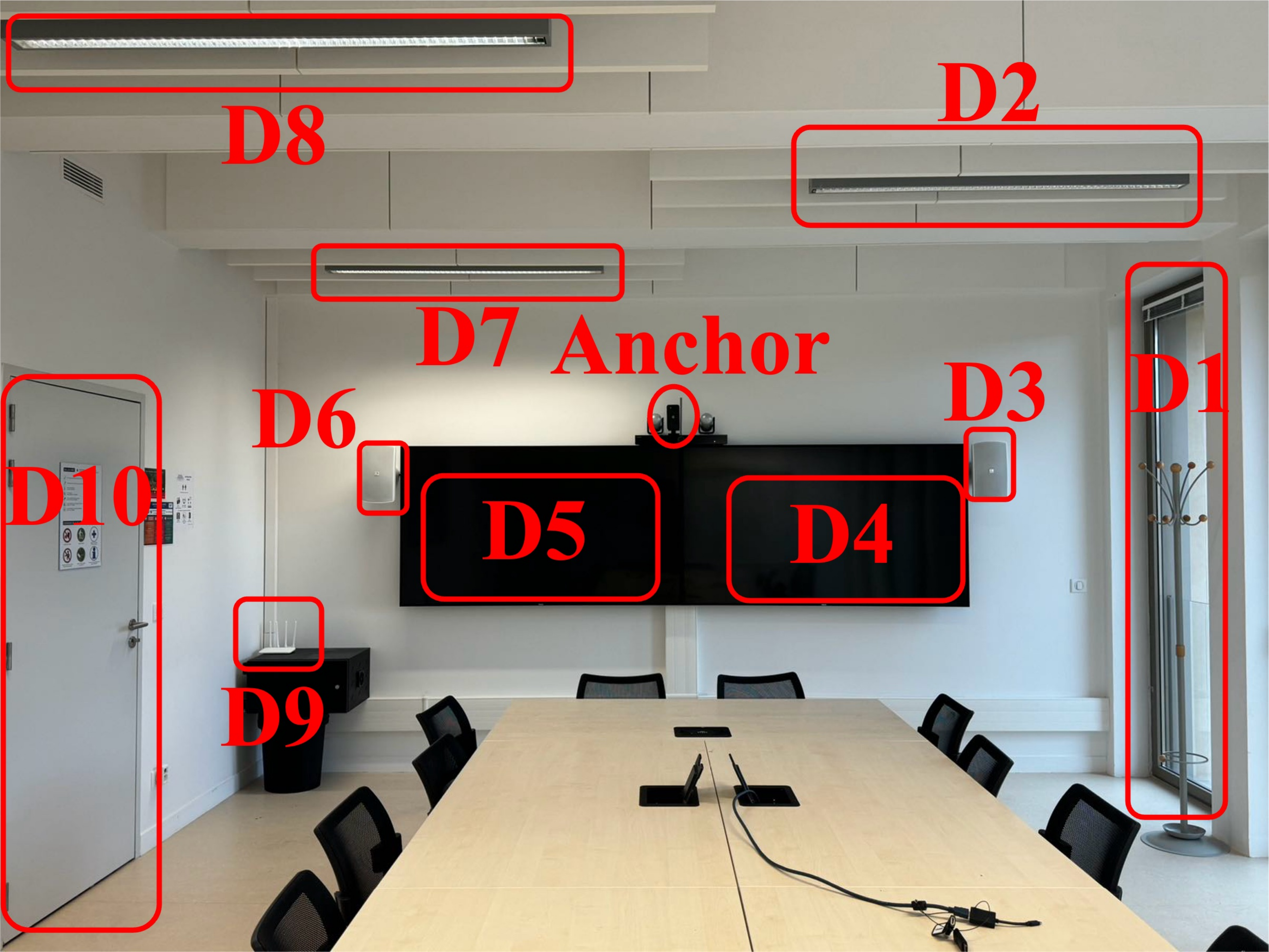}
        \label{fig5_7d}
        \hspace{0in}
    }
    \vspace{-0.0em}
    \caption{Real-world scenarios.}
    \label{fig5_7}
\end{figure*}

\subsubsection{Experiment setting}
The four environments are shown in Figure~\ref{fig5_7} and are denoted as S1, S2, S3, and S4, respectively.

\textbf{S1}: A small bedroom with a size of $2.8~m~\times~3.8~m$. There are ten devices in the bedroom, including an air conditioner~(D1), two curtains~(D2 and D5), two windows~(D3 and D4), two lights~(D6 and D10), a monitor~(D7), a heater~(D8), and a router~(D9).

\textbf{S2}: A large bedroom with a size of $4.1~m~\times~5.0~m$. There are ten devices in this bedroom, including a smart speaker~(D1), a telephone~(D2), five lamps~(D3, D4, D5, D8, and D9), a coffee machine~(D6), a refrigerator~(D7), and a TV~(D10).

\textbf{S3}: A classroom with a size of $3.0~m~\times~5.5~m$. There are ten devices in the classroom, including a monitor~(D1), a clock~(D2), two cameras~(D3 and D10), two lights~(D4 and D9), two curtains~(D5 and D7), and two windows~(D6 and D8).

\textbf{S4}: A meeting room with a size of $3.5~m~\times~6~m$. There are ten devices in the meeting room, including a window~(D1), three lights~(D2, D7, and D8), two speakers~(D3 and D6), two monitors~(D4 and D5), a router~(D9), and a door~(D10).

In each environment, we recruit six participants to point at all devices at ten different locations. At each location, each participant points at each device five times. Thus, we collect a number of 6 participants $\times$ 10 locations $\times$ 5 times = 300 data trials for each IoT device in an environment. For fair comparison across users, the ten user positions in each environment are predefined by the experimenters to cover representative standing locations within the usable area of the room, thereby reflecting typical positions from which users may interact with household appliances in practice. These positions are distributed throughout the room, rather than concentrated in a single region. During the experiments, the participants are asked to hold the device and perform the pointing gesture in their own most natural and comfortable manner, rather than following a strictly constrained posture or gesture style. At the same time, we provided two basic instructions to avoid severe self-induced NLoS conditions: participants were asked not to hold the upper part of the smartphone where the UWB antenna is located (i.e., near the camera module), and not to block the LoS path between the smartphone and the anchor with their body. These instructions were intended to ensure that the anchor could obtain stable UWB measurements during the experiments, while still allowing participants to perform natural pointing gestures.

\begin{figure*}[!t]
    \centering
        \subfloat[Impact of angle difference between two user locations.]{
        \includegraphics[height=1.15in]{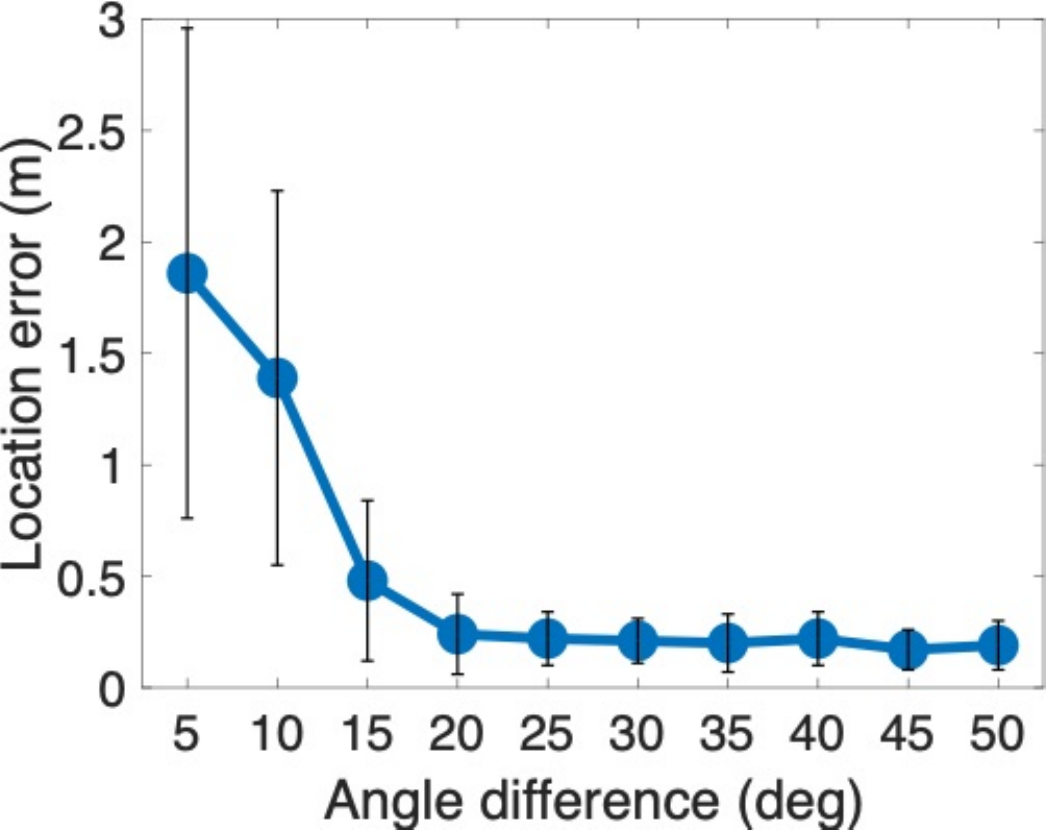}
        \label{fig:fig7_6b}
        \hspace{0in}
    }
        \subfloat[Small angle difference.]{
        \includegraphics[height=1.15in]{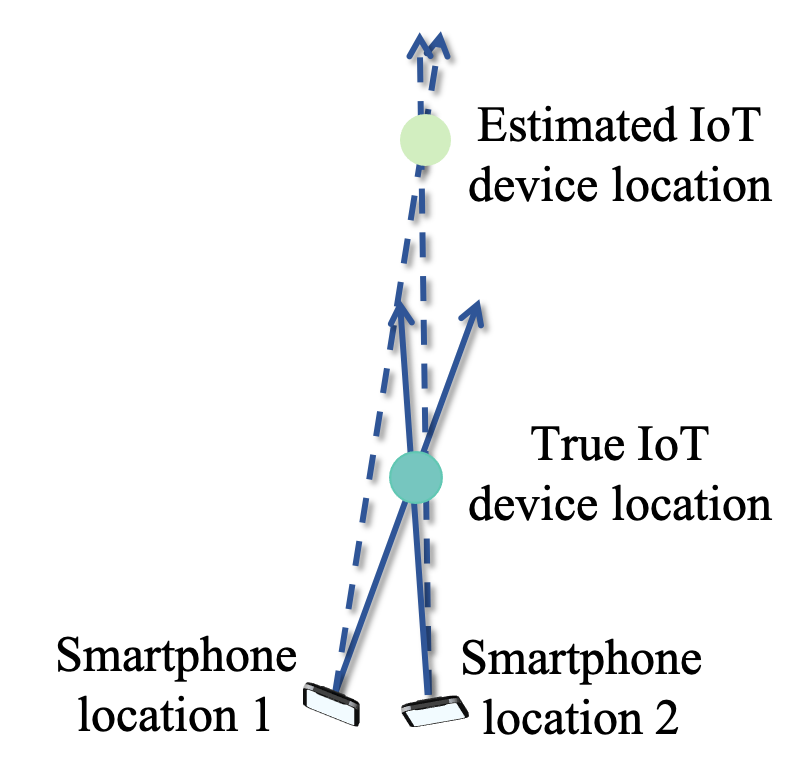}
        \label{fig:fig7_6d}
        \hspace{0in}
    }
        \subfloat[Large angle difference.]{
        \includegraphics[height=1.15in]{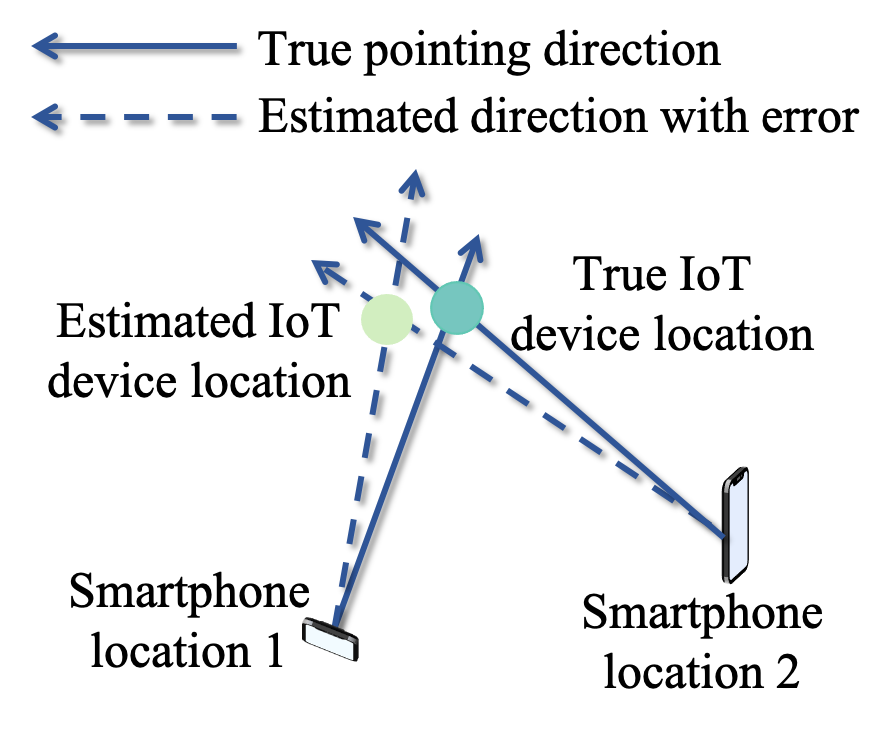}
        \label{fig:fig7_6e}
        \hspace{0in}
    }
    \subfloat[Results for different persons.]{
        \includegraphics[height=1.15in]{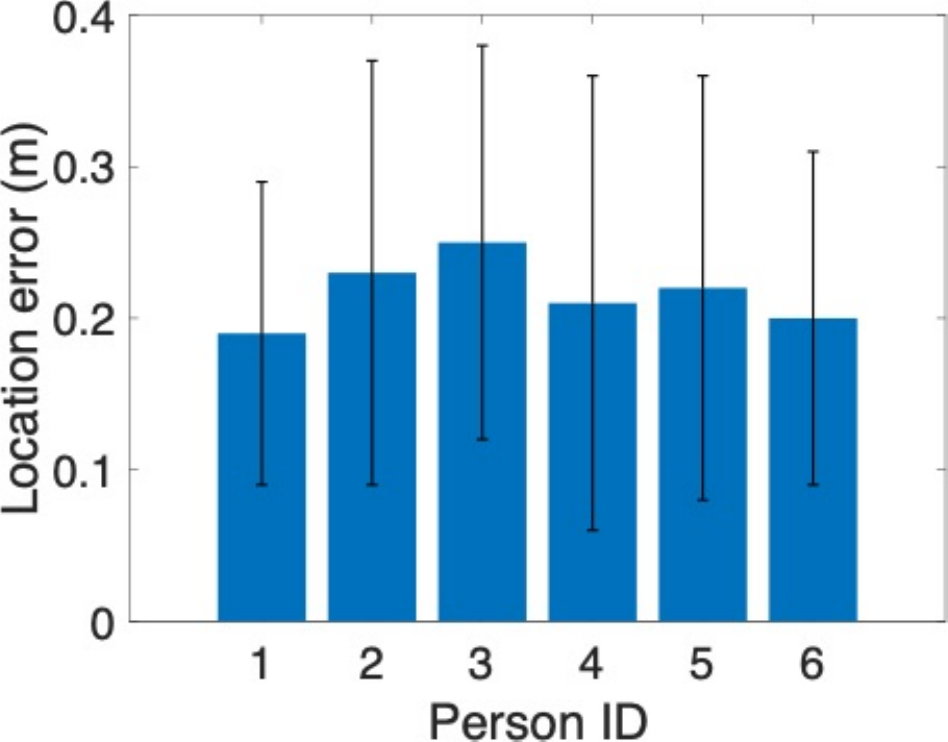}
        \label{fig:fig7_6c}
        \hspace{0in}
    }
    \caption{Evaluation of IoT device location estimation.}
    \label{fig:fig7_6}
\end{figure*}

\subsubsection{Evaluation on IoT Device Location Estimation} \label{sec532}
In this section, we conduct experiments to evaluate the effectiveness of the proposed IoT device location estimation scheme (Section~\ref{sec331}). We conduct this experiment in S1~(Figure~\ref{fig5_7a}). The ground truth location of each IoT device with respect to the anchor is measured manually. We use data collected at two locations to estimate the location of the IoT device.

First, we take a detailed observation of the effect of user location during IoT device location estimation. Specifically, we enumerate all possible pairs among the ten predefined user positions and use each pair to perform the two-position IoT device registration. For each pair of user positions, we estimate the locations of ten IoT devices and compare them with the manually measured ground-truth locations to compute the localization error. We also compute the angular difference between the two user positions with respect to each target IoT device, since this angular separation determines the geometric conditioning of the two-position localization problem. We group the enumerated position pairs into 5\degree{} angular-difference bins and compute the corresponding localization-error statistics for each bin. Figure~\ref{fig:fig7_6b} reports this binned empirical relationship between angular difference and localization error. It can be seen that the IoT device location estimation error decreases with the increase of angle difference. When the angle difference is larger than $20\degree$, the location estimation error tends to be stable and less than 0.23~m. Figure~\ref{fig:fig7_6d} and~\ref{fig:fig7_6e} illustrate the reason behind this observation. As shown in Figure~\ref{fig:fig7_6d}, when the angle difference between two smartphone locations is small, even a slight pointing direction estimation error will induce a large IoT device estimation error. In comparison, if the angle difference is significantly large~(i.e., greater than $20\degree$) as shown in Figure~\ref{fig:fig7_6e}, pointing direction error will not lead to large IoT device location estimation error. Note that when the distance between the user smartphone and the anchor is 4~m, as long as the distance between the two smartphone positions is greater than 1.4~m, the angle difference will be greater than $20\degree$, thereby ensuring a low IoT device location estimation error. Therefore, the user only needs to change a small location to achieve accurate device location estimation. We discuss potential solutions to enhance user experience while ensuring high IoT device location estimation accuracy in Section~\ref{sec62}. 

Figure~\ref{fig:fig7_6c} shows the IoT device location estimation errors for different participants. Note that we only report the results for cases where the angular difference between the two registration positions is larger than 20\degree{}. As shown in Figure~\ref{fig:fig7_6b}, when the angular difference is smaller than 20\degree{}, the localization error increases significantly, indicating that the two-position localization geometry becomes unreliable. Therefore, we regard such configurations as unsuitable for reliable device registration and do not recommend using them in practical deployments. When the angular difference is larger than 20\degree{}, the mean localization error is at most 0.23~m for all participants. 

\begin{figure*}[!t]
    \centering
        \subfloat[S1: a bedroom.]{
        \includegraphics[height=1.05in]{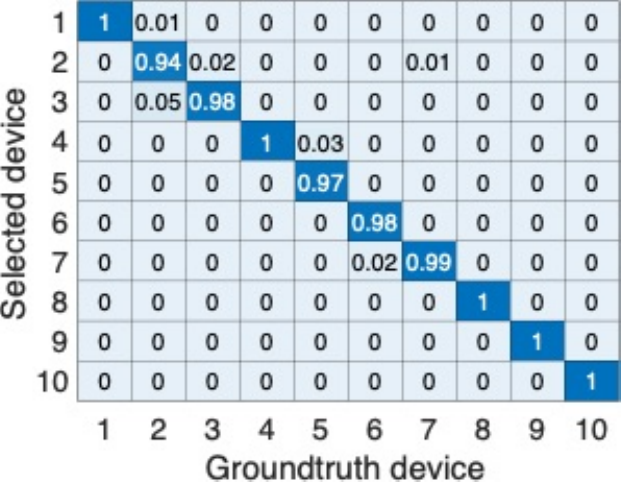}
        \label{fig5_8a}
        \hspace{0in}
    }
        \subfloat[S2: a bedroom.]{
        \includegraphics[height=1.05in]{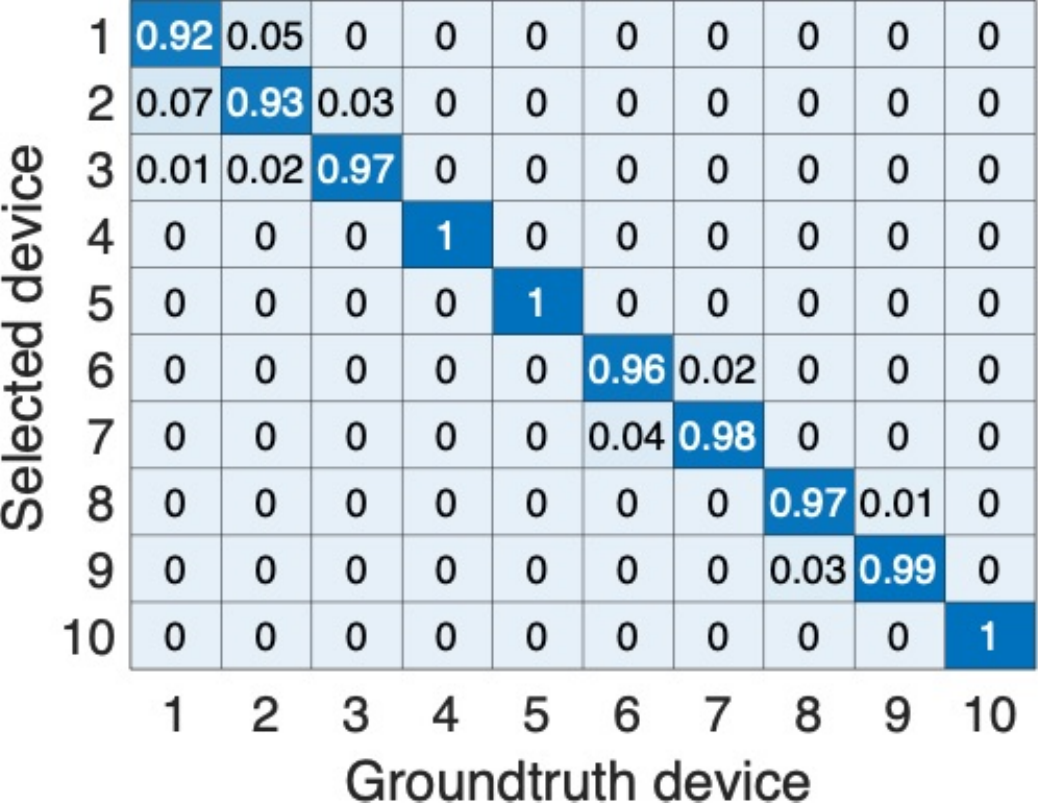}
        \captionsetup{justification=centering}
        \label{fig5_8b}
        \hspace{0in}
    }
        \subfloat[S3: a classroom.]{
        \includegraphics[height=1.05in]{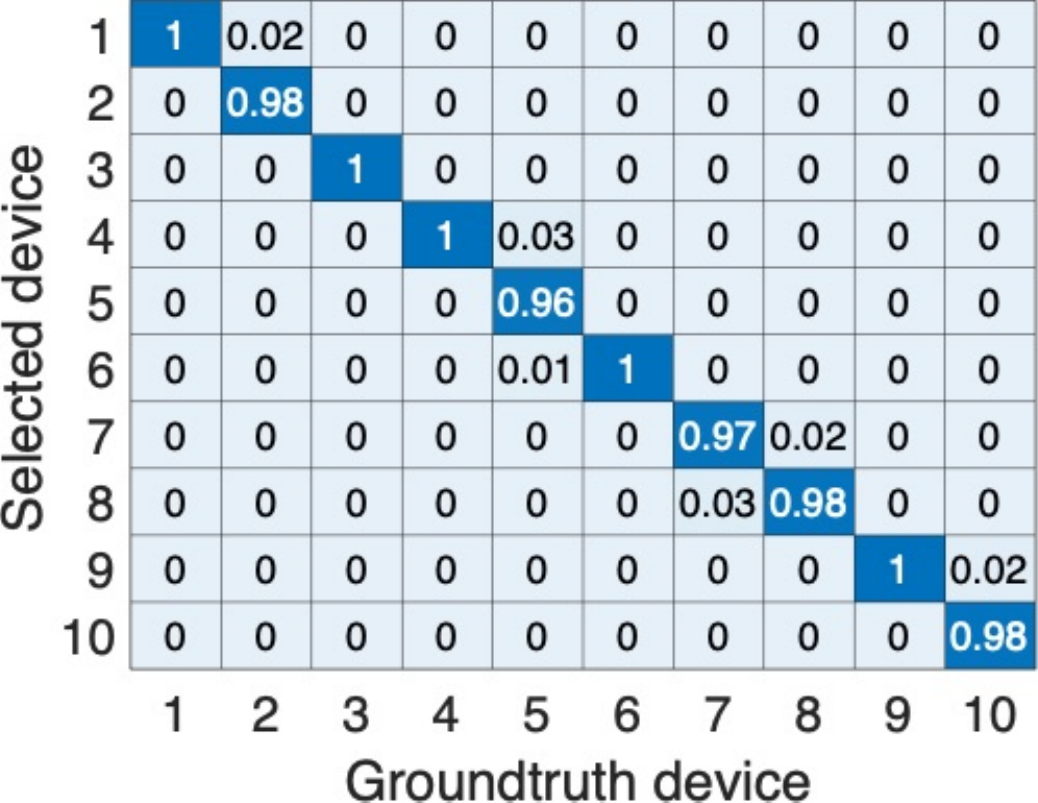}
        \label{fig5_8c}
        \hspace{0in}
    }
        \subfloat[S4: a meeting room.]{
        \includegraphics[height=1.05in]{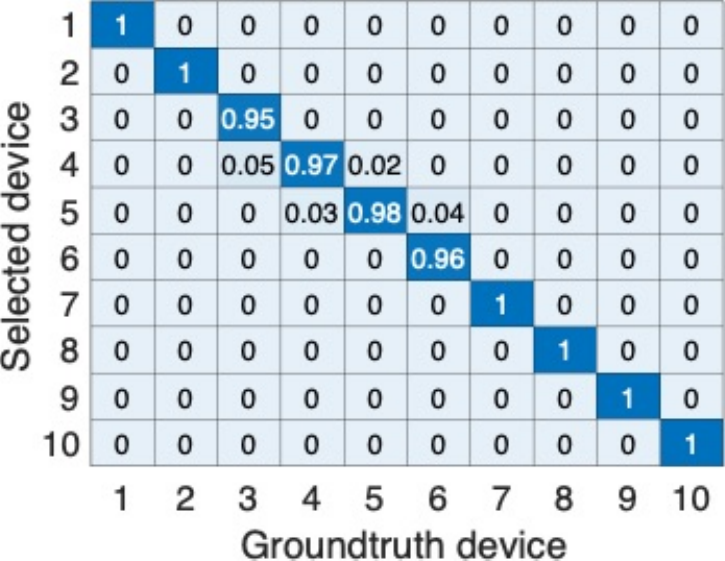}
        \label{fig5_8d}
        \hspace{0in}
    }
    \caption{Confusion matrices of IoT device selection in different environments.}
    \label{fig5_8}
\end{figure*}

\subsubsection{Overall Performance of IoT Device Selection}\label{sec533} 
We then conduct experiments to evaluate the accuracy of the proposed IoT device selection method. Note that in all environments, the location of each IoT device is estimated using our proposed scheme, rather than manual measurement. Figure~\ref{fig5_8} shows the confusion matrices of device selection in each environment. The average accuracy of device selection in four scenarios is 98.6\%, 97.2\%, 98.7\%, and 98.6\%, respectively. We further analyze the cases of incorrect selection. The reason for wrong selection is mainly due to the close proximity of IoT devices. Take D1 and D2 in S2 as an example. As shown in Figure 14~\ref{fig5_7b}, because D1 and D2 are spatially close, an estimated pointing direction intended for D1 may also pass near D2. We discuss potential solutions for this issue to enhance user experience in Section~\ref{sec62}.

\begin{figure*}[!t]
    \centering
        \subfloat[Experiment setup.]{
        \includegraphics[height=1.4in]{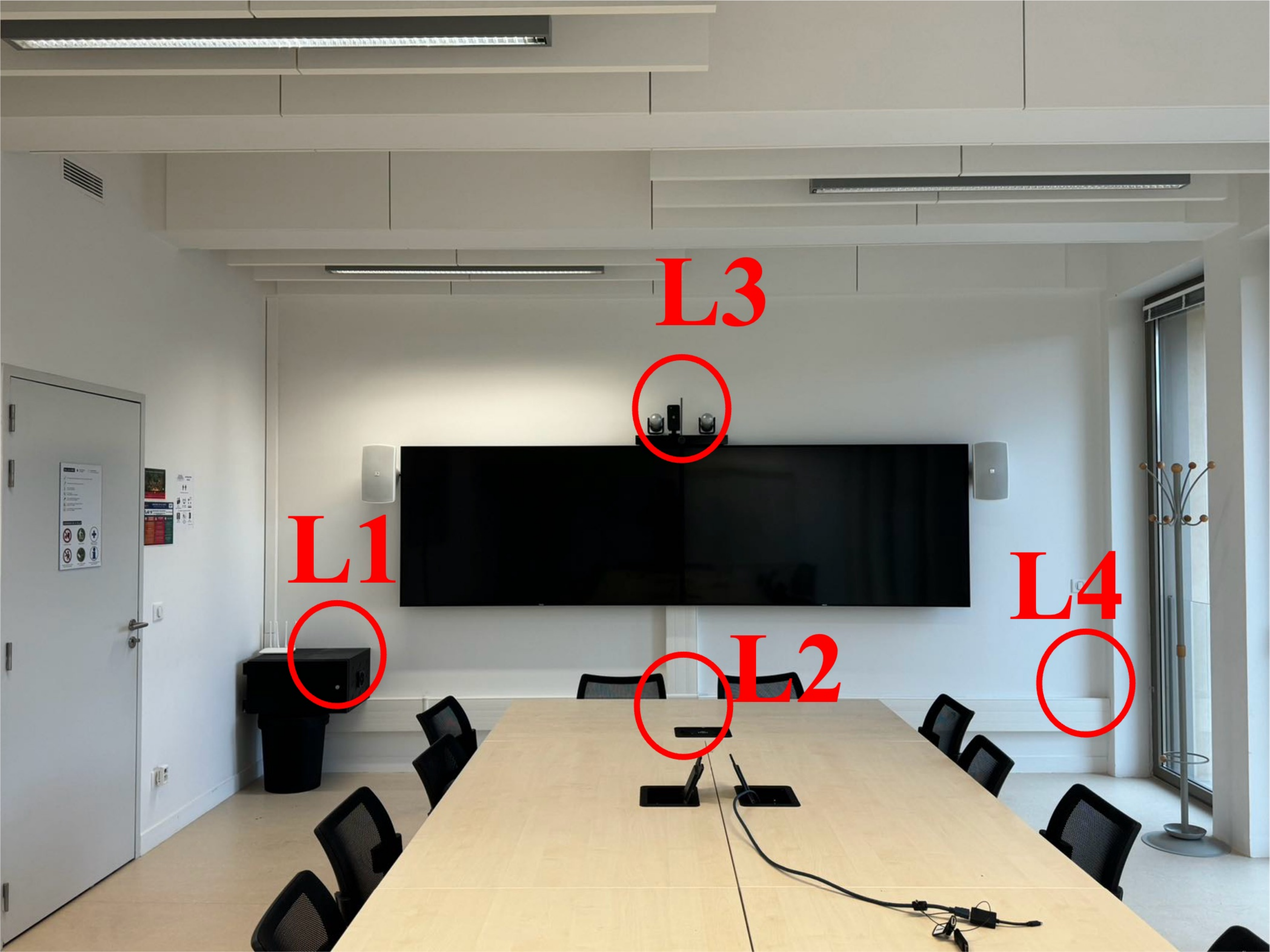}
        \label{fig5_9a}
        \hspace{0in}
    }
        \subfloat[Impact of anchor location.]{
        \includegraphics[height=1.4in]{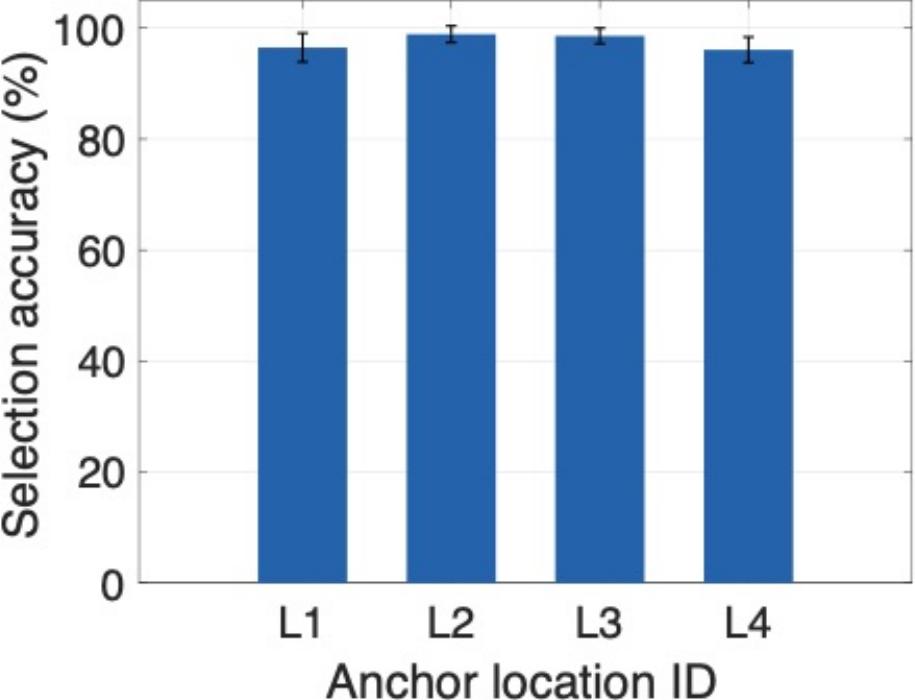}
        \label{fig5_9b}
        \hspace{0in}
    }
        \subfloat[Impact of user location.]{
        \includegraphics[height=1.4in]{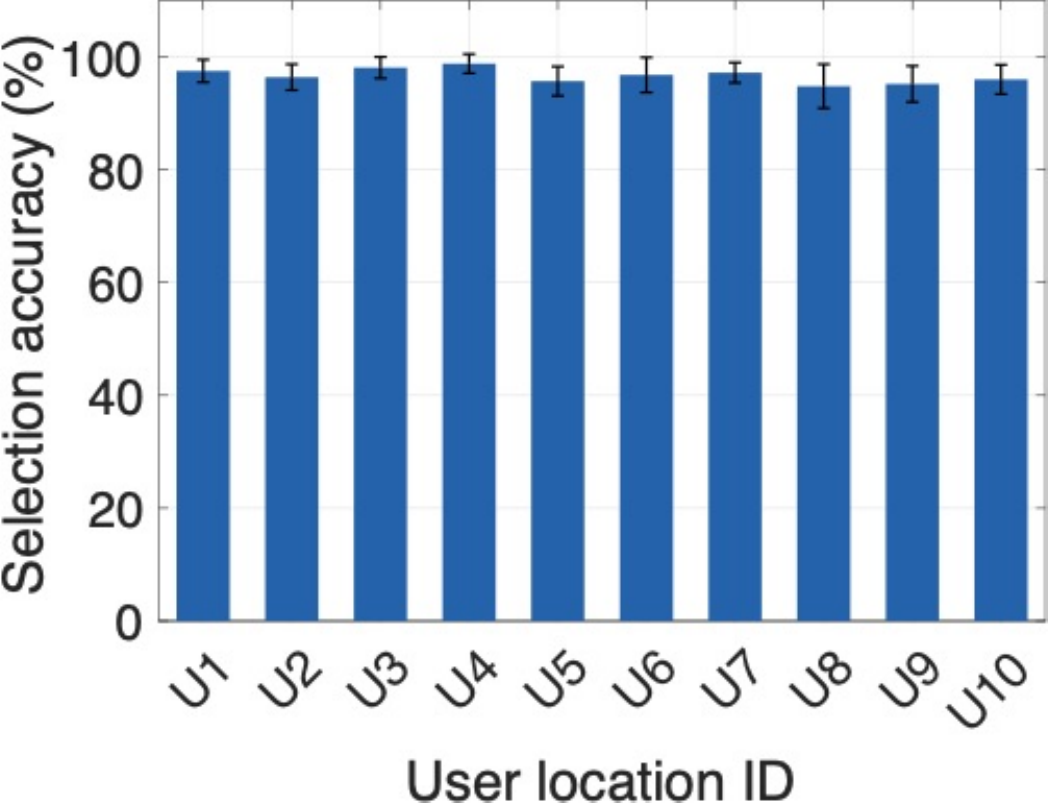}
        \captionsetup{justification=centering}
        \label{fig5_9c}
        \hspace{0in}
    }
    \vspace{-0.0em}
    \caption{Impact of anchor and user location.}
    \label{fig5_9}
\end{figure*}

\subsubsection{Impact of Anchor and User Location.} 
We further evaluate the impact of anchor location in S4~(the meeting room). As shown in Figure~\ref{fig5_9a}, we place the anchor at four different locations and repeat the above experiment. The result is shown in Figure~\ref{fig5_9b}. When the anchor is placed at L1 or L4, device-selection accuracy is slightly lower than at L2 or L3. These edge locations generally increase the user–anchor distance, which slightly increases the pointing-direction estimation error, as discussed in Section~\ref{sec513}. Nevertheless, selection accuracy remains above 96.1\% at all four anchor locations. Figure~\ref{fig5_9c} illustrates the device selection accuracy when the user is at ten different locations. Although we observe a slight decrease in accuracy when the user is farther from the anchor, the selection accuracy at each of the ten user locations remains above 94.8\%. These experiments demonstrate that our method maintains stable device selection performance across various anchor and user positions.

\subsubsection{Impact of Pointing Patterns}\label{sec535}
\begin{figure*}[!t]
    \centering
        \subfloat[Impact of user.]{
        \includegraphics[height=1.4in]{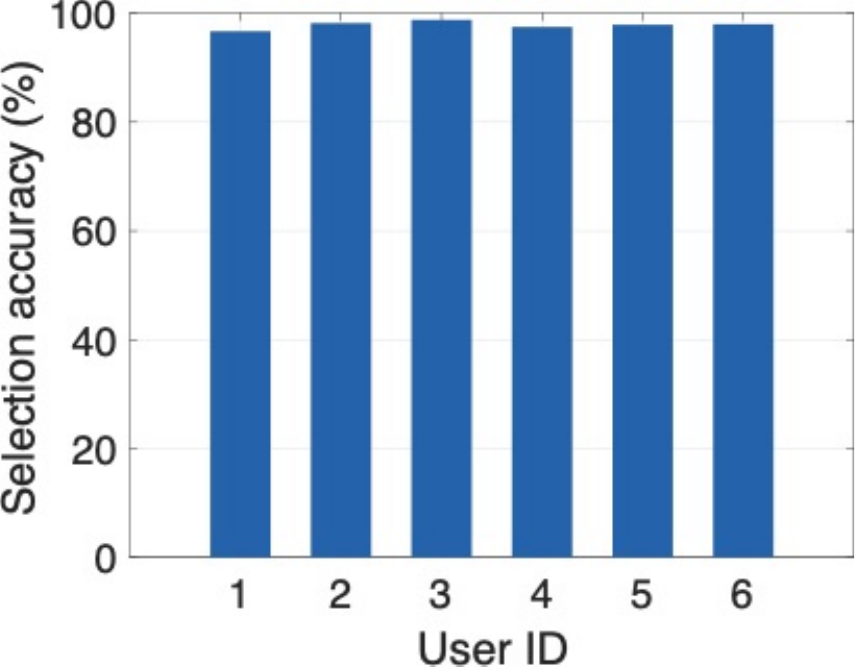}
        \label{fig5_10a}
        \hspace{0in}
    }
        \subfloat[Impact of pointing displacement.]{
        \includegraphics[height=1.4in]{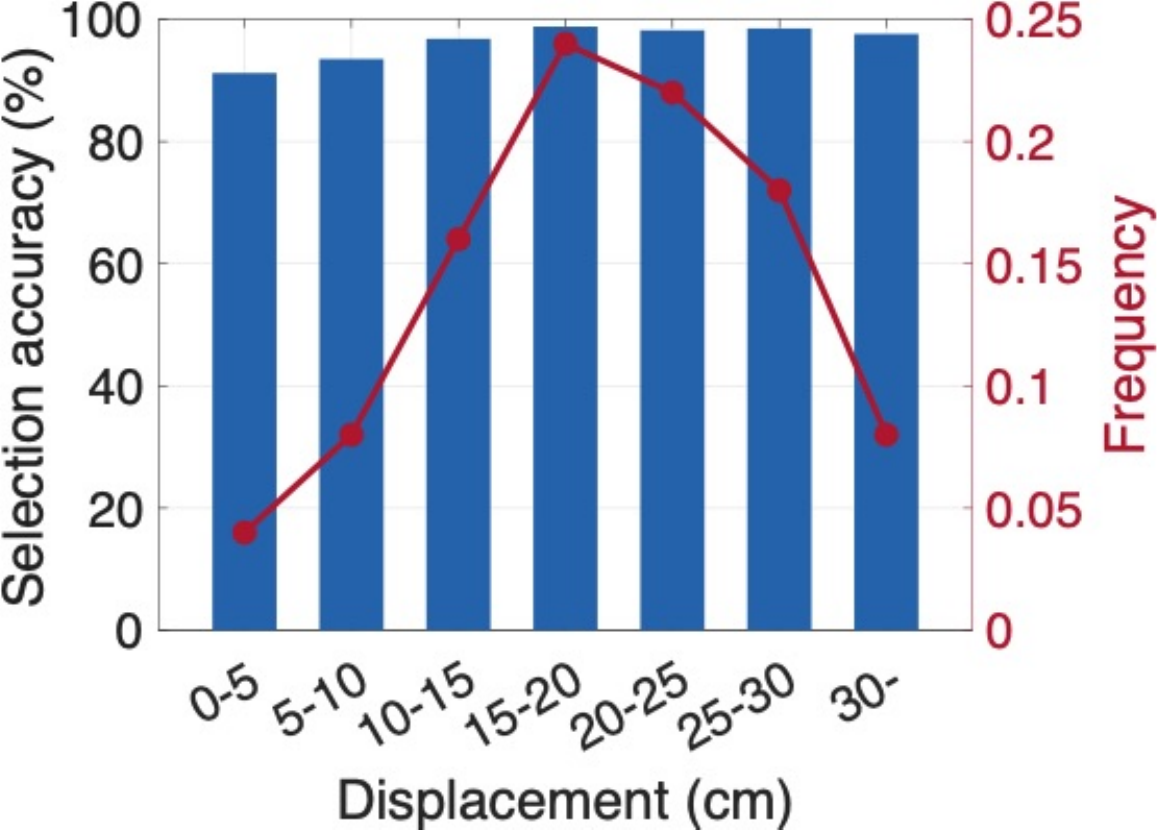}
        \label{fig5_10b}
        \hspace{0in}
    }
        \subfloat[Impact of pointing velocity.]{
        \includegraphics[height=1.4in]{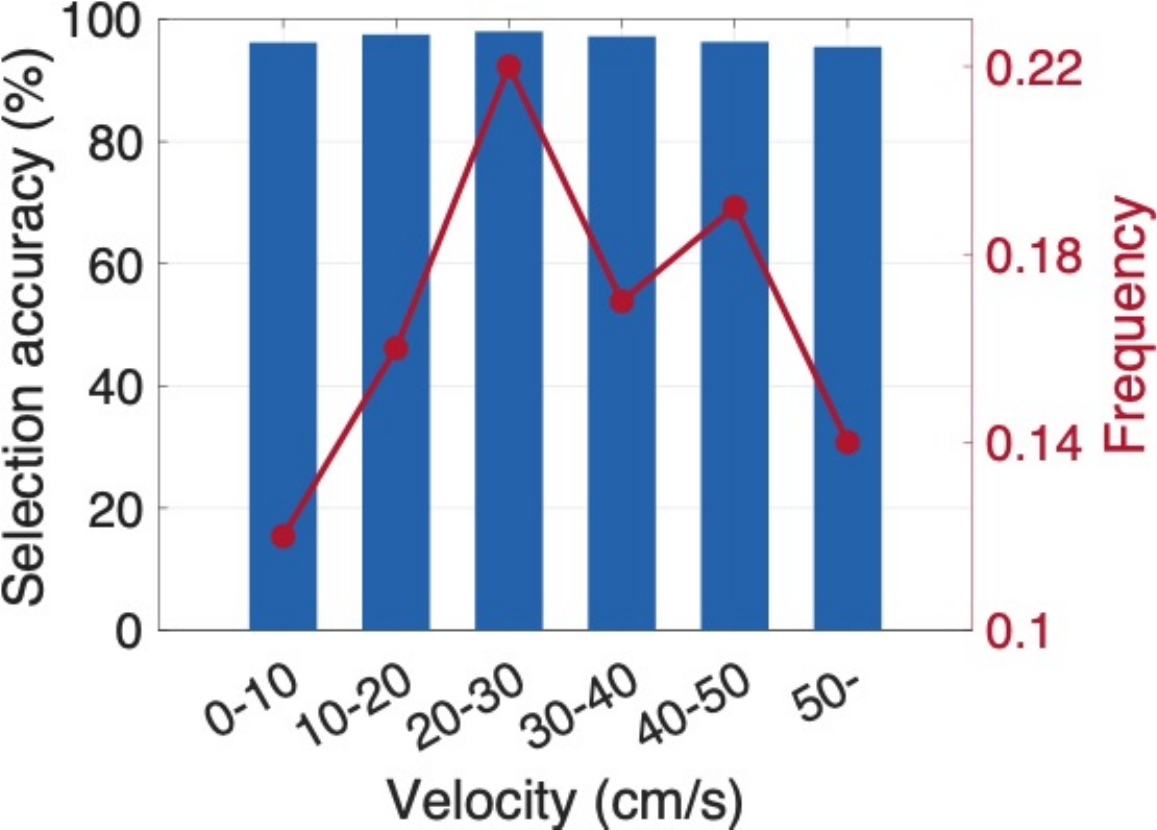}
        \captionsetup{justification=centering}
        \label{fig5_10c}
        \hspace{0in}
    }
    \vspace{-0.0em}
    \caption{Impact of user pointing patterns.}
    \label{fig5_10}
\end{figure*}

In practice, different users may exhibit various pointing patterns, including differences in displacement and velocity. Additionally, even the same individual's pointing behavior may vary at different times. Therefore, in this experiment, we evaluate the impact of different users, pointing displacements, and pointing velocities. In S1, we test the device selection accuracy of six participants (four males and two females, ranging in height from 160 cm to 180 cm, and aged between 25 to 55 years). Figure~\ref{fig5_10a} shows that the device-selection accuracy of every participant exceeds 95.4\%, with no clear differences across participants. Subsequently, we further analyze the data from all users to calculate the accuracy of selection across different pointing displacements and velocities. Displacements are divided into seven intervals, each 5 cm wide. The pointing velocity is computed from the recorded trajectory of the user device. Specifically, for each pointing trial, we calculate the average velocity as the distance between the starting and ending points of the trajectory divided by the total gesture duration. We then group the trials into six velocity intervals, each with a width of 10~cm/s, for statistical analysis. Figure~\ref{fig5_10b} shows the frequency of displacements occurring within each interval and the device selection accuracy for each. The majority of user displacements are concentrated between 10 cm and 30 cm, where the device selection accuracy remains above 95.4\%. We can observe that as displacements decrease below 10 cm, the accuracy of device selection declines (consistent with the findings in Section~\ref{sec514}). For velocities, as shown in Figure~\ref{fig5_10c}, there are no clear differences in device selection accuracy across different velocities.

\begin{figure}[!b]
    \centering
     \includegraphics[width=1.8in]{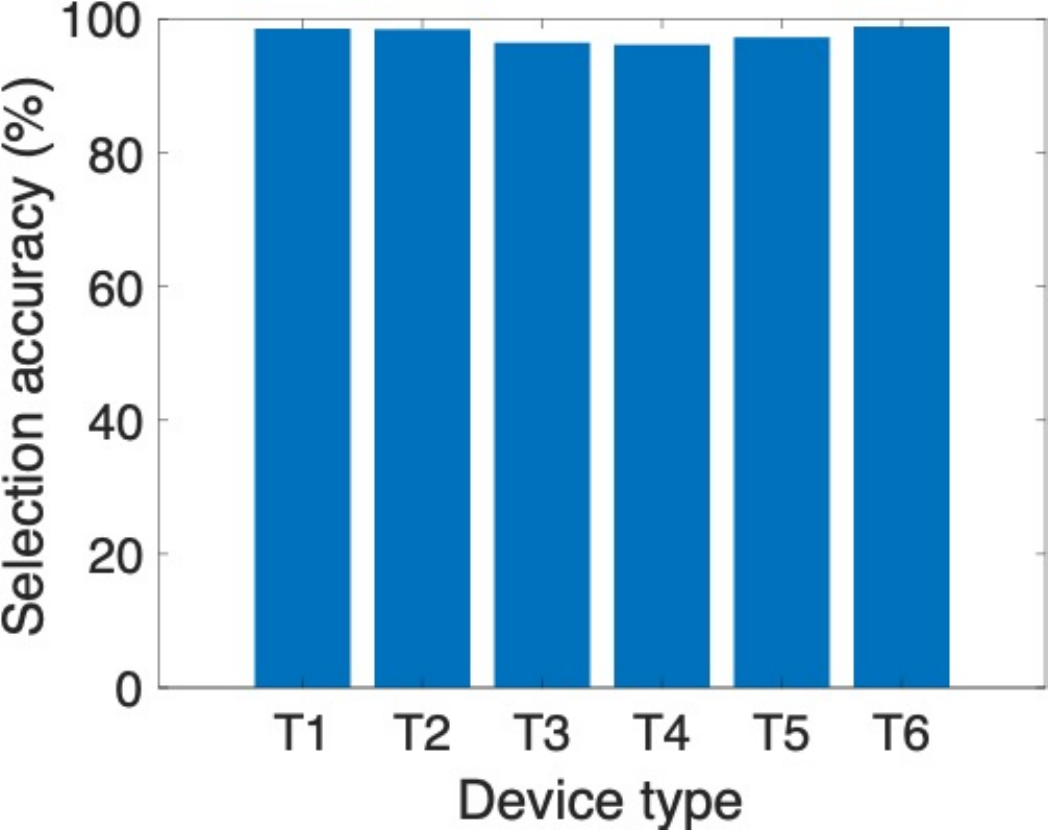}
    \caption{Impact of anchor and user-device types.}
    \label{fig5_11}
\end{figure}

\subsubsection{Impact of Anchor and User Device Types}
In this experiment, we evaluate the impact of different anchor and user-device types on IoT device selection accuracy. We use the same six participants as in Section~\ref{sec535} and conduct the experiment in S1. We test six anchor/user-device configurations: T1 uses an iPhone~11 as the anchor and an iPhone~12 Pro Max as the user device; T2 uses an iPhone~12 Pro Max as the anchor and an iPhone~11 as the user device; T3 uses an iPhone~11 as the anchor and an Apple Watch~S6 as the user device; T4 uses an iPhone~12 Pro Max as the anchor and an Apple Watch~S6 as the user device; T5 uses two Xiaomi Mix~4 smartphones as the anchor and user device; and T6 uses two NXP/Murata Type~2BP UWB evaluation boards as the anchor and user device. As shown in Figure~\ref{fig5_11}, the selection accuracy under all six configurations is above 96.2\%. These results indicate that PnPSelect can operate stably across different UWB hardware platforms and is not limited to a single anchor or user-device model.

\subsubsection{Computational Overhead}

We further evaluate the computational overhead of PnPSelect using a native implementation on an iPhone 12 Pro Max, processing recorded trajectories collected for the evaluation in Section~\ref{sec53}. For each complete trajectory, we measure the time required for UWB coordinate conversion, Kalman filtering, PCA-based pointing-direction estimation, and matching against ten candidate devices. These stages require an average of 0.0006 ms, 0.1216 ms, 0.0102 ms, and 0.0082 ms, respectively, with an average total processing time of approximately 0.141 ms. Thus, once the pointing gesture is complete and its trajectory measurements are available, PnPSelect can complete all processing stages and identify the target device in an average of 0.141 ms. In our tests, the phone provides position measurements in real time at approximately 50 Hz. Since the processing pipeline can run directly on the phone, transferring these measurements to an external computer is unnecessary, eliminating the associated data-transfer delay. These results demonstrate the computational feasibility of on-device processing. The measured time excludes gesture-completion detection and interface updates. A fully integrated deployment still requires connecting the live measurement stream to gesture segmentation and selection feedback, and evaluating the resulting end-to-end response latency.
\section{Discussion and Limitations}

\subsection{Design Trade-offs Between UWB and mmWave Sensing}
In this paper, we choose UWB as the sensing modality because our goal is to enable plug-and-play device selection with minimal infrastructure changes and direct sensing of the user’s intended target. We also note that mmWave sensing could be a promising direction for device selection in a contact-free manner, especially due to its high angular resolution and fine-grained motion sensitivity for gesture recognition~\cite{lien2016soli,liu2020real,palipana2021pantomime}. However, mmWave sensing systems are typically realized in a radar mode, where the transmitter and receiver are co-located, which is fundamentally different from our UWB setting. In such a radar mode, the received signals contain reflections not only from the user’s hand, but also from the entire body and surrounding objects in the environment. This introduces additional challenges, such as recognizing whether the user is intentionally performing a control gesture, isolating the hand motion relevant to pointing from other dynamic body movements, and suppressing environmental interference. In contrast, our UWB design directly tracks the user-held device in a separate transmitter-receiver configuration, making the problem formulation more direct and robust for plug-and-play device selection. We therefore believe that UWB provides a more suitable design point for the target scenario considered in this paper.

\subsection{System Generalization}

\subsubsection{Generalization Across User Devices} 

In this paper, we implement our system using the iPhone, Apple Watch, and Xiaomi MIX 4, due to the availability of detailed UWB reading extraction APIs. The latest UWB-equipped Google Pixel and Samsung smartphones also support the extraction of UWB data using Android APIs~\cite{androidapi}. Additionally, some wearable accessories also integrate UWB hardware. Looking ahead, we expect that in the future, an increasing number of commercial devices will incorporate UWB technology, including smart glasses, head-mounted displays, and smart rings, enabling the application of our design on a wide range of user devices. It should also be noted that the user device only needs one UWB antenna and does not need to support AoA estimation.

\subsubsection{Implementation of UWB Anchor} 
In this paper, we utilize a smartphone as the fixed UWB anchor in the environment. For real-world deployment, we envision two alternatives for the anchor. Firstly, the anchor can be realized by utilizing UWB-equipped smart devices (e.g., smart speakers and TVs). Alternatively, the UWB anchor can be implemented by leveraging third-party UWB chips and development boards~\cite{type2bp,type2hp,ubitraq}. These third-party UWB products can establish UWB connections with commercial smartphones and support data extraction, enabling our system design. Note that to estimate the 3D position of the user device, the anchor needs at least three UWB antennas arranged in an L-shaped layout. In this layout, one antenna pair is used to estimate the horizontal angle, while another antenna pair is used to estimate the elevation angle. By combining the two angle estimates with the ranging measurement, the anchor can obtain the 3D position of the user device. This is the minimum antenna requirement for the anchor in our system.

\subsection{Enhancing User Experience}\label{sec62}

During our experiments, we identified three limitations in our proposed method that could affect user experience in real-world applications. 

One limitation arises when estimating the location of a newly introduced IoT device. As introduced in Section~\ref{sec532}, the user is required to point at the device from two distinct positions to establish sufficient angular separation between the two pointing directions. Specifically, the angle difference between the two pointing directions should be greater than 20$\degree$. If the angular separation is too small, the system may struggle to accurately estimate the IoT device’s position. To improve user experience, we propose incorporating user guidance mechanisms that assist in ensuring an adequate separation between the two pointing positions. First, we can leverage the UWB anchor to track the user’s location change between the two pointing gestures. If the detected displacement is less than 1.4 m, the system can prompt the user to move farther before performing the second pointing action. Second, we believe that an alternative approach could involve introducing small artificial perturbations in the IoT device location estimation (Equation~\ref{eq:Eq4_3}). By analyzing the sensitivity of the IoT device’s estimated location to these perturbations, the system could determine whether the two pointing locations are too close. If minor errors in the input result in significant deviations in the estimated IoT device position~(Figure~\ref{fig:fig7_6d}), the system could infer that the two locations lack sufficient separation and notify the user to reposition themselves accordingly. We believe that incorporating these mechanisms can ensure the accuracy of IoT device location estimation while maintaining a positive user experience.

The second limitation occurs when multiple IoT devices are located close to each other, making it difficult for the system to unambiguously determine which device the user intends to select. This issue is particularly problematic when two IoT devices are positioned extremely close to each other, i.e., below the system’s spatial resolution, or when multiple devices lie along the user’s pointing direction. In such cases, relying solely on Equation~\ref{eq3_4} to select a single device may lead to incorrect selections, negatively impacting user experience. To address this, we propose an interactive selection mechanism that displays multiple candidate devices instead of automatically selecting one. When the system detects that more than one IoT device is aligned with the pointing direction, it does not immediately determine a single target device but instead presents all possible candidates on the user’s smartphone screen. This allows the user to manually confirm their selection, preventing accidental activations. In the example mentioned above, instead of automatically selecting a single device (D1 or D2), the application can display both D1 and D2, allowing the user to make a final selection. Compared to the traditional approach of displaying all IoT devices in the room (i.e., ten devices), presenting only the most probable two candidates can still reduce the complexity of the selection process. Additionally, this design helps prevent incorrect selections, thereby enhancing the overall user experience and minimizing frustration caused by unintended device activations.

The third limitation is that PnPSelect assumes reliable UWB localization between the anchor and the user device. It does not require perfect optical LoS at all times, and our evaluation in Section~5.1.6 shows that mild NLoS conditions can still be tolerated when stable UWB measurements are available. However, severe NLoS, such as complete body blockage or direct hand blockage of the phone’s UWB antenna area, can cause significant localization drift and make pointing-direction estimation unreliable. This is a common limitation of UWB-based and wireless-signal-based systems. In practice, this issue may be mitigated by placing the anchor to cover the main interaction area, avoiding covering the phone’s UWB antenna during use, monitoring UWB localization quality, and prompting the user to adjust posture or position when unstable measurements are detected.

\subsection{Potential Extension to Device Pairing and Commissioning}

Beyond device selection, \systemname{} may also be useful for device pairing and commissioning in smart home environments. One practical challenge in such workflows is to specify which physical device the user intends to set up, especially when multiple devices are present in the same space. In this context, \systemname{} could serve as a target specification mechanism: the user first points to the desired device from a distance, and the system then uses the identified target to bootstrap subsequent setup procedures, such as confirmation on the user device, authentication, network provisioning, or application-level binding. In this way, \systemname{} would simplify the initial device identification step, rather than replacing the entire pairing or commissioning pipeline. At the same time, we note that a complete pairing/commissioning workflow requires additional mechanisms beyond device identification, such as confirmation, security checks, and integration with existing setup protocols. Therefore, exploring how to integrate our design with practical setup protocols is an interesting direction for future work.

\subsection{Deployment in Smart Homes}
PnPSelect is primarily designed for room-level IoT device selection. In a typical household deployment, a natural configuration is to place one anchor in each room or functional zone, so that the anchor can provide reliable localization coverage for the user device within that local interaction area. For larger rooms or spaces with complex layouts, more than one anchor may be needed to maintain reliable coverage across the whole operating area. In such cases, the system can be extended to a zone-based deployment, where each anchor maintains the registered device set in its own coverage area. When the user moves across rooms or zones, the user device can automatically switch to the anchor that currently provides reliable localization coverage (e.g., based on the distance to each anchor or SNR), and PnPSelect can perform device selection using the corresponding room/zone-specific device map.

\subsection{Generalization to Other Gestures}

PnPSelect adopts a natural pointing gesture because it directly matches the interaction semantics of IoT device selection: the user moves the handheld device toward the intended target, and the dominant motion direction is expected to align with the target direction. However, the proposed framework is not fundamentally limited to this specific gesture, and some other works have shown this possibility~\cite{chenwandatch,zhang2023bleselect}. More generally, a predefined gesture can also be used if it produces a stable geometric feature that can be mapped to the intended pointing direction.

\begin{figure*}[!t]
    \centering
        \subfloat[Drawing a circle.]{
        \includegraphics[height=1.8in]{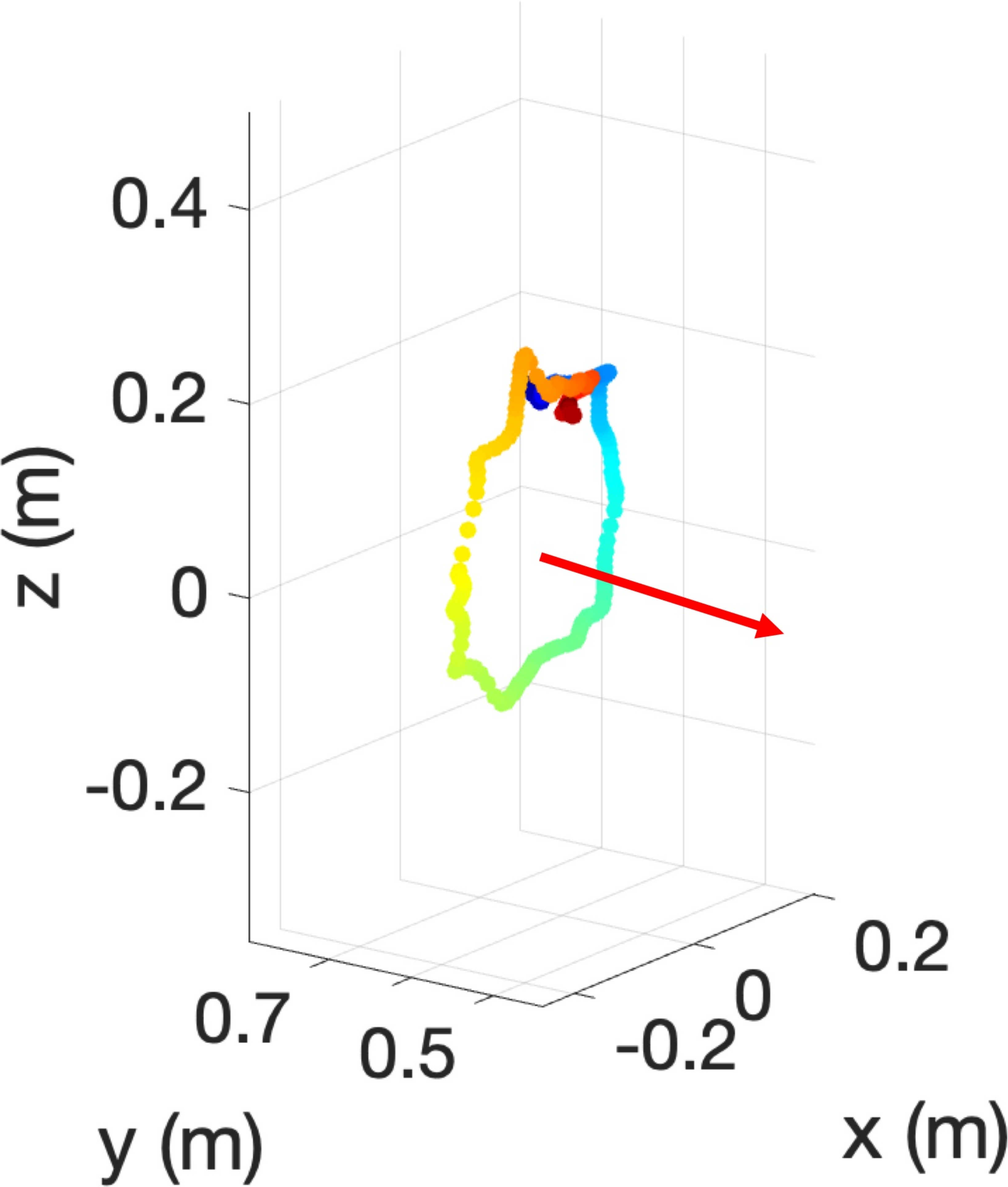}
        \label{fig5_n1}
        \hspace{0.2in}
    }
        \subfloat[Drawing a triangle.]{
        \includegraphics[height=1.8in]{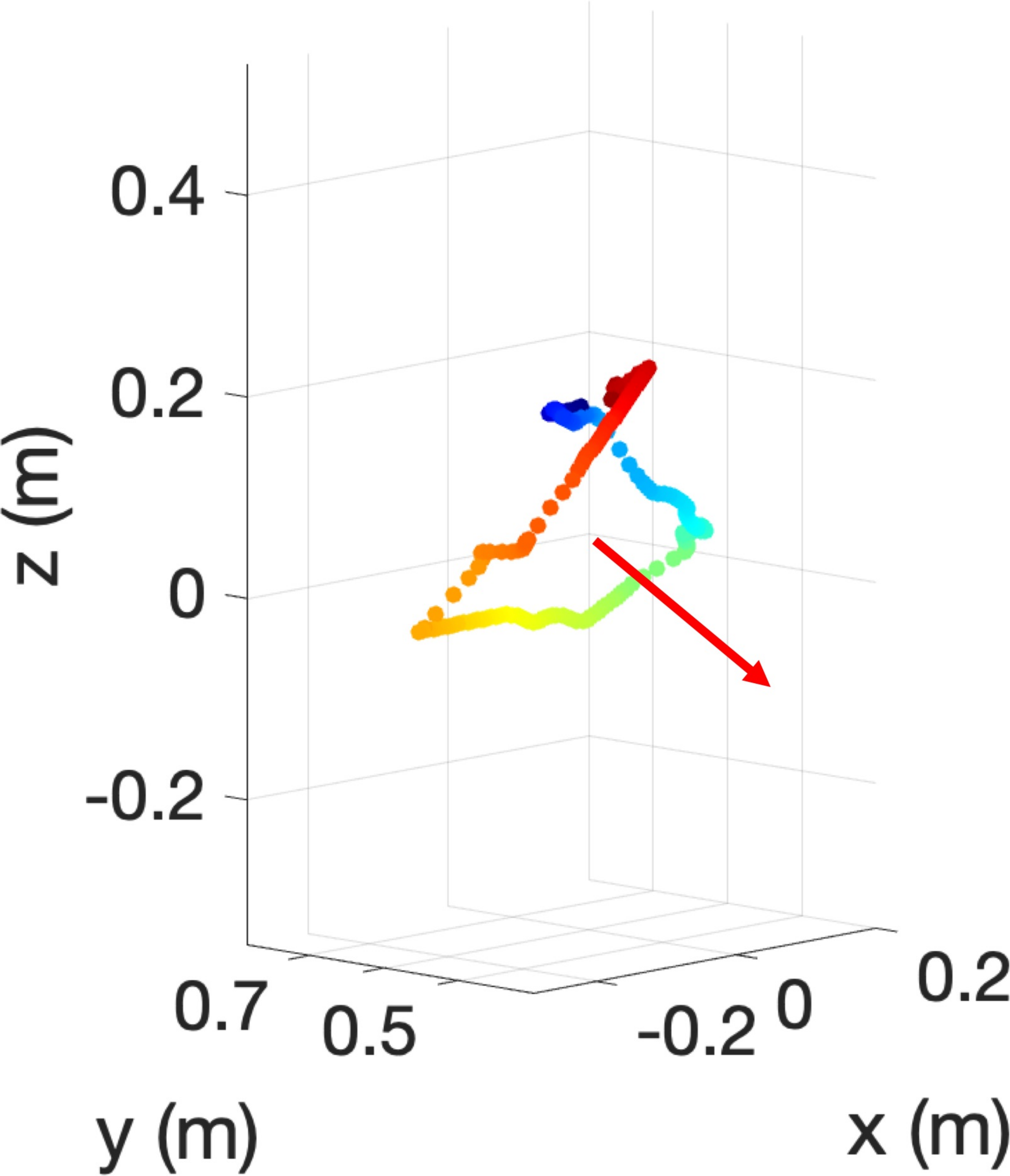}
        \captionsetup{justification=centering}
        \label{fig5_n2}
        \hspace{0in}
    }
    \caption{Illustration of pointing-direction estimation with alternative predefined gestures.}
    \label{fig6_1}
\end{figure*}

To examine this possibility, we conduct a preliminary study with two alternative predefined gestures: drawing a circle and drawing a triangle with the user device. Unlike the natural pointing gesture, whose direction can be estimated from the dominant trajectory direction, these gestures define a motion plane. For each gesture trajectory, we use the mean position of the trajectory points as the starting point of the pointing ray, and use PCA to estimate the principal motion plane. We then take the normal vector of this fitted plane as the pointing direction, as illustrated in Figure~\ref{fig6_1}. The results show that this approach can also recover a pointing direction and support device selection with reasonable accuracy. Therefore, we believe that extending PnPSelect to other predefined gestures is feasible when an appropriate gesture-to-direction mapping is defined. We also note a challenge, that is, alternative gestures require additional design to map the gesture geometry to the intended pointing direction. For example, when using the normal vector of a fitted trajectory plane, the system needs to distinguish between the two opposite normal directions. Therefore, we keep the natural pointing gesture as the default design because it provides a more direct and intuitive mapping to the target direction. At the same time, we view the design of alternative predefined gestures as a promising extension and leave a systematic exploration of this direction to future work.

\section{Conclusion}

In this paper, we present a novel design for IoT device selection using UWB-equipped commercial devices, eliminating the need to install dedicated hardware on each IoT device. The key innovation of our design lies in utilizing a single anchor to accurately estimate the user device's pointing direction. We also propose a novel design to address the practical challenge of registering the locations of IoT devices in the environment. To validate our system, we conduct comprehensive evaluations in both controlled lab settings and real-world environments. These results demonstrate the feasibility of PnPSelect for practical smart-home interaction.

\bibliographystyle{ACM-Reference-Format}
\bibliography{sample-base}

@String{Computing = "Computing" }

@String{Computer = "{IEEE} Computer" }

@String{Springer = "Springer-Verlag" }

@inproceedings{alanwar2017selecon,
  title={Selecon: Scalable iot device selection and control using hand gestures},
  author={Alanwar, Amr and Alzantot, Moustafa and Ho, Bo-Jhang and Martin, Paul and Srivastava, Mani},
  booktitle={Proceedings of the Second International Conference on Internet-of-Things Design and Implementation},
  pages={47--58},
  year={2017}
}

@inproceedings{wang2022faceori,
  title={FaceOri: Tracking Head Position and Orientation Using Ultrasonic Ranging on Earphones},
  author={Wang, Yuntao and Ding, Jiexin and Chatterjee, Ishan and Salemi Parizi, Farshid and Zhuang, Yuzhou and Yan, Yukang and Patel, Shwetak and Shi, Yuanchun},
  booktitle={Proceedings of the 2022 CHI Conference on Human Factors in Computing Systems},
  pages={1--12},
  year={2022}
}

@article{zhang2023bleselect,
  title={BLEselect: Gestural IoT Device Selection via Bluetooth Angle of Arrival Estimation from Smart Glasses},
  author={Zhang, Tengxiang and Lan, Zitong and Xu, Chenren and Li, Yanrong and Chen, Yiqiang},
  journal={Proceedings of the ACM on Interactive, Mobile, Wearable and Ubiquitous Technologies},
  volume={6},
  number={4},
  pages={1--28},
  year={2023},
  publisher={ACM New York, NY, USA}
}

@inproceedings{qin2023selecting,
  title={Selecting Real-World Objects via User-Perspective Phone Occlusion},
  author={Qin, Yue and Yu, Chun and Yao, Wentao and Yao, Jiachen and Liang, Chen and Weng, Yueting and Yan, Yukang and Shi, Yuanchun},
  booktitle={Proceedings of the 2023 CHI Conference on Human Factors in Computing Systems},
  pages={1--13},
  year={2023}
}

@article{sabath2005definition,
  title={Definition and classification of ultra-wideband signals and devices},
  author={Sabath, F and Mokole, EL and Samaddar, SN},
  journal={URSI Radio Science Bulletin},
  volume={2005},
  number={313},
  pages={12--26},
  year={2005},
  publisher={URSI}
}

@online{androidapi,
    year="2023",
    title="Ultra-wideband (UWB) communication",
url="https://developer.android.com/guide/topics/connectivity/uwb",
}

@online{xiaomi2020,
  year =         "2020",
  title =        "Xiaomi Introduces Groundbreaking UWB Technology",
  url =          "https://blog.mi.com/en/2020/10/13/xiaomi-introduces-groundbreaking-uwb-technology/",
}

@online{pixel6,
  year =         "2021",
  title =        "Google reiterates that Pixel 6 will have UWB as it works to expand support in Android 13",
  url =          "https://9to5google.com/2021/08/25/google-pixel-6-uwb-mention/",
}

@online{Samsung2021,
  year =         "2021",
  title =        "What is Ultra-Wideband (UWB) technology on Samsung Phones? How is it helpful?",
  url =          "https://www.smartprix.com/bytes/phones-with-uwb-ultrawideband-connectivity/",
}

@online{findmy,
    year="2021",
    title="Find your keys, wallet, and more with AirTag",
    url="https://support.apple.com/en-us/HT210967"
}

@online{nearbyapi,
    year="2023",
    title="Nearby Interaction",
    url="https://developer.apple.com/documentation/nearbyinteraction",
}

@online{qm33120w,
    year="2023",
    title="Qorvo QM33120W",
    url="https://www.qorvo.com/products/p/QM33120W",
}

@online{type2bp,
    year="2023",
    title="Murata Type2bp",
    url="https://www.murata.com/en-us/products/connectivitymodule/ultra-wide-band/nxp/type2bp",
}

@online{type2hp,
    year="2024",
    title="Murata Type2bp",
    url="https://www.murata.com/en-global/products/connectivitymodule/ultra-wide-band/nxp/type2hq",
}

@online{ubitraq,
    year="2020",
    title="Ubitraq",
    url="https://www.qorvo.com/innovation/ultra-wideband/partners/ubitraq",
}

@inproceedings{gao2022mom,
  title={MOM: Microphone based 3D Orientation Measurement},
  author={Gao, Zhihui and Li, Ang and Li, Dong and Liu, Jialin and Xiong, Jie and Wang, Yu and Li, Bing and Chen, Yiran},
  booktitle={2022 21st ACM/IEEE International Conference on Information Processing in Sensor Networks (IPSN)},
  pages={132--144},
  year={2022},
  organization={IEEE}
}

@article{heinrich2023smartphones,
  title={Smartphones with UWB: Evaluating the Accuracy and Reliability of UWB Ranging},
  author={Heinrich, Alexander and Krollmann, S{\"o}ren and Putz, Florentin and Hollick, Matthias},
  journal={arXiv preprint arXiv:2303.11220},
  year={2023}
}

@article{han2010nearest,
  title={Nearest approaches to multiple lines in n-dimensional space},
  author={Han, Lejia and Bancroft, John C},
  journal={Crewes Res. Rep},
  volume={22},
  pages={1--17},
  year={2010}
}

@inproceedings{xiao2017deus,
  title={Deus EM Machina: on-touch contextual functionality for smart IoT appliances},
  author={Xiao, Robert and Laput, Gierad and Zhang, Yang and Harrison, Chris},
  booktitle={Proceedings of the 2017 CHI Conference on Human Factors in Computing Systems},
  pages={4000--4008},
  year={2017}
}

@article{zhang2019facilitating,
  title={Facilitating Temporal Synchronous Target Selection through User Behavior Modeling},
  author={Zhang, Tengxiang and Yi, Xin and Wang, Ruolin and Gao, Jiayuan and Wang, Yuntao and Yu, Chun and Li, Simin and Shi, Yuanchun},
  journal={Proceedings of the ACM on Interactive, Mobile, Wearable and Ubiquitous Technologies},
  volume={3},
  number={4},
  pages={1--24},
  year={2019},
  publisher={ACM New York, NY, USA}
}

@article{zhang2018tap,
  title={Tap-to-pair: associating wireless devices with synchronous tapping},
  author={Zhang, Tengxiang and Yi, Xin and Wang, Ruolin and Wang, Yuntao and Yu, Chun and Lu, Yiqin and Shi, Yuanchun},
  journal={Proceedings of the ACM on Interactive, Mobile, Wearable and Ubiquitous Technologies},
  volume={2},
  number={4},
  pages={1--21},
  year={2018},
  publisher={ACM New York, NY, USA}
}

@inproceedings{verweij2017smart,
  title={Smart home control using motion matching and smart watches},
  author={Verweij, David and Esteves, Augusto and Khan, Vassilis-Javed and Bakker, Saskia},
  booktitle={Proceedings of the 2017 ACM International Conference on Interactive Surfaces and Spaces},
  pages={466--468},
  year={2017}
}

@inproceedings{aumi2013doplink,
  title={Doplink: Using the doppler effect for multi-device interaction},
  author={Aumi, Md Tanvir Islam and Gupta, Sidhant and Goel, Mayank and Larson, Eric and Patel, Shwetak},
  booktitle={Proceedings of the 2013 ACM international joint conference on Pervasive and ubiquitous computing},
  pages={583--586},
  year={2013}
}

@inproceedings{sun2013spartacus,
  title={Spartacus: spatially-aware interaction for mobile devices through energy-efficient audio sensing},
  author={Sun, Zheng and Purohit, Aveek and Bose, Raja and Zhang, Pei},
  booktitle={Proceeding of the 11th annual international conference on Mobile systems, applications, and services},
  pages={263--276},
  year={2013}
}

@article{chen2018snaplink,
  title={Snaplink: Fast and accurate vision-based appliance control in large commercial buildings},
  author={Chen, Kaifei and F{\"u}rst, Jonathan and Kolb, John and Kim, Hyung-Sin and Jin, Xin and Culler, David E and Katz, Randy H},
  journal={Proceedings of the ACM on Interactive, Mobile, Wearable and Ubiquitous Technologies},
  volume={1},
  number={4},
  pages={1--27},
  year={2018},
  publisher={ACM New York, NY, USA}
}

@inproceedings{de2016snap,
  title={Snap-to-it: A user-inspired platform for opportunistic device interactions},
  author={de Freitas, Adrian A and Nebeling, Michael and Chen, Xiang'Anthony' and Yang, Junrui and Karthikeyan Ranithangam, Akshaye Shreenithi Kirupa and Dey, Anind K},
  booktitle={Proceedings of the 2016 CHI Conference on Human Factors in Computing Systems},
  pages={5909--5920},
  year={2016}
}

@inproceedings{boring2010touch,
  title={Touch projector: mobile interaction through video},
  author={Boring, Sebastian and Baur, Dominikus and Butz, Andreas and Gustafson, Sean and Baudisch, Patrick},
  booktitle={Proceedings of the SIGCHI Conference on Human Factors in Computing Systems},
  pages={2287--2296},
  year={2010}
}

@inproceedings{vincent2013precise,
  title={Precise pointing techniques for handheld augmented reality},
  author={Vincent, Thomas and Nigay, Laurence and Kurata, Takeshi},
  booktitle={Human-Computer Interaction--INTERACT 2013: 14th IFIP TC 13 International Conference, Cape Town, South Africa, September 2-6, 2013, Proceedings, Part I 14},
  pages={122--139},
  year={2013},
  organization={Springer}
}

@inproceedings{mayer2020enhancing,
  title={Enhancing mobile voice assistants with worldgaze},
  author={Mayer, Sven and Laput, Gierad and Harrison, Chris},
  booktitle={Proceedings of the 2020 CHI Conference on Human Factors in Computing Systems},
  pages={1--10},
  year={2020}
}

@inproceedings{mayer2018effect,
  title={The effect of offset correction and cursor on mid-air pointing in real and virtual environments},
  author={Mayer, Sven and Schwind, Valentin and Schweigert, Robin and Henze, Niels},
  booktitle={Proceedings of the 2018 CHI Conference on Human Factors in Computing Systems},
  pages={1--13},
  year={2018}
}

@inproceedings{zhou2012ultra,
  title={Ultra low-power UWB-RFID system for precise location-aware applications},
  author={Zhou, Yuan and Law, Choi Look and Xia, Jingjing},
  booktitle={2012 IEEE Wireless Communications and Networking Conference Workshops (WCNCW)},
  pages={154--158},
  year={2012},
  organization={IEEE}
}

@article{zheng2023nn,
  title={NN-LCS: Neural Network and Linear Coordinate Solver Fusion Method for UWB Localization in Car Keyless Entry System},
  author={Zheng, Zengwei and Yan, Shuang and Sun, Lin and Shu, Hengxin and Zhou, Xiaowei},
  journal={Sensors},
  volume={23},
  number={5},
  pages={2694},
  year={2023},
  publisher={MDPI}
}

@inproceedings{ma2022involving,
  title={Involving ultra-wideband in consumer-level devices into the ecosystem of wireless sensing},
  author={Ma, Junqi and Chang, Zhaoxin and Zhang, Fusang and Xiong, Jie and Ni, Jiazhi and Jin, Beihong and Zhang, Daqing},
  booktitle={Proceedings of the 28th Annual International Conference on Mobile Computing And Networking},
  pages={758--760},
  year={2022}
}

@article{zhang2023embracing,
  title={Embracing Consumer-level UWB-equipped Devices for Fine-grained Wireless Sensing},
  author={Zhang, Fusang and Chang, Zhaoxin and Xiong, Jie and Ma, Junqi and Ni, Jiazhi and Zhang, Wenbo and Jin, Beihong and Zhang, Daqing},
  journal={Proceedings of the ACM on Interactive, Mobile, Wearable and Ubiquitous Technologies},
  volume={6},
  number={4},
  pages={1--27},
  year={2023},
  publisher={ACM New York, NY, USA}
}

@inproceedings{zhang2022mobi2sense,
  title={Mobi2Sense: empowering wireless sensing with mobility},
  author={Zhang, Fusang and Xiong, Jie and Chang, Zhaoxin and Ma, Junqi and Zhang, Daqing},
  booktitle={Proceedings of the 28th Annual International Conference on Mobile Computing And Networking},
  pages={268--281},
  year={2022}
}

@inproceedings{chen2021movi,
  title={MoVi-Fi: motion-robust vital signs waveform recovery via deep interpreted RF sensing},
  author={Chen, Zhe and Zheng, Tianyue and Cai, Chao and Luo, Jun},
  booktitle={Proceedings of the 27th Annual International Conference on Mobile Computing and Networking},
  pages={392--405},
  year={2021}
}

@inproceedings{chen2021octopus,
  title={Octopus: a practical and versatile wideband MIMO sensing platform},
  author={Chen, Zhe and Zheng, Tianyue and Luo, Jun},
  booktitle={Proceedings of the 27th Annual International Conference on Mobile Computing and Networking},
  pages={601--614},
  year={2021}
}

@inproceedings{li2021fine,
  title={Fine-grained respiration monitoring during overnight sleep using IR-UWB radar},
  author={Li, Siheng and Wang, Zhi and Zhang, Fusang and Jin, Beihong},
  booktitle={International Conference on Mobile and Ubiquitous Systems: Computing, Networking, and Services},
  pages={84--101},
  year={2021},
  organization={Springer}
}

@article{zheng2020v2ifi,
  title={V2iFi: In-vehicle vital sign monitoring via compact RF sensing},
  author={Zheng, Tianyue and Chen, Zhe and Cai, Chao and Luo, Jun and Zhang, Xu},
  journal={Proceedings of the ACM on Interactive, Mobile, Wearable and Ubiquitous Technologies},
  volume={4},
  number={2},
  pages={1--27},
  year={2020},
  publisher={ACM New York, NY, USA}
}

@article{yang2022vuloc,
  title={VULoc: Accurate UWB Localization for Countless Targets without Synchronization},
  author={Yang, Jing and Dong, BaiShun and Wang, Jiliang},
  journal={Proceedings of the ACM on Interactive, Mobile, Wearable and Ubiquitous Technologies},
  volume={6},
  number={3},
  pages={1--25},
  year={2022},
  publisher={ACM New York, NY, USA}
}

@article{wang2018research,
  title={Research on UWB positioning accuracy in warehouse environment},
  author={Wang, Zirui and Li, Shaoxian and Zhang, Zhengyuan and Lv, Fan and Hou, Yanzhao},
  journal={Procedia computer science},
  volume={131},
  pages={946--951},
  year={2018},
  publisher={Elsevier}
}

@inproceedings{sun2021idol,
  title={IDOL: Inertial deep orientation-estimation and localization},
  author={Sun, Scott and Melamed, Dennis and Kitani, Kris},
  booktitle={Proceedings of the AAAI Conference on Artificial Intelligence},
  volume={35},
  number={7},
  pages={6128--6137},
  year={2021}
}

@article{kalman1960new,
  title={A new approach to linear filtering and prediction problems},
  author={Kalman, Rudolph Emil},
  year={1960}
}

@article{strecker2023mr,
  title={MR Object Identification and Interaction: Fusing Object Situation Information from Heterogeneous Sources},
  author={Strecker, Jannis and Akhunov, Khakim and Carbone, Federico and Garc{\'\i}a, Kimberly and Bekta{\c{s}}, Kenan and Gomez, Andres and Mayer, Simon and Yildirim, Kasim Sinan},
  journal={Proceedings of the ACM on Interactive, Mobile, Wearable and Ubiquitous Technologies},
  volume={7},
  number={3},
  pages={1--26},
  year={2023},
  publisher={ACM New York, NY, USA}
}

@inproceedings{heydariaan2020anguloc,
  title={Anguloc: Concurrent angle of arrival estimation for indoor localization with uwb radios},
  author={Heydariaan, Milad and Dabirian, Hossein and Gnawali, Omprakash},
  booktitle={2020 16th International Conference on Distributed Computing in Sensor Systems (DCOSS)},
  pages={112--119},
  year={2020},
  organization={IEEE}
}

@article{zhao2021uloc,
  title={Uloc: Low-power, scalable and cm-accurate uwb-tag localization and tracking for indoor applications},
  author={Zhao, Minghui and Chang, Tyler and Arun, Aditya and Ayyalasomayajula, Roshan and Zhang, Chi and Bharadia, Dinesh},
  journal={Proceedings of the ACM on Interactive, Mobile, Wearable and Ubiquitous Technologies},
  volume={5},
  number={3},
  pages={1--31},
  year={2021},
  publisher={ACM New York, NY, USA}
}

@ARTICLE{6012487,
  author={},
  journal={IEEE Std 802.15.4-2011 (Revision of IEEE Std 802.15.4-2006)}, 
  title={IEEE Standard for Local and metropolitan area networks--Part 15.4: Low-Rate Wireless Personal Area Networks (LR-WPANs)}, 
  year={2011},
  volume={},
  number={},
  pages={1-314},
  doi={10.1109/IEEESTD.2011.6012487}}

@inproceedings{neirynck2016alternative,
  title={An alternative double-sided two-way ranging method},
  author={Neirynck, Dries and Luk, Eric and McLaughlin, Michael},
  booktitle={2016 13th workshop on positioning, navigation and communications (WPNC)},
  pages={1--4},
  year={2016},
  organization={IEEE}
}

@inproceedings{xu2006position,
  title={Position estimation using UWB TDOA measurements},
  author={Xu, Jun and Ma, Maode and Law, Choi Look},
  booktitle={2006 IEEE International Conference on Ultra-Wideband},
  pages={605--610},
  year={2006},
  organization={IEEE}
}

@inproceedings{ledergerber2015robot,
  title={A robot self-localization system using one-way ultra-wideband communication},
  author={Ledergerber, Anton and Hamer, Michael and D'Andrea, Raffaello},
  booktitle={2015 IEEE/RSJ International Conference on Intelligent Robots and Systems (IROS)},
  pages={3131--3137},
  year={2015},
  organization={IEEE}
}

@article{ma2024push,
  title={Push the Limit of Highly Accurate Ranging on Commercial UWB Devices},
  author={Ma, Junqi and Zhang, Fusang and Jin, Beihong and Su, Chang and Li, Siheng and Wang, Zhi and Ni, Jiazhi},
  journal={Proceedings of the ACM on Interactive, Mobile, Wearable and Ubiquitous Technologies},
  volume={8},
  number={2},
  pages={1--27},
  year={2024},
  publisher={ACM New York, NY, USA}
}

@inproceedings{kempke2016harmonium,
  title={Harmonium: Asymmetric, bandstitched UWB for fast, accurate, and robust indoor localization},
  author={Kempke, Benjamin and Pannuto, Pat and Dutta, Prabal},
  booktitle={2016 15th ACM/IEEE International Conference on Information Processing in Sensor Networks (IPSN)},
  pages={1--12},
  year={2016},
  organization={IEEE}
}

@inproceedings{arun2023xrloc,
  title={XRLoc: Accurate UWB localization to realize XR deployments},
  author={Arun, Aditya and Saruwatari, Shunsuke and Shah, Sureel and Bharadia, Dinesh},
  booktitle={Proceedings of the 21st ACM Conference on Embedded Networked Sensor Systems},
  pages={459--473},
  year={2023}
}

@article{palipana2021pantomime,
  title={Pantomime: Mid-air gesture recognition with sparse millimeter-wave radar point clouds},
  author={Palipana, Sameera and Salami, Dariush and Leiva, Luis A and Sigg, Stephan},
  journal={Proceedings of the ACM on interactive, mobile, wearable and ubiquitous technologies},
  volume={5},
  number={1},
  pages={1--27},
  year={2021},
  publisher={ACM New York, NY, USA}
}

@article{lien2016soli,
  title={Soli: Ubiquitous gesture sensing with millimeter wave radar},
  author={Lien, Jaime and Gillian, Nicholas and Karagozler, M Emre and Amihood, Patrick and Schwesig, Carsten and Olson, Erik and Raja, Hakim and Poupyrev, Ivan},
  journal={ACM Transactions on Graphics (ToG)},
  volume={35},
  number={4},
  pages={1--19},
  year={2016},
  publisher={ACM New York, NY, USA}
}

@article{liu2020real,
  title={Real-time arm gesture recognition in smart home scenarios via millimeter wave sensing},
  author={Liu, Haipeng and Wang, Yuheng and Zhou, Anfu and He, Hanyue and Wang, Wei and Wang, Kunpeng and Pan, Peilin and Lu, Yixuan and Liu, Liang and Ma, Huadong},
  journal={Proceedings of the ACM on interactive, mobile, wearable and ubiquitous technologies},
  volume={4},
  number={4},
  pages={1--28},
  year={2020},
  publisher={ACM New York, NY, USA}
}

@misc{huayuen_hybi305,
  author  = {{Shenzhen Huayuen Technology Co., Ltd.}},
  title   = {UWB Air-commanding Pen Tag | HY-BI305},
  url     = {https://www.huayuen.com/en/product/detail?id=55}
}

@misc{calterah_airmice_2025,
  author  = {{Calterah}},
  title   = {Control by Just Pointing! Calterah Dubhe UWB SoCs Enable Air Mice for Both Connectivity and Sensing},
  year    = {2025},
  month   = aug,
  day     = {19},
  url     = {https://www.calterah.com/en/company-news-detail/409},
  urldate = {2026-03-16}
}

@inproceedings{chang2024uwbori,
  title={Uwbori: Enabling accurate orientation estimation with ultra-wideband signals},
  author={Chang, Zhaoxin and Zhang, Fusang and Xiong, Jie and Xue, Xinyu and Wang, Zeyu and Jouaber, Badii and Zhang, Daqing},
  booktitle={2024 IEEE Smart World Congress (SWC)},
  pages={386--393},
  year={2024},
  organization={IEEE}
}

@article{chenwandatch,
  title={Wandatch: Infrastructure-Free Point-to-Command with Smartwatches and Speakers},
  author={Chen, Lin and Huang, Yandao and Qiu, Minghui and Zhong, Shuxin and Chen, Jun and Wu, Kaishun}
}

@online{gizchina_huawei_smart_screen_v5,
  author  = {{Gizchina}},
  title   = {Huawei Smart Screen V5},
  year    = {n.d.},
  url     = {https://www.gizchina.com/huawei/huawei-smart-screen-v5},
  urldate = {2026-05-09},
  note    = {Accessed online}
}

\end{document}